\documentclass[review]{elsarticle}

\usepackage{lineno,hyperref}
\modulolinenumbers[5]

\usepackage[T1]{fontenc}
\usepackage{ae,aecompl}
\usepackage{graphicx}   
\usepackage{amsmath}    
\usepackage{amssymb}    
\usepackage{soul}

\usepackage[utf8]{inputenc}
\usepackage{float}
\usepackage{booktabs}
\usepackage{multirow, threeparttable}

\usepackage{color}
\usepackage[table]{xcolor}
\definecolor{lightgray}{gray}{0.7}

\makeatletter
\def\convertto#1#2{\strip@pt\dimexpr #2*65536/\number\dimexpr 1#1}
\makeatother

\newcommand{\red}[1]{\textcolor{red}{#1}}
\newcommand{\ion}[2]{{\textrm{#1}}{\textrm{\sc #2}}}

\let\oldst\st
\renewcommand{\st}[1]{\textcolor{red}{\oldst{#1}}}

\journal{Astronomy and Computing}

\begin{document}

\begin{frontmatter}

\title{Photometric redshifts for the S-PLUS Survey: is machine learning up to the task?}





\author[IAGaddress]{E. V. R. Lima}
\cortext[mycorrespondingauthor]{Corresponding author}
\ead{erik.vini@usp.br}

\author[IAGaddress]{L. Sodré Jr.}
\author[CBPFaddress,Cefetaddress]{C. R. Bom}
\author[CBPFaddress]{G. S. M. Teixeira}

\author[IAGaddress]{L. Nakazono}
\author[IAGaddress]{M. L. Buzzo}

\author[IFaddress,UFRGSaddress]{C. Queiroz}
\author[IAGaddress]{F. R. Herpich}
\author[LaSerenaaddress]{J. L. Nilo Castellón}
\author[Polandaddress]{M. L. L. Dantas}
\author[UNIVAPaddress]{O. L. Dors}
\author[UEMaddress1,UEMaddress2,UFPRaddress]{R. C. T. Souza}
\author[Greeceaddress]{S. Akras}
\author[IAAaddress]{Y. Jiménez-Teja}

\author[UFSCadress]{A. Kanaan}
\author[NOAOaddress]{T. Ribeiro}
\author[GMTOaddress]{W. Schoennell}

\address[IAGaddress]{Universidade de São Paulo, Instituto de Astronomia, Geofísica e Ciências Atmosféricas, Rua do Matão 1226, CEP 05508-090, São Paulo, SP, Brazil}
\address[CBPFaddress]{Centro Brasileiro de Pesquisas Físicas, Rua Dr. Xavier Sigaud 150, CEP 22290-180, Rio de Janeiro, RJ, Brazil}
\address[Cefetaddress]{Centro Federal de Educação Tecnológica Celso Suckow da Fonseca, Rodovia Mário Covas, lote J2, quadra J Distrito Industrial de Itaguaí, CEP 23810-000, Itaguaí, RJ, Brazil}
\address[IFaddress]{Instituto de Física, Universidade de São Paulo, Rua do Matão 1371, CEP 05508-090, São Paulo, SP, Brazil}
\address[UFRGSaddress]{Departamento de Astronomia, Instituto de Física, Universidade Federal do Rio Grande do Sul, Av. Bento Gonçalves 9500, CEP 91501-970, Porto Alegre, RS, Brazil}
\address[LaSerenaaddress]{Astronomy Department, Universidad de La Serena. Avenida Juan Cisternas 1200. La Serena, Chile}
\address[Polandaddress]{Nicolaus Copernicus Astronomical Center, Polish Academy of Sciences, ul. Bartycka 18, 00-716, Warsaw, Poland}
\address[UNIVAPaddress]{Universidade do Vale do Paraíba, Av. Shishima Hifumi 2911, CEP 12244-000, São José dos Campos, SP, Brazil}
\address[UEMaddress1]{Universidade Estadual de Maringá, Programa de Pós-graduação em Ciência da Computação, Av. Colombo 5790, CEP 87020-900, Maringá, PR, Brazil}
\address[UEMaddress2]{Universidade Estadual de Maringá, Programa de Pós-graduação em Engenharia de Produção, Av. Colombo 5790, CEP 87020-900, Maringá, PR, Brazil}
\address[UFPRaddress]{Universidade Federal do Paraná, Campus Jandaia do Sul, Rua Doutor João Maximiano 426, CEP 86900-900, Jandaia do Sul, PR, Brazil}
\address[Greeceaddress]{Institute for Astronomy, Astrophysics, Space Applications and Remote Sensing, National Observatory of Athens, GR 15236 Penteli, Greece}
\address[IAAaddress]{Instituto de Astrofísica de Andalucía, Glorieta de la Astronomía s/n, 18008 Granada, Spain}

\address[UFSCadress]{Departamento de Física, Universidade Federal de Santa Catarina, CEP 88040-900, Florianópolis, SC, Brazil}
\address[NOAOaddress]{NOAO. 950 North Cherry Ave. Tucson, AZ 85719, United States}
\address[GMTOaddress]{GMTO Corporation, N. Halstead Street 465, Suite 250, Pasadena, CA 91107, United States}

\begin{abstract}
The Southern Photometric Local Universe Survey (S-PLUS) is a novel project that aims to map the Southern Hemisphere using a twelve filter system, comprising five broad-band SDSS-like filters and seven narrow-band filters optimized for important stellar features in the local universe. 

In this paper we use the photometry and morphological information from the first S-PLUS data release (S-PLUS DR1) cross-matched to unWISE data and spectroscopic redshifts from Sloan Digital Sky Survey DR15. We explore three different machine learning methods (Gaussian Processes with GPz and two Deep Learning models made with TensorFlow) and compare them with the currently used template-fitting method in the S-PLUS DR1 to address whether machine learning methods can take advantage of the twelve filter system for photometric redshift prediction. Using tests for accuracy for both single-point estimates such as the calculation of the scatter, bias, and outlier fraction, and probability distribution functions (PDFs) such as the Probability Integral Transform (PIT), the Continuous Ranked Probability Score (CRPS) and the Odds distribution, we conclude that a deep-learning method using a combination of a Bayesian Neural Network and a Mixture Density Network offers the most accurate photometric redshifts for the current test sample. It achieves single-point photometric redshifts with scatter ($\sigma_\text{NMAD}$) of 0.023, normalized bias of -0.001, and outlier fraction of 0.64\% for galaxies with \texttt{r\_auto} magnitudes between 16 and 21.

\end{abstract}


\end{frontmatter}


\newcommand\aap{A\&A}                
\let\astap=\aap                          
\newcommand\aapr{A\&ARv}             
\newcommand\aaps{A\&AS}              
\newcommand\actaa{Acta Astron.}      
\newcommand\afz{Afz}                 
\newcommand\aj{AJ}                   
\newcommand\ao{Appl. Opt.}           
\let\applopt=\ao                         
\newcommand\aplett{Astrophys.~Lett.} 
\newcommand\apj{ApJ}                 
\newcommand\apjl{ApJ}                
\let\apjlett=\apjl                       
\newcommand\apjs{ApJS}               
\let\apjsupp=\apjs                       
\newcommand\apss{Ap\&SS}             
\newcommand\araa{ARA\&A}             
\newcommand\arep{Astron. Rep.}       
\newcommand\aspc{ASP Conf. Ser.}     
\newcommand\azh{Azh}                 
\newcommand\baas{BAAS}               
\newcommand\bac{Bull. Astron. Inst. Czechoslovakia} 
\newcommand\bain{Bull. Astron. Inst. Netherlands} 
\newcommand\caa{Chinese Astron. Astrophys.} 
\newcommand\cjaa{Chinese J.~Astron. Astrophys.} 
\newcommand\fcp{Fundamentals Cosmic Phys.}  
\newcommand\gca{Geochimica Cosmochimica Acta}   
\newcommand\grl{Geophys. Res. Lett.} 
\newcommand\iaucirc{IAU~Circ.}       
\newcommand\icarus{Icarus}           
\newcommand\japa{J.~Astrophys. Astron.} 
\newcommand\jcap{J.~Cosmology Astropart. Phys.} 
\newcommand\jcp{J.~Chem.~Phys.}      
\newcommand\jgr{J.~Geophys.~Res.}    
\newcommand\jqsrt{J.~Quant. Spectrosc. Radiative Transfer} 
\newcommand\jrasc{J.~R.~Astron. Soc. Canada} 
\newcommand\memras{Mem.~RAS}         
\newcommand\memsai{Mem. Soc. Astron. Italiana} 
\newcommand\mnassa{MNASSA}           
\newcommand\mnras{MNRAS}             
\newcommand\na{New~Astron.}          
\newcommand\nar{New~Astron.~Rev.}    
\newcommand\nat{Nature}              
\newcommand\nphysa{Nuclear Phys.~A}  
\newcommand\pra{Phys. Rev.~A}        
\newcommand\prb{Phys. Rev.~B}        
\newcommand\prc{Phys. Rev.~C}        
\newcommand\prd{Phys. Rev.~D}        
\newcommand\pre{Phys. Rev.~E}        
\newcommand\prl{Phys. Rev.~Lett.}    
\newcommand\pasa{Publ. Astron. Soc. Australia}  
\newcommand\pasp{PASP}               
\newcommand\pasj{PASJ}               
\newcommand\physrep{Phys.~Rep.}      
\newcommand\physscr{Phys.~Scr.}      
\newcommand\planss{Planet. Space~Sci.} 
\newcommand\procspie{Proc.~SPIE}     
\newcommand\rmxaa{Rev. Mex. Astron. Astrofis.} 
\newcommand\qjras{QJRAS}             
\newcommand\sci{Science}             
\newcommand\skytel{Sky \& Telesc.}   
\newcommand\solphys{Sol.~Phys.}      
\newcommand\sovast{Soviet~Ast.}      
\newcommand\ssr{Space Sci. Rev.}     
\newcommand\zap{Z.~Astrophys.}       

\section{Introduction} \label{sec:introduction}

Recent developments in both computer science and engineering allowed astronomers to create projects with sky area coverage and photometric depths never seen before. Current surveys, such as the Dark Energy Survey \citep[DES,][]{DES} and the Kilo-Degree Survey \citep[KiDS,][]{KIDS},  and future wide field surveys, such as the Vera C. Rubin Observatory \citep[LSST,][]{LSST}, Nancy Grace Roman Space Telescope \citep{WFIRST} and Euclid survey \citep{EUCLID}, among others, are gathering and/or will be able to gather information on hundreds of millions of objects, and more efficient methods to determine redshifts need to be developed. Due to time and cost constraints, and limited spectroscopic capabilities for faint objects, the determination of spectroscopic redshifts ($z_\text{spec}$) in the current generation of spectroscopic surveys for most objects in the aforementioned surveys is unfeasible. Consequently, a new way to estimate distances to celestial objects has been developed in the last decades which rely only on much faster and cheaper photometric observations. These estimates are named photometric redshifts ($z_\text{phot}$).

Several ongoing projects, such as the Advanced Large Homogeneous Area Medium Band Redshift Astronomical Survey \citep[ALHAMBRA,][]{Molino2014}, the Cosmic Evolution Survey \citep[COSMOS,][]{Laigle2016}, Survey for High-z Absorption Red and Dead Sources \citep[SHARDS,][]{Barro2019}, Physics of the Accelerating Universe Survey \citep[PAUS,][]{Alarcon2021}, Javalambre Physics of the Accelerating Universe Survey (J-PAS, \citealt{JPAS}, \citealt{MiniJPAS}), the Javalambre-Photometric Local Universe Survey \citep[J-PLUS,][]{JPLUS}, and the Southern Photometric Local Universe Survey \citep[S-PLUS,][]{SPLUS}, are able to obtain accurate photometric redshifts due to the use of narrow-band filters. The filter systems of these surveys provide an increase in $z_\text{phot}$ accuracy by a factor of 4 when compared to the use of broad-band filters only \citep{AlbertoSPLUS}.

Redshifts are a measurement of fundamental importance for astronomy, allowing researchers to have insight on different fields of cosmology and astrophysics. Although $z_\text{phot}$s are not as precise as the spectroscopic counterpart, they still allow several studies, and previous works have shown what accuracy is necessary for different science cases. For instance, an error of less than $0.004(1+z_\text{spec})$ is necessary to constrain the Baryonic Acoustic Oscillations (BAO) (\citealt{BenitezBAO}, \citealt{Weinberg2013}), while photometric redshifts with $<1\%$ error can also be used on weak-lensing studies (\citealt{Hildebrandt2012}, \citealt{Kilbinger2013}, \citealt{Rozo2009}). This approach can also be used to find overdensities of faint galaxies \citep{Pharo2018}, or to constrain cosmological parameters using clusters \citep{Battye2003}.

Photometric redshifts can be estimated using three different approaches: template fitting, machine learning, and hybrid methods. Template fitting (TF) methods make use of the spectral energy distribution of an object and fit it to a template library, which can be synthetic or empirical, containing spectra of different types of galaxies at different redshifts. Methods that apply this approach are BPZ \citep{Benitez2000}, Le Phare (\citealt{Arnouts1999}, \citealt{Ilbert2006}), and EAZY \citep{Brammer2008}, among others.

Machine learning (ML) methods are based on the assumption that colours and/or magnitudes of galaxies are dependent of their redshift. Using some sort of data (tabular or images) split into training, validation and testing samples, it is possible to train a model to map a relation between these quantities. After a model is trained, it can be applied to a sample of galaxies to estimate their redshifts. Methods that use this approach are GPz \citep{Almosallam2016}, TPZ \citep{CarrascoKind2013}, ANNz \citep{Firth2003}, ANNz2 \citep{Sadeh2016} and deep learning models created with TensorFlow \citep{TensorFlow}, PyTorch \citep{PyTorch}, or Theano \citep{Theano}.

Both TF and ML methods have their advantages and disadvantages. The former needs, for example, well-calibrated, unbiased templates, as well as explicit assumptions about the dust extinction, and are more computationally expensive. The latter requires a training sample with a reasonable number of objects that is both complete and representative of the sample of interest and it is reliable in the range of input parameters of the training sample only. To explore the advantages of both approaches, a few hybrid methods have been developed, combining features from both procedures, such as data augmentation using model templates for machine learning codes \citep{Hoyle2015}, improving template fitting codes by the use of unsupervised self-organizing maps \citep{Speagle2017}, and the ZEBRA photometric redshift code, which uses a training set of spectroscopic redshifts to improve the robustness of template fitting $z_\text{phot}$ estimates \citep{Feldmann2006}.

In this work, we explore the performance of the machine learning methods for the determination of photometric redshifts for galaxies using the information of the twelve bands of the S-PLUS filter system, the ellipticity, the maximum surface brightness, and the full width at half maximum from its catalogue. We have chosen to compare Gaussian Processes with GPz and two Deep Learning models made with TensorFlow. We also compare our results with the template fitting method currently used for the determination of $z_\text{phot}$s in S-PLUS, the BPZ2 code \citep{AlbertoSPLUS}. With this work, we examine how these  machine learning methods perform with the inclusion of narrow-band filters and non-photometric information. 

This paper is organized as follows: in \S\ref{sec:data} we describe the S-PLUS survey and the data that will be used in this study. In \S\ref{sec:methods} we describe briefly the methods and the metrics applied to evaluate the quality of the estimates. In \S\ref{sec:results} we present our results and discussions and give our final remarks in \S\ref{sec:conclusions}.

\section{Data} \label{sec:data}

\begin{figure}
	\centering
		\includegraphics[width=\columnwidth]{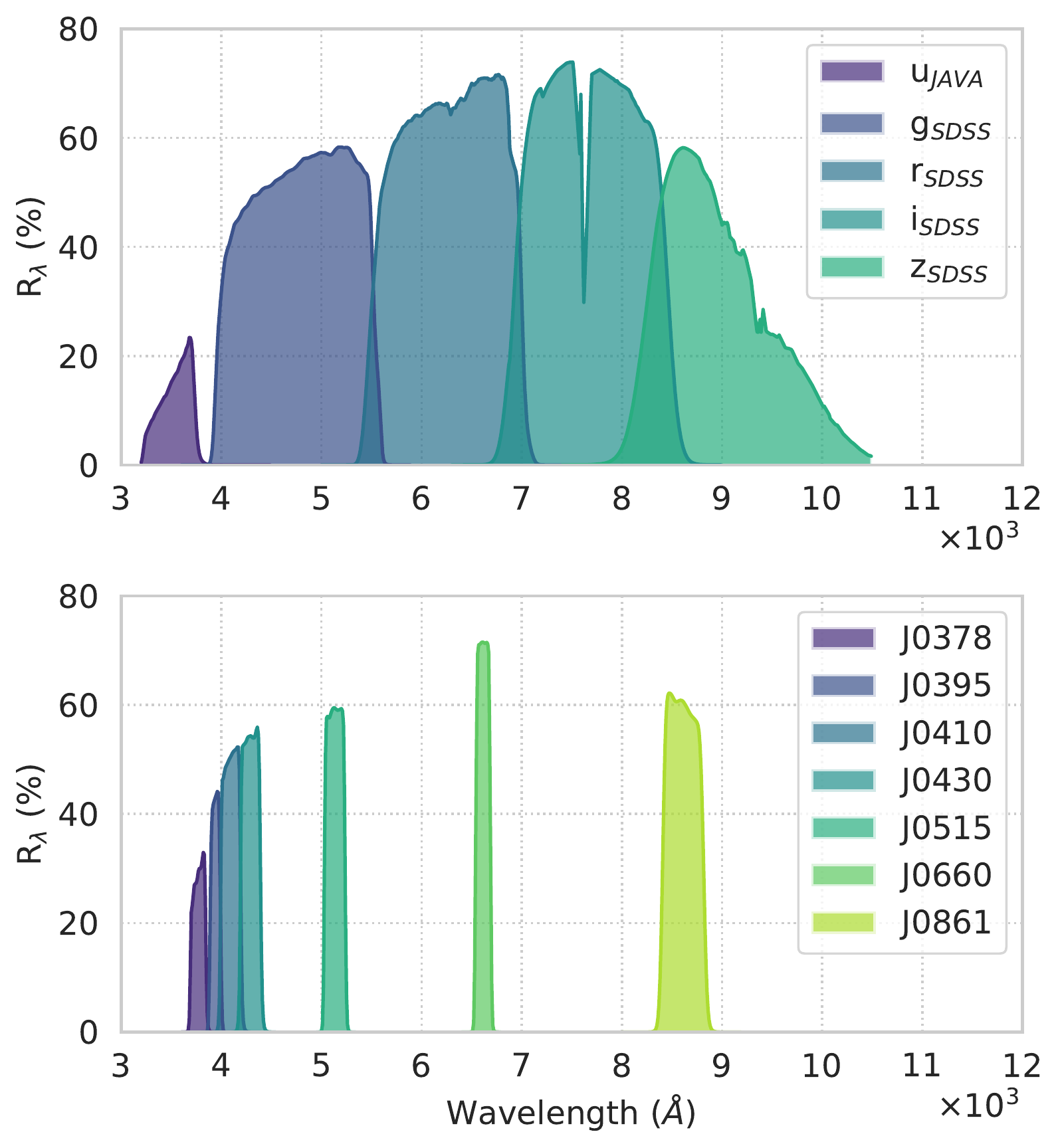}
		\caption{Transmission curves ($\text{R}_\lambda$) for the S-PLUS filter system. \textit{Top}: broad-band filters. \textit{Bottom}: narrow-band filters.}
		\label{fig:filters}
\end{figure}

In this work we used the photometric data from the S-PLUS Data Release 1 \citep{SPLUS} and the unWISE project \citep{unWISE}, cross-matched to the spectroscopic information from SDSS Data Release 15 \citep{SDSS15}. Further details about their cross-matching are given in Section \S\ref{sub:sample_construction}. 
The Southern Photometric Local Universe Survey \citep{SPLUS} is an ongoing project which aims to cover over 9000 square degrees in the southern sky, which will overlap with existing and future surveys (like DES, KiDS and LSST), using the same twelve bands adopted by the J-PLUS survey \citep{JPLUS}, covering the wavelength range between 3700{\AA} and 9000{\AA}. The S-PLUS survey started in August 2016 and is expected to finish in 2025 using a robotic 0.8 metres diameter telescope located at Cerro Tololo, Chile, named T80-S. The camera used in this project has a 2 square degrees field-of-view and is equipped with a 9k $\times$ 9k CCD with a 0.55 arcsec/pixel scale. The photometric system has 5 SDSS-like broadband filters ($u_\text{JAVA}$, $g_\text{SDSS}$, $r_\text{SDSS}$, $i_\text{SDSS}$ and $z_\text{SDSS}$) and 7 narrow-band filters (J0378, J0395, J0410, J0430, J0515, J0660 and J0861) designed specifically to explore important spectral features in the local universe, such as [\ion{O}{ii}] 3727,3729 {\AA}, Ca H+K, H$\delta$, G-band, Mgb triplet, H$\alpha$, and the Ca triplet, respectively \citep{Gruel2012,Marin-Franch2012}. This number of filters offers a great opportunity to estimate high-quality photometric redshifts in the local universe. The filter transmission curves\footnote{Obtained from \url{http://www.splus.iag.usp.br/instrumentation/}} are presented in Figure \ref{fig:filters}.

The S-PLUS DR1 covers 336 square degrees in Stripe-82. This region overlaps with other surveys, presenting a great opportunity to train supervised machine learning models due to the availability of spectroscopic redshifts. The typical depth of the survey is  \mbox{\texttt{r} $\sim 21$ AB} magnitudes for objects with signal-to-noise greater than 3.

The survey calibration pipeline provides 3 different types of magnitudes for each filter, following the output of \texttt{SExtractor} \citep{SExtractor}: the \texttt{auto\_restricted} magnitudes, which have the advantage of high signal-to-noise while integrating most of the light from galaxies; the petrosian (\texttt{petro}) magnitudes, with an aperture size defined by the distance from the centre of the galaxy to a region where the signal-to-noise drops to zero in the detection images, reducing possible contamination from neighbouring galaxies; and the \texttt{aper} magnitudes, computed in a fixed 3 arcsecond diameter aperture to match the SDSS fibre size. As discussed in \citep{Molino2017}, the \texttt{auto\_restricted}\footnote{Throughout this paper we will denote the \texttt{auto\_restricted} magnitudes as \texttt{auto} for simplicity.} magnitudes provide the best properties for photometric redshift determination due to the high signal-to-noise while containing most of the galaxy light, and so it was chosen to be used in this work. For further details on the calibration pipeline, please refer to \citet{SPLUS}.

We also use infrared photometry information from unWISE \citep{unWISE}. This project improves upon the products of AllWISE by taking advantage of deeper imaging and improved source detection in crowded regions. This allows unWISE photometry to be 0.7 mag deeper in the \texttt{W1} (3.4 $\mu$m) band and to detect three times more sources than AllWISE.

\subsection{Sample construction and pre-processing} \label{sub:sample_construction}

The supervised machine learning methods used in this work require a dataset containing the photometric and morphological information (the inputs) and the spectroscopic redshift (our target variable) to train the models. Since S-PLUS does not provide spectroscopic data, a cross-match with other catalogues is necessary to obtain this information.

We construct the samples for training, validation and testing in two steps:
\begin{enumerate}
	\item A cross-match between unWISE photometric data and SDSS DR15 data in the Stripe-82 is done. Since both catalogues share the same identification number for objects, a simple search between corresponding IDs is made with the \texttt{TOPCAT} \citep{TOPCAT} software. This creates a sample containing the spectroscopic redshift from SDSS and the \texttt{W1} (3.4 $\mu$m) and \texttt{W2} (4.6 $\mu$m) magnitudes from unWISE;
	\item The sample created in the previous step is cross-matched with S-PLUS. This match is done using equatorial coordinates (RA and Dec) with a maximum tolerance radius of 1 arcsecond. The resulting sample contains the information from S-PLUS, the \texttt{W1} and \texttt{W2} magnitudes of unWISE and the spectroscopic redshift of SDSS. Furthermore, since a comparison with the currently used method is required, this sample also contains the photometric redshifts obtained with BPZ2 \citep{AlbertoSPLUS}, available in the S-PLUS DR1.
\end{enumerate}

After these steps we have a sample containing all input features needed for training. In addition to S-PLUS \texttt{auto} magnitudes, we included in our input set the two unWISE magnitudes \texttt{W1} and \texttt{W2}, the colours calculated with respect to the \texttt{r\_auto} magnitude, and the \texttt{SExtractor} parameters full width at half maximum (\texttt{FWHM\_n}), maximum surface brightness (\texttt{MUMAX}), and the ellipticity of the objects. In total we use 30 different input features but do not take into account the uncertainties of the observations at this moment. We test which input configuration offers the most accurate photometric redshifts in Section \S\ref{sub:input_features_analysis}.

The quality of the results of a given machine learning method is dependent on the quality of the data itself. With this in mind, a pre-processing step is done to assure that only high quality data is used during the training process of the models. The constraints on different parameters are listed in Table \ref{tab:restric}. The parameter \texttt{nDet\_auto} refers to the number of filters in which a given object is detected and is used to study the effect of missing bands in $z_\text{phot}$ determination. \texttt{PhotoFlag} is related to the quality flag given by \texttt{SExtractor} and is important to avoid poor photometry, such as due to blended objects. \texttt{PROB\_GAL} is a value related to a star-galaxy separation described in \citet{MarcusSep} and we use it for the selection of galaxies. $z_\text{spec}$ is the spectroscopic redshift and it is also used to remove remaining stars misclassified as galaxies after \texttt{PROB\_GAL} constraints. \texttt{$z_\text{err}$} is the error in $z_\text{spec}$ and is used to assure redshift estimates of good quality. \texttt{class\_SDSS} is the spectroscopic class determined by SDSS, used to remove quasars. After applying these constraints, the total number of galaxies in our sample is 37559 when \mbox{\texttt{nDet\_auto} = 12} and increases to 57158 when \texttt{nDet\_auto} $\geqslant 8$ (see Figure \ref{fig:ndetsizeandmissperc}). 
\begin{table*}
\centering
\caption{Parameters, range of values, and source surveys.}
\label{tab:restric}
\vspace{0.2cm}
    \begin{tabular}{@{}llc@{}}\toprule
        Parameter          & Value                   & Survey                \\ \midrule
        Magnitude (\texttt{r\_auto}) & [16, 21]                & S-PLUS      \\ 
        \texttt{nDet\_auto}         & [8, 12]                 & S-PLUS       \\ 
        \texttt{PhotoFlag}          & $= 0$                   & S-PLUS       \\ 
        \texttt{PROB\_GAL}          & $\geqslant 0.5$              & S-PLUS  \\ 
        $z_\text{spec}$             & $\geqslant 1 \times 10^{-4}$ & SDSS    \\ 
        $z_\text{err}$     & $\leqslant 0.4$              & SDSS    \\ 
        \texttt{class\_SDSS}        & $\neq \text{QSO}$       & SDSS         \\
        \bottomrule
    \end{tabular}
\end{table*}

The sample is split randomly into training and testing samples, containing 70\% and 30\% of the total number of objects, respectively. These fractions are chosen to allow reasonable sample sizes to map the inputs to the output of the model and to allow a robust estimate of statistics as shown in Section \ref{sec:results}. Afterwards, during the training process, the initial training sample is further split into training and validation samples, for cross-validation.

A $K$-fold cross-validation, with $K=4$, consists on splitting a sample randomly into $K$ parts. During the training process, the sample containing 70\% of all objects is split into 75\% for training and 25\% for validation. With this approach we have a more robust estimate of the performance of our models while making a compromise between training time and the validation dataset size. It should be noticed that, for all methods, the same training, validation and testing samples are used so we can have a fair comparison.

\section{Methods} \label{sec:methods}

Is this section, we describe the methods and the choice of parameters used in this work, as well as the metrics adopted to evaluate the results.

\subsection{Gaussian Processes for photometric redshifts (GPz)} \label{sub:gaussian_processes_for_photometric_redshifts}

GPz \citep{Almosallam2016} is a photometric redshift estimation code based on Sparse Gaussian Processes, developed to overcome the limitations related to the application of Gaussian Processes to large samples and improve upon simple regression methods.

One of the disadvantages of simpler methods, such as a polynomial regression, is the poor adaptiveness to non-linear data. One way of overcoming this problem is using Gaussian Processes \citep[GPs,][]{GP1995}. GPs are a Bayesian, non-parametric and non-linear regression method that can adapt to the data. 

It works by finding a distribution of functions that can map the input space (e.g. magnitudes) to the output space (e.g. redshifts) based on the observations present in the training sample. This method relies on the concept of locality, in which two nearby objects in the input space will also be nearby in the output space (e.g. two galaxies with similar colours will likely have similar redshifts). This increases the prediction power of the method but also increases its complexity and computational demand (see \citealt{Gomes2017}).

One way to maintain the advantages of Gaussian Processes and reduce its computational cost was introduced in \citet{Almosallam2016}, through Sparse Gaussian Processes (SGPs). SGPs assumes that one can map the input-output relation using a reduced number of Gaussian kernel functions without losing accuracy. In the GPz algorithm, each kernel function can have its own hyperparameters in order to describe different densities and patterns of the training sample. The position of each kernel is defined in a way to maintain the distribution of the input data (for a more detailed discussion, see \citealt{Gomes2017}).

The GPz code can be adapted to a number of scientific problems by changing its configuration parameters. \citet{Almosallam2015} present a detailed description of each parameter and their effects on the model. After extensive testing, we choose the following set of parameters:
\begin{itemize}
	\item 2000 iterations;
	\item 200 basis functions;
	\item Joint: True;
	\item Heteroscedastic: True;
	\item Cost-sensitive learning method: Normalized;
	\item Decorrelate: False.
\end{itemize}

Both the iterations and basis functions are self-explanatory. ``Joint'' indicates whether the method should learn a simple linear regression prior from the data. The ``Heteroscedastic'' parameter is set to \texttt{True} so that the method will model the uncertainty in the $z_\text{phot}$ estimates as a function of the input (input-dependent noise). ``Cost-sensitive learning'' is related to the weight of each sample in the dataset, here defined as $1/(1+z_\text{sample})$. The final parameter, ``Decorrelate'', is set to \texttt{False}, so the method does not do a PCA pre-processing step. Further details about these and other parameters are presented in \citet{AlmosallamThesis}.

The optimization method is also changed from ``BFGS'' to ``L-BFGS-B'', for lower memory consumption, less time per iteration, and no loss in accuracy.

\subsection{Deep Learning} \label{sub:deep_learning}

In this work we compared two different deep-learning architectures, both developed using Tensorflow \citep{TensorFlow} and the Keras \citep{Keras} Application Programming Interface.

Both deep-learning models used in this work are based on Multi-Layer Perceptrons \citep[MLPs,][]{Rosenblatt}. In this approach, called Dense, all units in a given layer are connected to all units in the previous and following layer. The training process occurs via changes of the weights connecting those neurons based on feedback related to the errors in the predictions at each step of feedforward and backpropagation.

\subsubsection{Dense Network (DN)} \label{ssub:dense_network}

\begin{figure*}
    \centering
    \includegraphics[width=0.9\columnwidth]{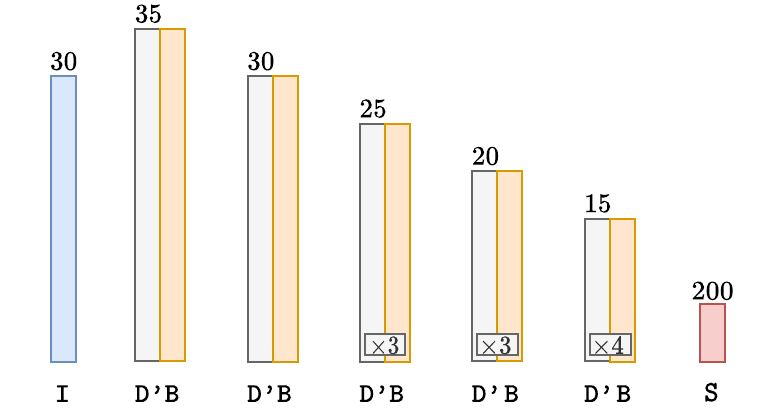}
    \caption{Architecture of the Dense Network. The input layer (\texttt{I}, in blue) is followed by blocks of \texttt{Dense} (\texttt{D'}, white) and \texttt{BatchNorm} (\texttt{B}, orange) layers, and the last layer (\texttt{S}, red) has a Softmax activation. The numbers represent the quantity of neurons in that layer. The numbers in the 3 last hidden blocks represent how many times these blocks are repeated.}
    \label{fig:densenet}
\end{figure*}

Our Dense Network (DN) architecture consists of an input layer with 30 neurons, one for each input feature (as described in Section \ref{sec:data}), followed by 12 hidden layers with 35 to 15 neurons (see Figure \ref{fig:densenet}), with Batch Normalizations \citep{BatchNorm} in between, and an output layer containing 200 neurons. All activations are ReLU \citep{ReLU} and the Categorical Cross-entropy loss is optimized with the \texttt{Nadam} \citep{Nadam} algorithm. This architecture has a total of 10081 parameters and was trained for 200 epochs.

The output layer has a Softmax \citep{Goodfellow-et-al-2016} function and contains 200 neurons to represent the photometric redshift PDFs, i.e., the redshift space is split into 200 bins of width 0.0067 between 0 and 1.34 (the maximum spectroscopic redshift in our sample). This approach converts the regression task into a classification task that gives the probability that a galaxy redshift is in a given bin.

\subsubsection{Mixture Density Network (MDN)} \label{ssub:mixture_density_network}
For this specific model, we have used Tensorflow Probability \citep{TFP}, since it allows the use of probabilistic reasoning in the Tensorflow framework, making the creation of Bayesian Neural Networks easier. They differ from traditional Neural Networks by the description of the network weights. While the latter uses trainable single values for the weights, the former uses trainable distributions. This change allows the estimation of the epistemic and aleatoric uncertainties \citep{Hullermeier2019}. The epistemic error is related to model uncertainty and its predictions, and can be reduced to zero as the sample size increases. The aleatoric uncertainty is due to measurement errors and cannot be reduced to zero as more samples are observed.

Mixture Density Networks (MDN) were introduced by \citet{MDN1994} and are defined as a combination of a conventional neural network and a mixture density model. They can represent arbitrary probability distributions in the same way that a neural network can represent arbitrary functions. This is interesting for photometric redshift prediction because this problem may have many solutions due to the color-morphology-redshift degeneracy (objects with different magnitudes having the same redshift) which is more apparent when using only broad-band photometry to estimate photometric redshifts, since they cannot provide a detailed sampling of the spectral features, and  the input-output mapping is better expressed with distributions than with single-point estimates. The use of narrow-band filters, such as those in S-PLUS, allow for a better sampling of the spectra and thus lessens the effect of this degeneracy, improving the $z_\text{phot}$ estimates.

Our MDN outputs 20 Gaussian functions, each described by three values: the mean, standard deviation and weight. The PDFs are the combination of these Gaussian functions (see Section \ref{sub:pdf_analysis}). The number of components was chosen as the one providing the best single-point and PDF metrics, described in Section \ref{sec:metrics}, after several tests.

\begin{figure}
    \centering
    \includegraphics[width=0.9\columnwidth]{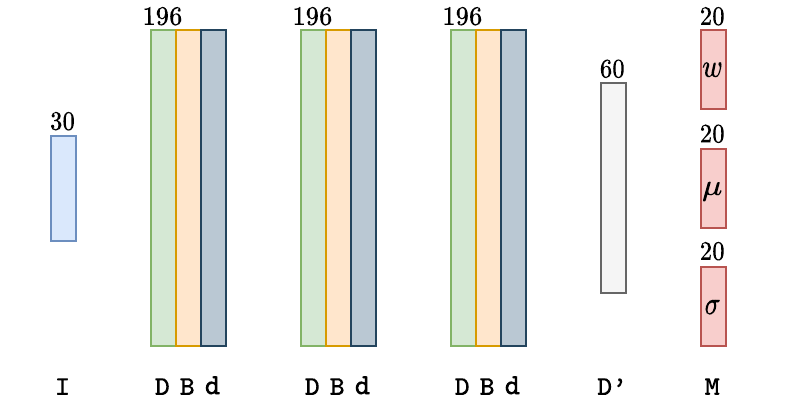}
    \caption{Architecture of the Mixture Density Network. The input layer (\texttt{I}) is represented in blue and is followed by blocks of \texttt{DenseVariational} (\texttt{D}, white), \texttt{BatchNorm} (\texttt{B}, orange), and \texttt{Dropout} (\texttt{d}, dark grey) layers, a Dense (\texttt{D'}, grey) layer, and the output layer (\texttt{M}, a \texttt{MixtureNormal}) is in red. The numbers represent the quantity of neurons in that layer.}
    \label{fig:mdnnet}
\end{figure}

The network  is composed of 3 blocks containing \texttt{DenseVariational} layers (provided by Tensorflow Probability) with 196 units, a Batch Normalization, and a Dropout layer with 10\% rate. The output layer is a \texttt{MixtureNormal}, which is responsible for returning the mean ($\mu$), standard deviation ($\sigma$) and weight ($w$) for 20 Gaussian distributions. The architecture is illustrated in Figure \ref{fig:mdnnet}. The Negative-Log Likelihood loss was minimized using the \texttt{Nadam} optimizer, with a learning rate of 0.001. All hidden layers have a Leaky ReLU activation and this model was trained for 500 epochs.

A \texttt{DenseVariational} layer is similar to the \texttt{Dense} layer, but replaces the description of the neuron weights with distributions. The \texttt{MixtureNormal} layer generates $N = 20$ independent distributions (the user can choose the number and type of distributions) to generate the PDF output. Further details can be found in the Tensorflow Probability documentation\footnote{\url{https://www.tensorflow.org/probability/api_docs/python/tfp}.}.

We also tested a shallower architecture for the DN model and a deeper architecture for the MDN model but we did not obtain satisfactory results.

\subsection{Evaluation Metrics} \label{sec:metrics}
\subsubsection{Single-point estimates}
To assess the single-point estimate accuracy of each method, we need to analyse three main quantities: the scatter of the $z_\text{phot}$ predictions, their bias, and the number of catastrophic outliers in our results. We present these metrics per bin of r-band magnitudes and spectroscopic redshift, as well as for the complete sample (without binning).

The scatter is computed as in \citet{Brammer2008}. It is robust, with low sensitivity to outliers:
\begin{align}
	\sigma_\text{NMAD} = 1.48 \times \text{median} \left( \frac{{\delta z} - \text{median}(\delta z)}{1+z_\text{spec}} \right),
	\label{eq:nmad}
\end{align}
where  $\delta z = (z_\text{phot} - z_\text{spec})$, $z_\text{phot}$ is the photometric redshift calculated by the code, and $z_\text{spec}$ is the true redshift of the object.

The systematic bias of the results, representing whether our predictions have a tendency to over- or underestimate the redshifts, is defined as:
\begin{align}
 	\overline{\mu} = \left\langle{\frac{\delta z}{1+z_\text{spec}}}\right\rangle,
 	\label{eq:bias}
 \end{align}
and the outliers are defined as the objects for which the condition
\begin{align}
	\frac{|\delta z|}{1+z_\text{spec}} > 0.15,
	\label{eq:outfrac}
\end{align} 
is satisfied \citep{Ilbert2006}. We will represent the outlier fraction by $\eta$.

Since we analyse these metrics as a function of bins of magnitude and/or redshift, it is worth mentioning that these quantities are calculated for all objects inside that bin and we take the median of the values. The magnitude bins have a width of 0.5 and the redshift bins have a width of 0.05. These values are chosen in a way to accommodate a reasonable number of objects inside each bin, so the statistics calculated in Section \ref{sec:results} are reliable.

\subsubsection{Probability distribution functions}
In order to verify the quality of the generated PDFs, we use three different metrics.

The Odds value, as introduced by \citet{Benitez2000}, can be used as an estimation of the accuracy of a PDF:
\begin{align}
  \label{eq:Odds}
  \text{Odds} = \int_{z_\text{peak}-\Delta z}^{z_\text{peak}+\Delta z} p(z) \text{d}z,
\end{align}
where $z_\text{peak}$ is the photometric redshift corresponding to the peak of the PDF, $\Delta z$ is the typical $z_\text{phot}$ accuracy in the local universe (see below), and $p(z)$ is the PDF itself. Since the Odds value represents the fraction of the PDF contained in the interval $z_\text{peak} \pm \Delta z$, a lower value will represent a broad distribution, while a higher value translates to a narrow PDF.

For this work, the value of $\Delta z$ is chosen as 0.02, which is the typical  $z_\text{phot}$ standard deviation for methods such as BPZ2 \citep{AlbertoSPLUS}. This particular value was adopted in order to provide a fair comparison between our our models and BPZ2.

The second tool used to check the quality of the PDFs is the Probability Integral Transform  histogram  \citep[PIT,][]{Polsterer2016}. Each PDF has a PIT value calculated as:
\begin{align}
  \label{eq:PIT}
  \text{PIT} = \int_{0}^{z_\text{spec}} p(z) \text{d}z,
\end{align}
where the shape of its distribution reveals details about the PDFs. The ideal case would be a uniform distribution. If the PDFs are too narrow, the PIT histogram will be concave (u-shaped). In the opposite situation, the histogram will be convex. Also, any bias in the PDFs appears as a slope in the distribution, positive in the case of underestimation and negative in the case of overestimation of the redshift.

We also calculate the Continuous Ranked Probability Score \citep[CRPS, ][]{CRPS, Polsterer2016}. The CRPS is a metric that measures the distance between the Cumulative Distribution Function (CDF) of a given PDF and its spectroscopic redshift, described as a step function. It is calculated as:
\begin{align*}
    \text{CRPS}_i = \int_{-\infty}^{+\infty} \left[ \text{CDF}_i(z) - \text{CDF}_\text{spec}(z) \right]^2 \text{d}z,
\end{align*}
where the subscript $i$ denotes a specific galaxy, $\text{CDF}_i$ is the cumulative distribution function of the PDF of $i$, and $\text{CDF}_\text{spec}$ is the cumulative distribution function of spectroscopic redshift of $i$ and is defined based on the Heaviside step-function:
\begin{align*}
    \text{CDF}_\text{spec}(z) = H(z - z_\text{spec}) = 
    \begin{cases}
      0 \text{ for } z < z_\text{spec}, \\
      1 \text{ for } z \geqslant z_\text{spec}.\\
    \end{cases} 
\end{align*}

\section{Results} \label{sec:results}

Now we present and discuss the results for $z_\text{phot}$ point estimates and PDFs for GPz, DN, and MDN, comparing them with BPZ2. First we verify the impact of missing-band values, due to non-detections or non-observations, on the training process. Then we analyse how the choice of input features affects our results. After selecting the optimal input features, we compare the single-point estimate results from all methods and check the quality of their PDFs. When available, the shaded regions in the plots shown in this section represent the standard deviation from the \mbox{4-Fold} Cross Validation, except for the BPZ2 method, for which the uncertainties were calculated using bootstrapping.

\subsection{Missing feature analysis} \label{sub:missing_feature_analysis}

\begin{figure}
    \centering
    \includegraphics[width=\columnwidth]{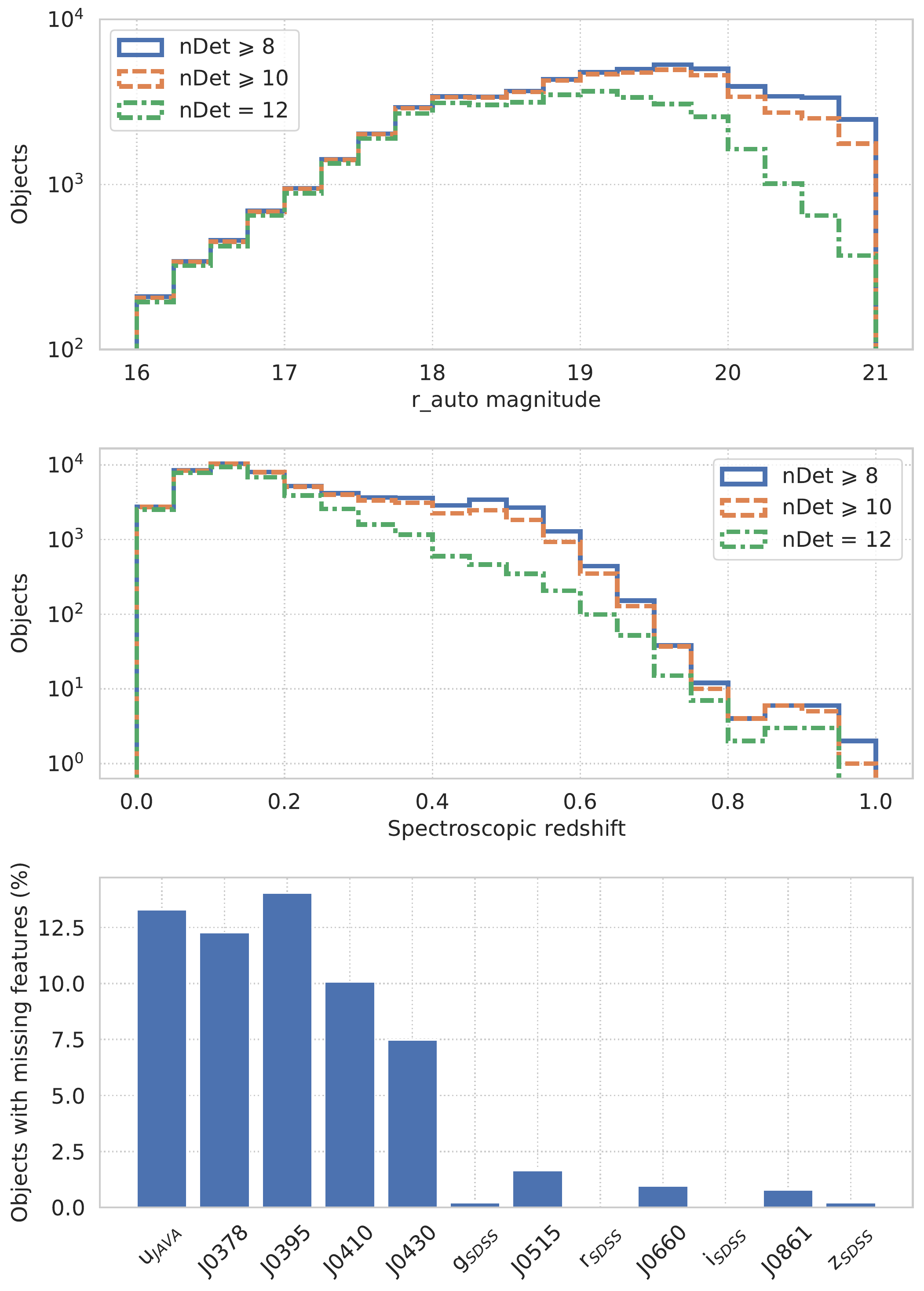}
    \caption{\textit{Top}: \texttt{r\_auto} distribution for three values of \texttt{nDet\_auto}. \textit{Centre}: $z_\text{spec}$ distribution for three values of \texttt{nDet\_auto}. \textit{Bottom}: Fraction of objects with missing measurements for each filter in our training sample.}
    \label{fig:ndetsizeandmissperc}
\end{figure}

In multiband surveys, such as S-PLUS, not all objects are always detected in all filters, and the impact of objects with missing photometry in the $z_\text{phot}$ determination should be addressed.  Figure \ref{fig:ndetsizeandmissperc} (bottom) shows the fraction of galaxies in our sample with missing photometry for the 12 bands of S-PLUS photometric system. Due to galaxy spectra and filter sensitivity, most of the missing bands are in the blue side of the spectrum.

For this analysis, we trained the models with constrained   values  of \texttt{nDet\_auto}, i.e., considering samples with \texttt{nDet\_auto} $\geqslant 8,~\cdots,~12$. Following  \citet{CholletDL}, we replace missing values by zeroes. 

\begin{figure*}
    \centering
    \includegraphics[width=\textwidth]{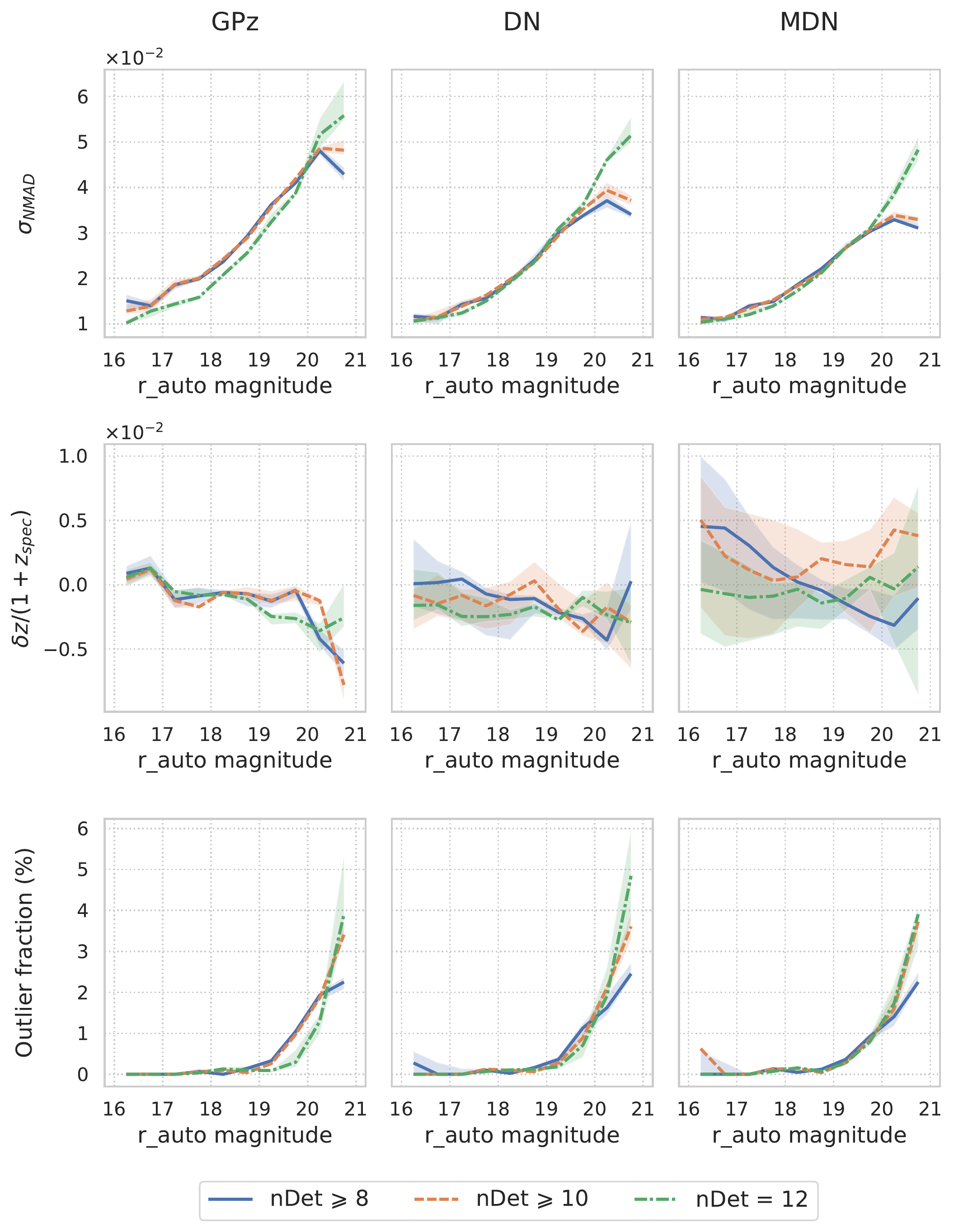}
    \caption{Results for the scatter, bias, and outlier fraction for each method (columns) as a function of the magnitude in the \texttt{r}-band for different \texttt{nDet\_auto} values. The shaded areas represent the uncertainties obtained from the 4-fold cross validation.}
    \label{fig:ndetmags}
\end{figure*}

Figure \ref{fig:ndetmags} presents the performance of the machine learning methods as a function of \texttt{r\_auto} for different values of \texttt{nDet\_auto}. We can notice, somewhat surprisingly, that the performance increases as \texttt{nDet\_auto} decreases. The improvement is more noticeable in the faint-end, where most of the objects with missing information are located. This can be explained by two major factors: first, as we ease the restrictions on \texttt{nDet\_auto} more objects are included in the training sample, as described in Section \ref{sub:sample_construction}. Second, even though these objects were not detected in some filters (mostly in the blue; Figure \ref{fig:ndetsizeandmissperc}), the information contained in the redder bands allow for a better input-output mapping during the training process, increasing the generalization capacity of the models \citep{CholletDL}. Notice that, in this section, the training, validation, and testing samples change due to  different  (\texttt{nDet\_auto}) values, but all models shared the same datasets in each case. Due to these results, the analyses in the following sections are done using all objects that have \texttt{nDet\_auto} $\geqslant$ 8.

\subsection{Input features analysis} \label{sub:input_features_analysis}

We now analyse which combination of input features provide the most accurate photometric redshifts. 
 
\begin{figure*}
    \centering
    \includegraphics[height=.80\textheight]{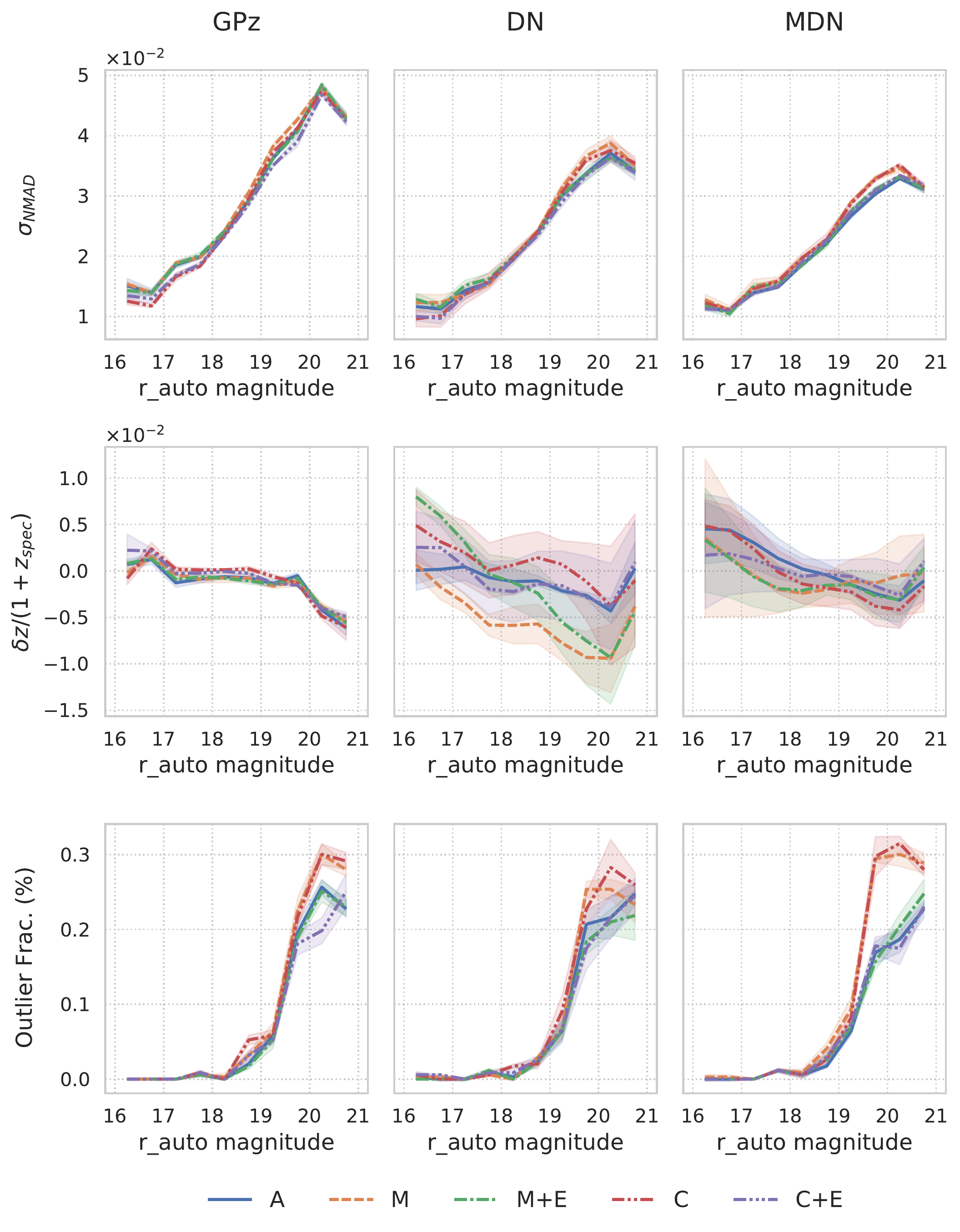}
    \caption{Comparison of the results as a function of the \texttt{r\_auto} magnitude for the scatter, bias, and outlier fraction for the three  machine learning models, using different input feature configurations as described in Table \ref{tab:inputs}.}
    \label{fig:inputfeatR}
\end{figure*}

The configuration for each input combination is described in Table \ref{tab:inputs}. The results of this section are shown in Figure \ref{fig:inputfeatR} as a function of \texttt{r\_auto} and in Table \ref{tab:inputresults} for the entire sample without binning. The DN, MDN, and GPz models perform better when trained using all input features available. For this reason, and to maintain consistency between the methods, we present only the results for the models trained using the ``A'' configuration in the following sections.

\begin{table*}
\centering
\caption{Name of samples and the input features considered for the tests described in Section \ref{sub:input_features_analysis}.}
\label{tab:inputs}
\vspace{0.2cm}
\begin{threeparttable}
    \begin{tabular}{@{}lp{2cm}p{6cm}@{}}
        \toprule
        Sample                                     & Notation                   & Input Features \tnote{a}                                                                  \\ \midrule
        \multirow[t]{3}{*}{All}               & \multirow[t]{3}{*}{A} & \texttt{u, J0378, J0395, J0410, J0430, g, J0515, r, J0660, i, J0861, z, W1, W2}      \\ \arrayrulecolor{lightgray}\cmidrule(l{6pt}){3-3}\arrayrulecolor{black}
        ~                                          & ~                            & \texttt{u$-$r, 378$-$r, 395$-$r, 410$-$r, 430$-$r, g$-$r, 515$-$r, r$-$660, r$-$i, r$-$861, r$-$z, r$-$W1, r$-$W2} \\ \arrayrulecolor{lightgray}\cmidrule(l{6pt}){3-3}\arrayrulecolor{black}
        ~                                          & ~                            & \texttt{FWHM\_n, MUMAX, Ellipticity}                                                     \\ \cmidrule{1-3}
        \multirow[t]{1}{*}{Magnitudes}                & \multirow[t]{1}{*}{M}           & \texttt{u, J0378, J0395, J0410, J0430, g, J0515, r, J0660, i, J0861, z, W1, W2}      \\ \cmidrule{1-3}
        \multirow[t]{2}{*}{Magnitudes+Extras} & \multirow[t]{2}{*}{M+E} & \texttt{u, J0378, J0395, J0410, J0430, g, J0515, r, J0660, i, J0861, z, W1, W2}      \\ \arrayrulecolor{lightgray}\cmidrule(l{6pt}){3-3}\arrayrulecolor{black}
        ~                                          & ~                            & \texttt{FWHM\_n, MUMAX, Ellipticity}                                                     \\ \cmidrule{1-3}
        \multirow[t]{2}{*}{Colours}           & \multirow[t]{2}{*}{C}   & \texttt{r}                                                                               \\ \arrayrulecolor{lightgray}\cmidrule(l{6pt}){3-3}\arrayrulecolor{black}
        ~                                          & ~                            & \texttt{u$-$r, 378$-$r, 395$-$r, 410$-$r, 430$-$r, g$-$r, 515$-$r, r$-$660, r$-$i, r$-$861, r$-$z, r$-$W1, r$-$W2} \\ \cmidrule{1-3}
        \multirow[t]{3}{*}{Colours+Extras}    & \multirow[t]{3}{*}{C+E} & \texttt{r}                                                                               \\ \arrayrulecolor{lightgray}\cmidrule(l{6pt}){3-3}\arrayrulecolor{black}
        ~                                          & ~                            & \texttt{u$-$r, 378$-$r, 395$-$r, 410$-$r, 430$-$r, g$-$r, 515$-$r, r$-$660, r$-$i, r$-$861, r$-$z, r$-$W1, r$-$W2} \\ \arrayrulecolor{lightgray}\cmidrule(l{6pt}){3-3}\arrayrulecolor{black}
        ~                                          & ~                            & \texttt{FWHM\_n, MUMAX, Ellipticity}                                                  \\
        \bottomrule
    \end{tabular}
    \begin{tablenotes}[para,flushleft]
        \item[a] Some filter names were abbreviated for simplicity but are the same as described in Section \ref{sec:data}.
    \end{tablenotes}
\end{threeparttable}
\end{table*}

\begin{table}
\centering
\caption{Results for both Deep Learning methods and GPz using the samples described in Table \ref{tab:inputs} for nDet\_auto $\geqslant$ 8.}
\label{tab:inputresults}
\vspace{0.2cm}
    \begin{tabular}{@{}lccc@{}}
        \toprule
        Configuration & $\sigma_\text{NMAD}$ ($\times 10^{-2}$) & $\overline{\mu}$ ($\times 10^{-2}$) & $\eta$ (\%) \\ \midrule
        \multicolumn{4}{c}{\textbf{GPz}} \\
        A             & 3.159                     & -0.178                & 0.717      \\ 
        M             & 3.239                     & -0.182                & 0.857      \\ 
        M+E           & 3.139                     & -0.180                & 0.717      \\ 
        C             & 3.114                     & -0.159                & 0.904      \\ 
        C+E           & 3.017                     & -0.160                & 0.682      \\ 
        \midrule
        \multicolumn{4}{c}{\textbf{DN}} \\
        A             & 2.487                     & -0.182                & 0.636      \\ 
        M             & 2.550                     & -0.629                & 0.723      \\ 
        M+E           & 2.494                     & -0.376                & 0.548      \\ 
        C             & 2.509                     & -0.017                & 0.717      \\ 
        C+E           & 2.426                     & -0.220                & 0.571      \\ 
        \midrule
        \multicolumn{4}{c}{\textbf{MDN}} \\
        A             & 2.343                     & -0.104                & 0.636      \\ 
        M             & 2.457                     & -0.141                & 0.904      \\ 
        M+E           & 2.354                     & -0.125                & 0.665      \\ 
        C             & 2.437                     & -0.223                & 0.934      \\ 
        C+E           & 2.338                     & -0.078                & 0.583      \\ 
        \bottomrule
    \end{tabular}
\end{table}

We also evaluated the training and validation loss of the models and verified that there is no under- or overfitting, indicating that the results are reliable and the models have a good generalization capacity.

\subsection{Single-point estimates} \label{sub:single_point_estimates}

Now we analyse the single-point estimates (SPE) of each model in comparison with BPZ2, the method currently used in S-PLUS for $z_\text{phot}$ estimation. Each method obtains SPE in different ways. GPz provides the mean (used as SPE) and uncertainty, the SPE of the DN is a weighted average of the possible $z_\text{phot}$s and their probabilities (the expected value), the $z_\text{phot}$ of the MDN is chosen as the value corresponding to the peak of the PDF (maximum-a-posteriori), and BPZ2 single-point estimate is given by $z_b = \sum_T \int p(z,T)z~\text{d}z$, where $p(z,T)$ is the PDF and $T$ is a template spectral type at redshift $z$ \citep{Benitez2000}.

\begin{figure*}
    \centering
    \includegraphics[height=.75\textheight]{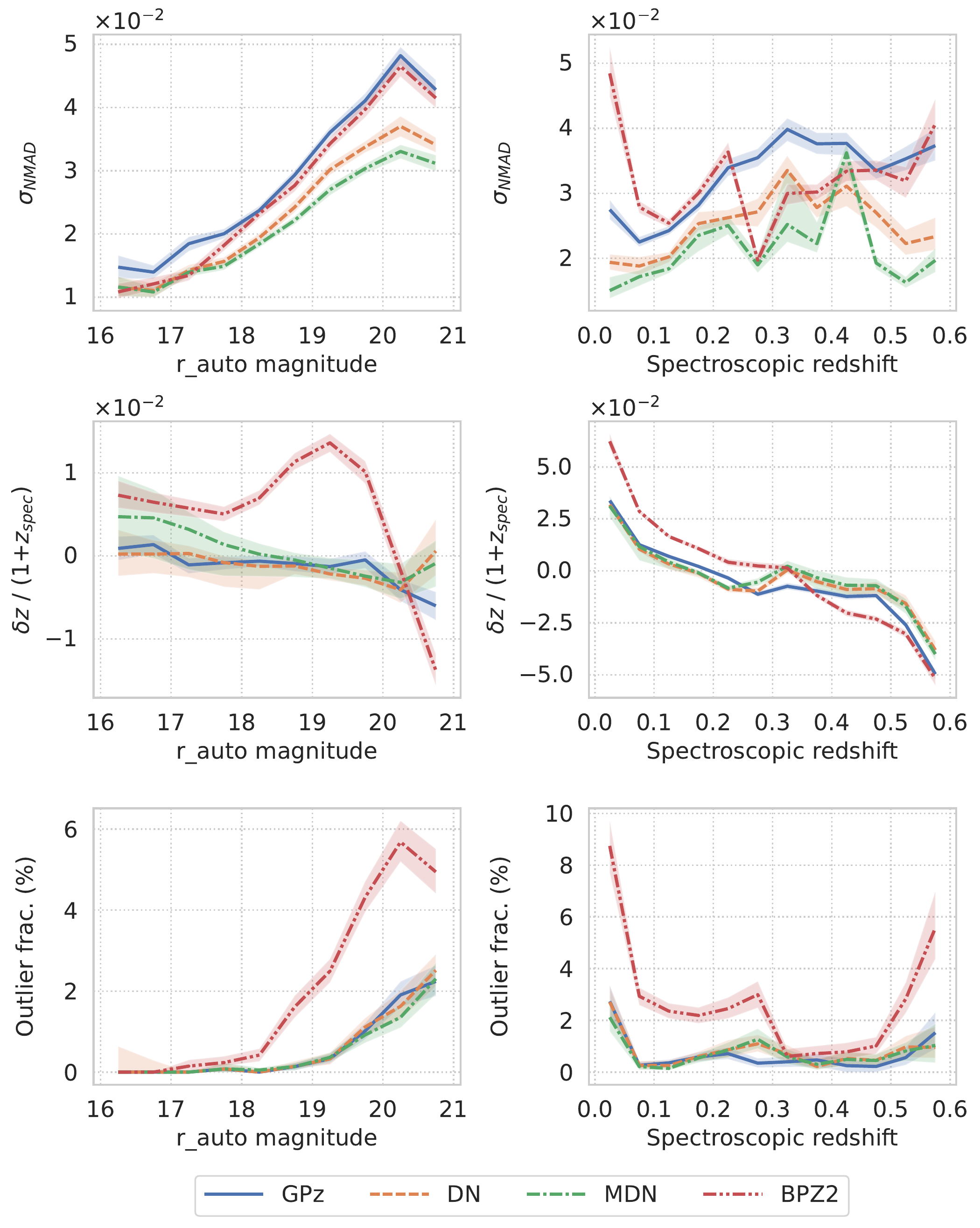}
    \caption{Comparison between the metrics obtained for all methods. Left column, from top to bottom: scatter, bias and outlier fraction as a function of the \texttt{r\_auto} magnitude. Right column, from top to bottom: same metrics as a function of the spectroscopic redshift. The GPz results are in blue, DN are in orange, MDN are in green, and BPZ2 results are in red. The shaded areas represent the uncertainties obtained from the 4-fold cross validation (or bootstrapping in the case of BPZ2).}
    \label{fig:methodcomp}
\end{figure*}

The results obtained for all methods are present in Figure \ref{fig:methodcomp}. The deep learning-based model MDN provides the most accurate $z_\text{phot}$s both as a function of magnitude and spectroscopic redshift. The currently used standard for $z_\text{phot}$ determination in S-PLUS, the BPZ2 method, underperformed in this sample in comparison to the other methods, mainly due to higher bias and outlier fraction. The results for the entire sample (without binning in magnitude or redshift) are shown in Table \ref{tab:Summaryspe}.

It should be pointed out that the deep learning-based methods were the only ones sensitive to the ``redshift window of opportunity'' \citep{AlbertoSPLUS}. This feature is characterized by two emission lines entering two narrow band filters of S-PLUS (namely the [\ion{O}{iii}] and H$\alpha$ lines in the J0660 and J0861 filters, respectively) in the redshift interval of \mbox{$0.26 \leqslant z_\text{spec} \leqslant 0.32$} and also greatly improves the accuracy of template-fitting methods.
\begin{table}
\centering
\caption{Summary of model results for \texttt{nDet\_auto} $\geqslant$ 8.}
\label{tab:Summaryspe}
\vspace{0.2cm}
    \begin{tabular}{@{}lccc@{}}
        \toprule
        Method         & $\sigma_\text{NMAD}$ ($\times 10^{-2}$) & $\mu$ ($\times 10^{-2}$) & $\eta$ (\%) \\ \midrule
        GPz            & 3.159                     & -0.178                & 0.717      \\ 
        DN             & 2.487                                   & -0.182                   & 0.636      \\ 
        \textbf{MDN}            & \textbf{2.343}                                   & \textbf{-0.104}                   & \textbf{0.636}      \\ 
        BPZ2           & 3.080                                   & \hphantom{-}0.580        & 2.758      \\ 
        \bottomrule
    \end{tabular}
\end{table}

In summary, we verified that the deep learning-based MDN model provided the most accurate single-point estimate $z_\text{phot}$s. This model also had the best performance for faint objects, while having low bias and a negligible outlier fraction. When it is compared with the current standard in S-PLUS, BPZ2, the MDN model achieved 24\% lower scatter, 82\% lower bias, and 77\% lower outlier fraction in the same testing sample.

\subsection{PDF analysis} \label{sub:pdf_analysis}

Besides providing a probabilistic description of results, probability distribution functions (PDFs) are useful to understand and assess the uncertainties of a model. The PDFs here represent all possible values of an object photometric redshift, and include errors in both the data measurement process (the aleatoric uncertainty) and in the model itself (the epistemic uncertainty).

Each method generates PDFs in different ways:
\begin{itemize}
    \item The DN produces PDFs due to the Softmax activation used in the output layer. The PDF is constructed by the probability of the galaxy belonging to each of the 200 bins of redshift between \mbox{$0 \leqslant z_\text{phot} \leqslant 1.34$}.
    \item For the MDN, the PDF is obtained as a combination of 20 Gaussian distributions, each with a different mean, standard deviation, and weight. A combination of these distributions allows the generation of a multi-peak PDF.
    \item In the case of GPz, each $z_\text{phot}$ prediction also generates a predictive variance, modelled as a function of input (heteroscedastic noise), so that the PDF is a Gaussian function.
\end{itemize}

We use three benchmarks to assess the quality of the PDFs: the Odds values, the Probability Integral Transform (PIT) histogram and the Continuous Ranked Probability Score (CRPS), all described in Section \ref{sec:metrics}.

\begin{figure}
    \centering
    \includegraphics[width=\textwidth]{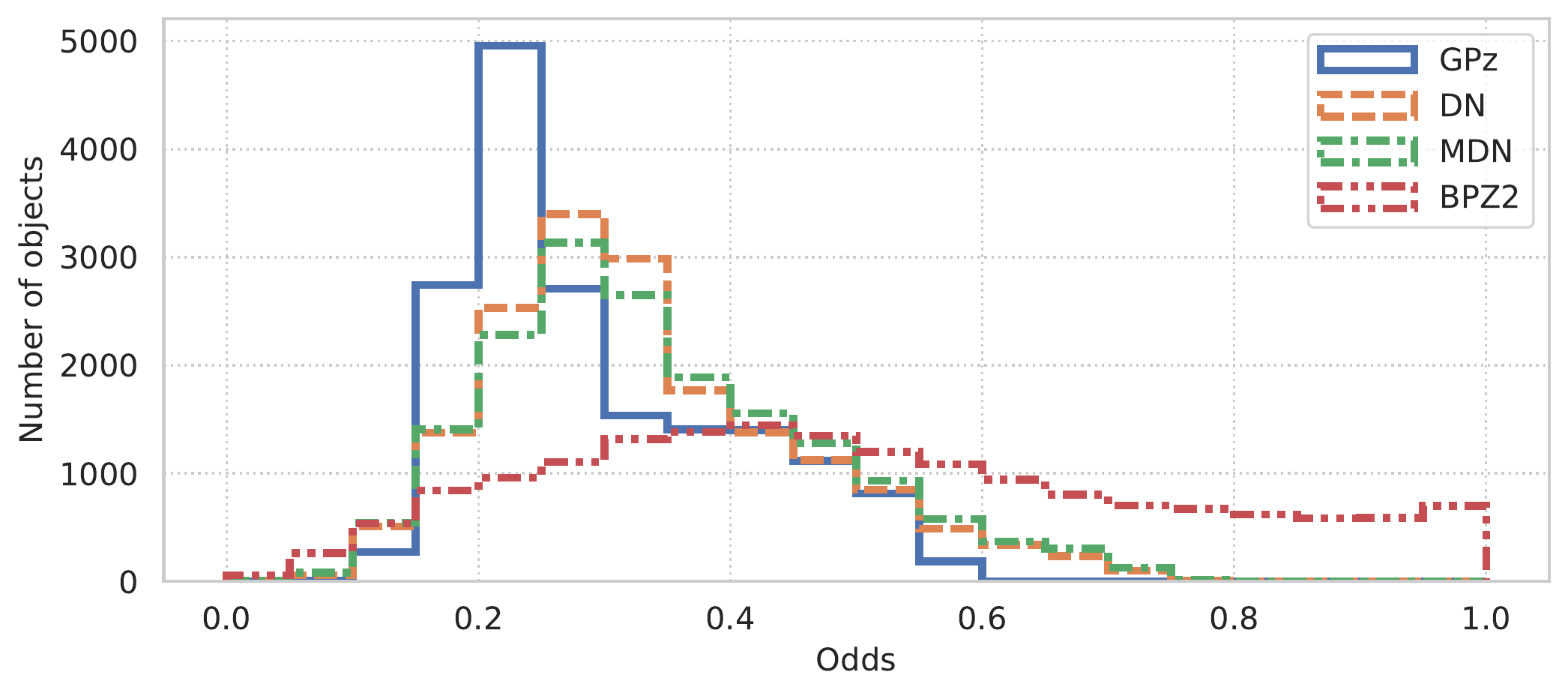}
    \caption{Odds value distributions for all methods. Values closer to 1 indicate narrower PDFs.}
    \label{fig:odds}
\end{figure}

As shown in Figure \ref{fig:odds}, the machine learning methods have broader PDFs than BPZ2, with peaks in the distribution of Odds between 0.20 and 0.25 for GPz, and between 0.25 and 0.30 for both the DN and MDN. BPZ2 provides PDFs which are narrower in general, as can be verified by the number of objects with \mbox{$\text{Odds} \geqslant 0.6$}.

\begin{figure}
    \centering
    \includegraphics[width=\textwidth]{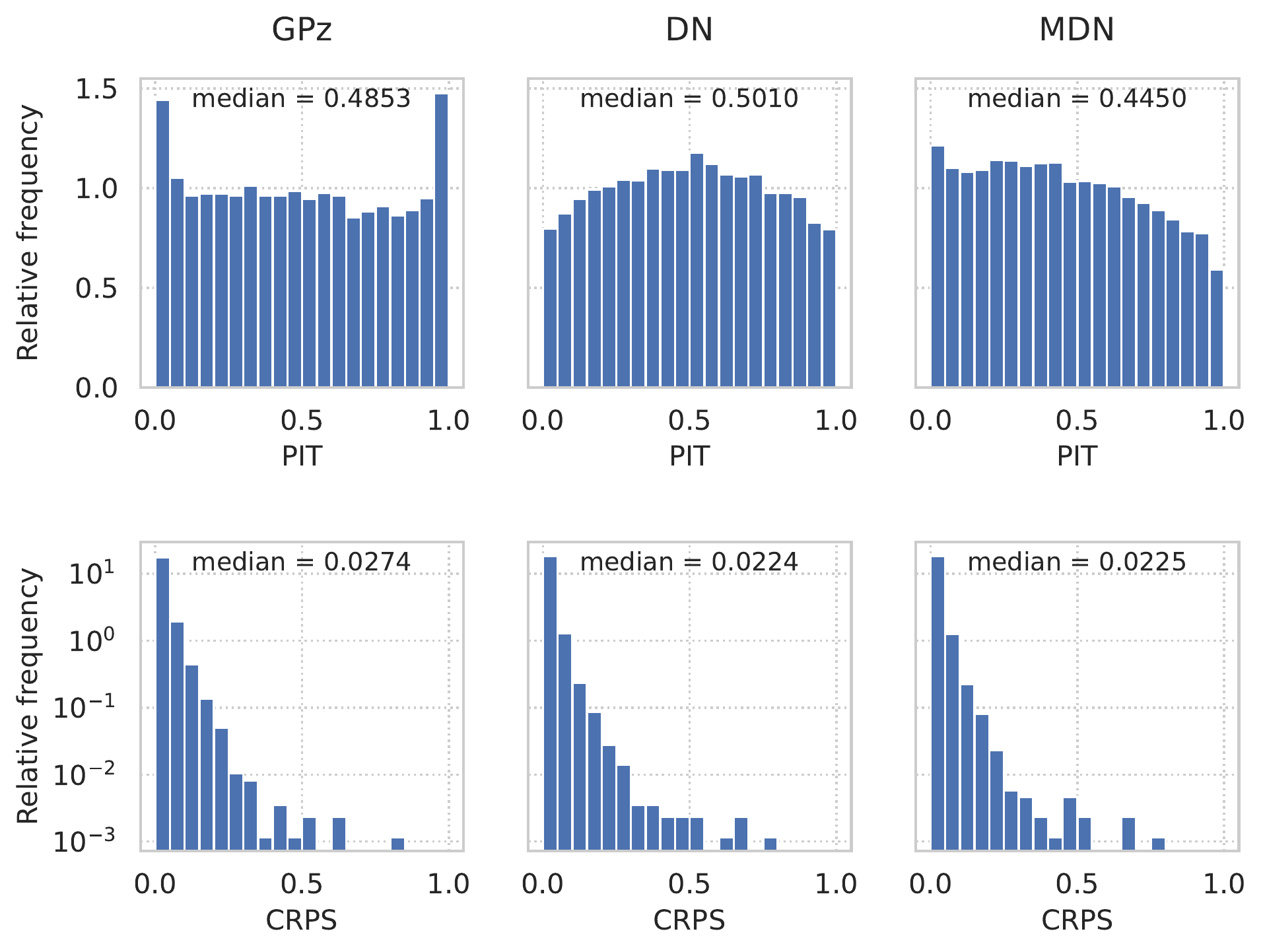}
    \caption{PIT and CRPS distributions for GPz, DN, and the MDN models.}
    \label{fig:pit}
\end{figure}

The results obtained for the PIT test are presented in Figure \ref{fig:pit}. BPZ2 is not included in this comparison because the PDFs of this method were not available. From this figure, we see that GPz PDFs have a near ideal distribution but have a higher quantity of outliers with PIT close to 0 or 1, which can be related to the way this model describes the PDFs (as unimodal Gaussians). The deep learning methods PDFs are better calibrated in relation to the width but present a negative slope in the case of the MDN, indicating that this method overestimates $z_\text{phot}$s (this can also be seen in Figure \ref{fig:methodcomp}), whereas PDFs are slightly broader for the DN and does not show any signs of bias. 

The CRPS histograms are very similar between the different models, but the median values (presented inside the plots in Figure \ref{fig:pit}) indicate that the DN and MDN models provide better PDFs in relation to the true redshift values, but with no significant difference between both.

\section{Conclusions} \label{sec:conclusions}

With the advent of all-sky photometric surveys, the need for photometric redshifts to study the large scale universe has increased. Recent advances in computational resources and software are making possible to extract accurate distance estimates from the immense volume of data produced by surveys.

In this work, we have explored three different machine learning methods (GPz, Dense Network and Mixture Density Network), evaluating their performances on the S-PLUS twelve-filter system to generate accurate photometric redshifts. We also compared our results with the method currently used in the S-PLUS DR1, the template-fitting code BPZ2.

We started our analysis by testing how the inclusion of objects with different numbers of missing bands impact photometric redshift estimates. Our findings, as shown in Section \ref{sub:missing_feature_analysis}, indicate that even objects with four missing bands provide enough information to improve the results. This can be explained by two main factors: they are generally faint, and adding them to the training sample improves model performance mostly at high magnitudes; and by forcing the model to learn with incomplete information, missing-band objects help increasing the model generalization capacity.

In Section \ref{sub:input_features_analysis} we verified how the choice of input features impacts the photometric redshift accuracy. According to our tests for all models, the \texttt{photo-z} prediction scatter ($\sigma_\text{NMAD}$) changed less than 5\% between the different input choices described in Table \ref{tab:inputs}. Also, as can be seen in Figure \ref{fig:inputfeatR}, although the scatter does not change significantly, the bias and outlier fractions are, in general, lower when we use all 30 input features for training.

Figure \ref{fig:methodcomp} shows that, for the dataset used in this work, all machine learning methods outperform the template-fitting approach currently used in S-PLUS DR1. A special highlight should be given to the deep learning-based methods, which are able to reach a higher accuracy than BPZ2 for faint objects ($\sigma_\text{NMAD}$ 30\% lower for \mbox{$20 \leqslant \texttt{r\_auto} \leqslant 21$}). For all objects in our sample, the MDN model achieved 24\% lower scatter, 82\% lower bias, and 77\% lower outlier fraction when compared to BPZ2 results on the same sample. Also, when analysing the results as a function of spectroscopic redshift, we see an increase in photometric redshift accuracy in the interval $0.26 \leqslant z_\text{spec} \leqslant 0.32$. This only happens due to the [\ion{O}{iii}] and H$\alpha$ emission lines entering two narrow-band filters of the survey at this redshift range. This is an evidence of how important the twelve-filter system of S-PLUS is. 

The photometric redshifts have a systematic bias as a function of redshift, as shown in Figure \ref{fig:methodcomp}. A similar trend is also present in other works, such as \citet[p.11]{Kitching2016} and \citet[p.11]{Banerji2008}. A possible explanation is the lower amount of training samples in the low- and high-redshift regions. A possible solution is the implementation of sample weights, which will be done in future work.

When comparing the PDFs, we verified that none of the methods provided well-calibrated distributions, even for the deep learning-based methods with the best single-point estimate results. A second version of the MDN is under development aiming to solve those issues. 

In summary, we concluded that, between the methods compared in this work, the best-performing  for this sample is the Mixture Density Network, providing accurate photometric redshifts both as a function of \texttt{r\_auto} and \mbox{\texttt{spec-z}}. It also provides PDFs that are able to take into account the redshift-colour degeneracy due to its multi-peaked nature. Our conclusions can be extended to the remainder of the S-PLUS survey, so that we are able to provide accurate photometric redshifts for several science cases in the Collaboration.

This method can be easily applied to future S-PLUS data releases offering improved photometric redshift performance when compared to the current template-fitting based standard \citep{AlbertoSPLUS}. As more regions are observed and the overlaps with spectroscopic data increases, we also expect further improvements of our model.

\section{Acknowledgments} \label{sec:acknowledgments}
The author acknowledges the financial support given by Coordenação de Aperfeiçoamento de Pessoal de Nível Superior (CAPES, grant 88887.470064/2019-00) and Conselho Nacional de Desenvolvimento Científico e Tecnológico (CNPq, grant 169181/2017-0) during the development of this research. L. S. J. acknowledges support from Brazilian agencies FAPESP (2019/10923-5) and CNPq (304819/2017-4). L. N. acknowledges the financial support provided by the São Paulo Research Foundation (FAPESP) (grant 2019/01312-2 and 2014/10566-4). M. L. B. acknowledges the financial support provided by the São Paulo Research Foundation (FAPESP) (grant 2018/09165-6 and 2019/23388-0). C. Q. acknowledges the financial support provided by the São Paulo Research Foundation (grant 2015/11442-0 and 2019/06766-1) and Coordenação de Aperfeiçoamento de Pessoal de Nível Superior (CAPES). F. R. H. thanks FAPESP for the financial support through the program 2018/21661-9. J. L. N. C. is grateful for the financial support received from the Southern Office of Aerospace Research and development (SOARD), grants FA9550-18-1-0018 and FA9550-22-1-0037, of the Air Force Office of the Scientific Research International Office of the United States (AFOSR/IO). M. L. L. D. acknowledges the polish NCN grant number 2019/34/E/ST9/00133. S. A. acknowledges support under the grant 5077 financed by IAASARS/NOA. Y. J-T has received funding from the European Union’s Horizon 2020 research and innovation programme under the Marie Skłodowska-Curie grant agreement No 898633. Y. J-T. also acknowledges financial support from the State Agency for Research of the Spanish MCIU through the ‘‘Center of Excellence Severo Ochoa’’ award to the Instituto de Astrofísica de Andalucía (SEV-2017-0709).

This work has made use of the computing facilities of the Laboratory of Astroinformatics (IAG/USP, NAT/Unicsul), whose purchase was made possible by the Brazilian agency FAPESP (grant 2009/54006-4) and the INCT-A.

The S-PLUS project, including the T80S robotic telescope and the S-PLUS scientific survey, was founded as a partnership between the Fundação de Amparo à Pesquisa do Estado de São Paulo (FAPESP), the Observatório Nacional (ON), the Federal University of Sergipe (UFS), and the Federal University of Santa Catarina (UFSC), with important financial and practical contributions from other collaborating institutes in Brazil, Chile (Universidad de La Serena), and Spain (Centro de Estudios de F ısica del Cosmos de Aragón, CEFCA).

The unWISE coadded images and catalogue are based on data products from the Wide-field Infrared Survey Explorer, which is a joint project of the University of California, Los Angeles, and the Jet Propulsion Laboratory/California Institute of Technology, and NEOWISE, which is a project of the Jet Propulsion Laboratory/California Institute of Technology. WISE and NEOWISE are funded by the National Aeronautics and Space Administration. 

Funding for the Sloan Digital Sky Survey IV has been provided by the Alfred P. Sloan Foundation, the U.S. Department of Energy Office of Science, and the Participating Institutions. SDSS-IV acknowledges support and resources from the Center for High Performance Computing  at the University of Utah. The SDSS website is www.sdss.org. SDSS-IV is managed by the Astrophysical Research Consortium for the Participating Institutions of the SDSS Collaboration including the Brazilian Participation Group, the Carnegie Institution for Science, Carnegie Mellon University, Center for Astrophysics | Harvard \& Smithsonian, the Chilean Participation Group, the French Participation Group, Instituto de Astrof\'isica de Canarias, The Johns Hopkins University, Kavli Institute for the Physics and Mathematics of the Universe (IPMU) / University of Tokyo, the Korean Participation Group, Lawrence Berkeley National Laboratory, Leibniz Institut f\"ur Astrophysik Potsdam (AIP),  Max-Planck-Institut f\"ur Astronomie (MPIA Heidelberg), Max-Planck-Institut f\"ur Astrophysik (MPA Garching), Max-Planck-Institut f\"ur Extraterrestrische Physik (MPE), National Astronomical Observatories of China, New Mexico State University, New York University, University of Notre Dame, Observat\'ario Nacional / MCTI, The Ohio State University, Pennsylvania State University, Shanghai Astronomical Observatory, United Kingdom Participation Group, Universidad Nacional Aut\'onoma de M\'exico, University of Arizona, University of Colorado Boulder, University of Oxford, University of Portsmouth, University of Utah, University of Virginia, University of Washington, University of Wisconsin, Vanderbilt University, and Yale University.

\bibliography{Texs/Artigo.bib}

\begin{thebibliography}{61}
\expandafter\ifx\csname natexlab\endcsname\relax\def\natexlab#1{#1}\fi
\providecommand{\url}[1]{\texttt{#1}}
\providecommand{\href}[2]{#2}
\providecommand{\path}[1]{#1}
\providecommand{\DOIprefix}{doi:}
\providecommand{\ArXivprefix}{arXiv:}
\providecommand{\URLprefix}{URL: }
\providecommand{\Pubmedprefix}{pmid:}
\providecommand{\doi}[1]{\href{http://dx.doi.org/#1}{\path{#1}}}
\providecommand{\Pubmed}[1]{\href{pmid:#1}{\path{#1}}}
\providecommand{\bibinfo}[2]{#2}
\ifx\xfnm\relax \def\xfnm[#1]{\unskip,\space#1}\fi
\bibitem[{{Abadi} et~al.(2016){Abadi}, {Barham}, {Chen}, {Chen}, {Davis},
  {Dean}, {Devin}, {Ghemawat}, {Irving}, {Isard}, {Kudlur}, {Levenberg},
  {Monga}, {Moore}, {Murray}, {Steiner}, {Tucker}, {Vasudevan}, {Warden},
  {Wicke}, {Yu} and {Zheng}}]{TensorFlow}
\bibinfo{author}{{Abadi}, M.}, \bibinfo{author}{{Barham}, P.},
  \bibinfo{author}{{Chen}, J.}, \bibinfo{author}{{Chen}, Z.},
  \bibinfo{author}{{Davis}, A.}, \bibinfo{author}{{Dean}, J.},
  \bibinfo{author}{{Devin}, M.}, \bibinfo{author}{{Ghemawat}, S.},
  \bibinfo{author}{{Irving}, G.}, \bibinfo{author}{{Isard}, M.},
  \bibinfo{author}{{Kudlur}, M.}, \bibinfo{author}{{Levenberg}, J.},
  \bibinfo{author}{{Monga}, R.}, \bibinfo{author}{{Moore}, S.},
  \bibinfo{author}{{Murray}, D.G.}, \bibinfo{author}{{Steiner}, B.},
  \bibinfo{author}{{Tucker}, P.}, \bibinfo{author}{{Vasudevan}, V.},
  \bibinfo{author}{{Warden}, P.}, \bibinfo{author}{{Wicke}, M.},
  \bibinfo{author}{{Yu}, Y.}, \bibinfo{author}{{Zheng}, X.},
  \bibinfo{year}{2016}.
\newblock \bibinfo{title}{{TensorFlow: A system for large-scale machine
  learning}}.
\newblock \bibinfo{journal}{arXiv e-prints} ,
  \bibinfo{pages}{arXiv:1605.08695}\href{http://arxiv.org/abs/1605.08695}{\tt
  arXiv:1605.08695}.
\bibitem[{{Alarcon} et~al.(2021){Alarcon}, {Gaztanaga}, {Eriksen}, {Baugh},
  {Cabayol}, {Casas}, {Carretero}, {Castander}, {De Vicente}, {Fernandez},
  {Garcia-Bellido}, {Hildebrandt}, {Hoekstra}, {Joachimi}, {Manzoni}, {Miquel},
  {Norberg}, {Padilla}, {Renard}, {Sanchez}, {Serrano}, {Sevilla-Noarbe},
  {Siudek} and {Tallada-Cresp{\'\i}}}]{Alarcon2021}
\bibinfo{author}{{Alarcon}, A.}, \bibinfo{author}{{Gaztanaga}, E.},
  \bibinfo{author}{{Eriksen}, M.}, \bibinfo{author}{{Baugh}, C.M.},
  \bibinfo{author}{{Cabayol}, L.}, \bibinfo{author}{{Casas}, R.},
  \bibinfo{author}{{Carretero}, J.}, \bibinfo{author}{{Castander}, F.J.},
  \bibinfo{author}{{De Vicente}, J.}, \bibinfo{author}{{Fernandez}, E.},
  \bibinfo{author}{{Garcia-Bellido}, J.}, \bibinfo{author}{{Hildebrandt}, H.},
  \bibinfo{author}{{Hoekstra}, H.}, \bibinfo{author}{{Joachimi}, B.},
  \bibinfo{author}{{Manzoni}, G.}, \bibinfo{author}{{Miquel}, R.},
  \bibinfo{author}{{Norberg}, P.}, \bibinfo{author}{{Padilla}, C.},
  \bibinfo{author}{{Renard}, P.}, \bibinfo{author}{{Sanchez}, E.},
  \bibinfo{author}{{Serrano}, S.}, \bibinfo{author}{{Sevilla-Noarbe}, I.},
  \bibinfo{author}{{Siudek}, M.}, \bibinfo{author}{{Tallada-Cresp{\'\i}}, P.},
  \bibinfo{year}{2021}.
\newblock \bibinfo{title}{{The PAU Survey: an improved photo-z sample in the
  COSMOS field}}.
\newblock \bibinfo{journal}{\mnras} \bibinfo{volume}{501},
  \bibinfo{pages}{6103--6122}.
\newblock \DOIprefix\doi{10.1093/mnras/staa3659},
  \href{http://arxiv.org/abs/2007.11132}{\tt arXiv:2007.11132}.
\bibitem[{Almosallam(2017)}]{AlmosallamThesis}
\bibinfo{author}{Almosallam, I.}, \bibinfo{year}{2017}.
\newblock \bibinfo{title}{Heteroscedastic Gaussian processes for uncertain and
  incomplete data}.
\newblock Ph.D. thesis. University of Oxford.
\bibitem[{Almosallam et~al.(2016)Almosallam, Jarvis and
  Roberts}]{Almosallam2016}
\bibinfo{author}{Almosallam, I.A.}, \bibinfo{author}{Jarvis, M.J.},
  \bibinfo{author}{Roberts, S.J.}, \bibinfo{year}{2016}.
\newblock \bibinfo{title}{{GPz}: non-stationary sparse gaussian processes for
  heteroscedastic uncertainty estimation in photometric redshifts}.
\newblock \bibinfo{journal}{Monthly Notices of the Royal Astronomical Society}
  \bibinfo{volume}{462}, \bibinfo{pages}{726--739}.
\newblock \URLprefix \url{https://doi.org/10.1093/mnras/stw1618},
  \DOIprefix\doi{10.1093/mnras/stw1618}.
\bibitem[{Almosallam et~al.(2015)Almosallam, Lindsay, Jarvis and
  Roberts}]{Almosallam2015}
\bibinfo{author}{Almosallam, I.A.}, \bibinfo{author}{Lindsay, S.N.},
  \bibinfo{author}{Jarvis, M.J.}, \bibinfo{author}{Roberts, S.J.},
  \bibinfo{year}{2015}.
\newblock \bibinfo{title}{A sparse gaussian process framework for photometric
  redshift estimation}.
\newblock \bibinfo{journal}{Monthly Notices of the Royal Astronomical Society}
  \bibinfo{volume}{455}, \bibinfo{pages}{2387--2401}.
\newblock \URLprefix \url{https://doi.org/10.1093/mnras/stv2425},
  \DOIprefix\doi{10.1093/mnras/stv2425}.
\bibitem[{Arnouts et~al.(1999)Arnouts, Cristiani, Moscardini, Matarrese,
  Lucchin, Fontana and Giallongo}]{Arnouts1999}
\bibinfo{author}{Arnouts, S.}, \bibinfo{author}{Cristiani, S.},
  \bibinfo{author}{Moscardini, L.}, \bibinfo{author}{Matarrese, S.},
  \bibinfo{author}{Lucchin, F.}, \bibinfo{author}{Fontana, A.},
  \bibinfo{author}{Giallongo, E.}, \bibinfo{year}{1999}.
\newblock \bibinfo{title}{Measuring and modelling the redshift evolution of
  clustering: the hubble deep field north}.
\newblock \bibinfo{journal}{Monthly Notices of the Royal Astronomical Society}
  \bibinfo{volume}{310}, \bibinfo{pages}{540--556}.
\newblock \URLprefix \url{https://doi.org/10.1046/j.1365-8711.1999.02978.x},
  \DOIprefix\doi{10.1046/j.1365-8711.1999.02978.x}.
\bibitem[{{Banerji} et~al.(2008){Banerji}, {Abdalla}, {Lahav} and
  {Lin}}]{Banerji2008}
\bibinfo{author}{{Banerji}, M.}, \bibinfo{author}{{Abdalla}, F.B.},
  \bibinfo{author}{{Lahav}, O.}, \bibinfo{author}{{Lin}, H.},
  \bibinfo{year}{2008}.
\newblock \bibinfo{title}{{Photometric redshifts for the Dark Energy Survey and
  VISTA and implications for large-scale structure}}.
\newblock \bibinfo{journal}{\mnras} \bibinfo{volume}{386},
  \bibinfo{pages}{1219--1233}.
\newblock \DOIprefix\doi{10.1111/j.1365-2966.2008.13095.x},
  \href{http://arxiv.org/abs/0711.1059}{\tt arXiv:0711.1059}.
\bibitem[{{Barro} et~al.(2019){Barro}, {P{\'e}rez-Gonz{\'a}lez}, {Cava},
  {Brammer}, {Pandya}, {Eliche Moral}, {Esquej}, {Dom{\'\i}nguez-S{\'a}nchez},
  {Alcalde Pampliega}, {Guo}, {Koekemoer}, {Trump}, {Ashby}, {Cardiel},
  {Castellano}, {Conselice}, {Dickinson}, {Dolch}, {Donley}, {Espino Briones},
  {Faber}, {Fazio}, {Ferguson}, {Finkelstein}, {Fontana}, {Galametz},
  {Gardner}, {Gawiser}, {Giavalisco}, {Grazian}, {Grogin}, {Hathi}, {Hemmati},
  {Hern{\'a}n-Caballero}, {Kocevski}, {Koo}, {Kodra}, {Lee}, {Lin}, {Lucas},
  {Mobasher}, {McGrath}, {Nandra}, {Nayyeri}, {Newman}, {Pforr}, {Peth},
  {Rafelski}, {Rodr{\'\i}guez-Munoz}, {Salvato}, {Stefanon}, {van der Wel},
  {Willner}, {Wiklind} and {Wuyts}}]{Barro2019}
\bibinfo{author}{{Barro}, G.}, \bibinfo{author}{{P{\'e}rez-Gonz{\'a}lez},
  P.G.}, \bibinfo{author}{{Cava}, A.}, \bibinfo{author}{{Brammer}, G.},
  \bibinfo{author}{{Pandya}, V.}, \bibinfo{author}{{Eliche Moral}, C.},
  \bibinfo{author}{{Esquej}, P.},
  \bibinfo{author}{{Dom{\'\i}nguez-S{\'a}nchez}, H.}, \bibinfo{author}{{Alcalde
  Pampliega}, B.}, \bibinfo{author}{{Guo}, Y.}, \bibinfo{author}{{Koekemoer},
  A.M.}, \bibinfo{author}{{Trump}, J.R.}, \bibinfo{author}{{Ashby}, M.L.N.},
  \bibinfo{author}{{Cardiel}, N.}, \bibinfo{author}{{Castellano}, M.},
  \bibinfo{author}{{Conselice}, C.J.}, \bibinfo{author}{{Dickinson}, M.E.},
  \bibinfo{author}{{Dolch}, T.}, \bibinfo{author}{{Donley}, J.L.},
  \bibinfo{author}{{Espino Briones}, N.}, \bibinfo{author}{{Faber}, S.M.},
  \bibinfo{author}{{Fazio}, G.G.}, \bibinfo{author}{{Ferguson}, H.},
  \bibinfo{author}{{Finkelstein}, S.}, \bibinfo{author}{{Fontana}, A.},
  \bibinfo{author}{{Galametz}, A.}, \bibinfo{author}{{Gardner}, J.P.},
  \bibinfo{author}{{Gawiser}, E.}, \bibinfo{author}{{Giavalisco}, M.},
  \bibinfo{author}{{Grazian}, A.}, \bibinfo{author}{{Grogin}, N.A.},
  \bibinfo{author}{{Hathi}, N.P.}, \bibinfo{author}{{Hemmati}, S.},
  \bibinfo{author}{{Hern{\'a}n-Caballero}, A.}, \bibinfo{author}{{Kocevski},
  D.}, \bibinfo{author}{{Koo}, D.C.}, \bibinfo{author}{{Kodra}, D.},
  \bibinfo{author}{{Lee}, K.S.}, \bibinfo{author}{{Lin}, L.},
  \bibinfo{author}{{Lucas}, R.A.}, \bibinfo{author}{{Mobasher}, B.},
  \bibinfo{author}{{McGrath}, E.J.}, \bibinfo{author}{{Nandra}, K.},
  \bibinfo{author}{{Nayyeri}, H.}, \bibinfo{author}{{Newman}, J.A.},
  \bibinfo{author}{{Pforr}, J.}, \bibinfo{author}{{Peth}, M.},
  \bibinfo{author}{{Rafelski}, M.}, \bibinfo{author}{{Rodr{\'\i}guez-Munoz},
  L.}, \bibinfo{author}{{Salvato}, M.}, \bibinfo{author}{{Stefanon}, M.},
  \bibinfo{author}{{van der Wel}, A.}, \bibinfo{author}{{Willner}, S.P.},
  \bibinfo{author}{{Wiklind}, T.}, \bibinfo{author}{{Wuyts}, S.},
  \bibinfo{year}{2019}.
\newblock \bibinfo{title}{{The CANDELS/SHARDS Multiwavelength Catalog in
  GOODS-N: Photometry, Photometric Redshifts, Stellar Masses, Emission-line
  Fluxes, and Star Formation Rates}}.
\newblock \bibinfo{journal}{\apjs} \bibinfo{volume}{243}, \bibinfo{pages}{22}.
\newblock \DOIprefix\doi{10.3847/1538-4365/ab23f2},
  \href{http://arxiv.org/abs/1908.00569}{\tt arXiv:1908.00569}.
\bibitem[{Battye and Weller(2003)}]{Battye2003}
\bibinfo{author}{Battye, R.A.}, \bibinfo{author}{Weller, J.},
  \bibinfo{year}{2003}.
\newblock \bibinfo{title}{Constraining cosmological parameters using
  sunyaev-zel'dovich cluster surveys}.
\newblock \bibinfo{journal}{Physical Review D} \bibinfo{volume}{68}.
\newblock \URLprefix \url{https://doi.org/10.1103/physrevd.68.083506},
  \DOIprefix\doi{10.1103/physrevd.68.083506}.
\bibitem[{Benitez(2000)}]{Benitez2000}
\bibinfo{author}{Benitez, N.}, \bibinfo{year}{2000}.
\newblock \bibinfo{title}{Bayesian photometric redshift estimation}.
\newblock \bibinfo{journal}{The Astrophysical Journal} \bibinfo{volume}{536},
  \bibinfo{pages}{571--583}.
\newblock \URLprefix \url{https://doi.org/10.1086/308947},
  \DOIprefix\doi{10.1086/308947}.
\bibitem[{{Benitez} et~al.(2014){Benitez}, {Dupke}, {Moles}, {Sodre},
  {Cenarro}, {Marin-Franch}, {Taylor}, {Cristobal}, {Fernandez-Soto}, {Mendes
  de Oliveira}, {Cepa-Nogue}, {Abramo}, {Alcaniz}, {Overzier},
  {Hernandez-Monteagudo}, {Alfaro}, {Kanaan}, {Carvano}, {Reis}, {Martinez
  Gonzalez}, {Ascaso}, {Ballesteros}, {Xavier}, {Varela}, {Ederoclite},
  {Vazquez Ramio}, {Broadhurst}, {Cypriano}, {Angulo}, {Diego}, {Zandivarez},
  {Diaz}, {Melchior}, {Umetsu}, {Spinelli}, {Zitrin}, {Coe}, {Yepes}, {Vielva},
  {Sahni}, {Marcos-Caballero}, {Shu Kitaura}, {Maroto}, {Masip}, {Tsujikawa},
  {Carneiro}, {Gonzalez Nuevo}, {Carvalho}, {Reboucas}, {Carvalho}, {Abdalla},
  {Bernui}, {Pigozzo}, {Ferreira}, {Chandrachani Devi}, {Bengaly}, {Campista},
  {Amorim}, {Asari}, {Bongiovanni}, {Bonoli}, {Bruzual}, {Cardiel}, {Cava},
  {Cid Fernandes}, {Coelho}, {Cortesi}, {Delgado}, {Diaz Garcia}, {Espinosa},
  {Galliano}, {Gonzalez-Serrano}, {Falcon-Barroso}, {Fritz}, {Fernandes},
  {Gorgas}, {Hoyos}, {Jimenez-Teja}, {Lopez-Aguerri}, {Lopez-San Juan},
  {Mateus}, {Molino}, {Novais}, {OMill}, {Oteo}, {Perez-Gonzalez}, {Poggianti},
  {Proctor}, {Ricciardelli}, {Sanchez-Blazquez}, {Storchi-Bergmann}, {Telles},
  {Schoennell}, {Trujillo}, {Vazdekis}, {Viironen}, {Daflon},
  {Aparicio-Villegas}, {Rocha}, {Ribeiro}, {Borges}, {Martins}, {Marcolino},
  {Martinez-Delgado}, {Perez-Torres}, {Siffert}, {Calvao}, {Sako}, {Kessler},
  {Alvarez-Candal}, {De Pra}, {Roig}, {Lazzaro}, {Gorosabel}, {Lopes de
  Oliveira}, {Lima-Neto}, {Irwin}, {Liu}, {Alvarez}, {Balmes}, {Chueca},
  {Costa-Duarte}, {da Costa}, {Dantas}, {Diaz}, {Fabregat}, {Ferrari},
  {Gavela}, {Gracia}, {Gruel}, {Gutierrez}, {Guzman}, {Hernandez-Fernandez},
  {Herranz}, {Hurtado-Gil}, {Jablonsky}, {Laporte}, {Le Tiran}, {Licandro},
  {Lima}, {Martin}, {Martinez}, {Montero}, {Penteado}, {Pereira}, {Peris},
  {Quilis}, {Sanchez-Portal}, {Soja}, {Solano}, {Torra} and
  {Valdivielso}}]{JPAS}
\bibinfo{author}{{Benitez}, N.}, \bibinfo{author}{{Dupke}, R.},
  \bibinfo{author}{{Moles}, M.}, \bibinfo{author}{{Sodre}, L.},
  \bibinfo{author}{{Cenarro}, J.}, \bibinfo{author}{{Marin-Franch}, A.},
  \bibinfo{author}{{Taylor}, K.}, \bibinfo{author}{{Cristobal}, D.},
  \bibinfo{author}{{Fernandez-Soto}, A.}, \bibinfo{author}{{Mendes de
  Oliveira}, C.}, \bibinfo{author}{{Cepa-Nogue}, J.},
  \bibinfo{author}{{Abramo}, L.R.}, \bibinfo{author}{{Alcaniz}, J.S.},
  \bibinfo{author}{{Overzier}, R.}, \bibinfo{author}{{Hernandez-Monteagudo},
  C.}, \bibinfo{author}{{Alfaro}, E.J.}, \bibinfo{author}{{Kanaan}, A.},
  \bibinfo{author}{{Carvano}, J.M.}, \bibinfo{author}{{Reis}, R.R.R.},
  \bibinfo{author}{{Martinez Gonzalez}, E.}, \bibinfo{author}{{Ascaso}, B.},
  \bibinfo{author}{{Ballesteros}, F.}, \bibinfo{author}{{Xavier}, H.S.},
  \bibinfo{author}{{Varela}, J.}, \bibinfo{author}{{Ederoclite}, A.},
  \bibinfo{author}{{Vazquez Ramio}, H.}, \bibinfo{author}{{Broadhurst}, T.},
  \bibinfo{author}{{Cypriano}, E.}, \bibinfo{author}{{Angulo}, R.},
  \bibinfo{author}{{Diego}, J.M.}, \bibinfo{author}{{Zandivarez}, A.},
  \bibinfo{author}{{Diaz}, E.}, \bibinfo{author}{{Melchior}, P.},
  \bibinfo{author}{{Umetsu}, K.}, \bibinfo{author}{{Spinelli}, P.F.},
  \bibinfo{author}{{Zitrin}, A.}, \bibinfo{author}{{Coe}, D.},
  \bibinfo{author}{{Yepes}, G.}, \bibinfo{author}{{Vielva}, P.},
  \bibinfo{author}{{Sahni}, V.}, \bibinfo{author}{{Marcos-Caballero}, A.},
  \bibinfo{author}{{Shu Kitaura}, F.}, \bibinfo{author}{{Maroto}, A.L.},
  \bibinfo{author}{{Masip}, M.}, \bibinfo{author}{{Tsujikawa}, S.},
  \bibinfo{author}{{Carneiro}, S.}, \bibinfo{author}{{Gonzalez Nuevo}, J.},
  \bibinfo{author}{{Carvalho}, G.C.}, \bibinfo{author}{{Reboucas}, M.J.},
  \bibinfo{author}{{Carvalho}, J.C.}, \bibinfo{author}{{Abdalla}, E.},
  \bibinfo{author}{{Bernui}, A.}, \bibinfo{author}{{Pigozzo}, C.},
  \bibinfo{author}{{Ferreira}, E.G.M.}, \bibinfo{author}{{Chandrachani Devi},
  N.}, \bibinfo{author}{{Bengaly}, Jr., C.A.P.}, \bibinfo{author}{{Campista},
  M.}, \bibinfo{author}{{Amorim}, A.}, \bibinfo{author}{{Asari}, N.V.},
  \bibinfo{author}{{Bongiovanni}, A.}, \bibinfo{author}{{Bonoli}, S.},
  \bibinfo{author}{{Bruzual}, G.}, \bibinfo{author}{{Cardiel}, N.},
  \bibinfo{author}{{Cava}, A.}, \bibinfo{author}{{Cid Fernandes}, R.},
  \bibinfo{author}{{Coelho}, P.}, \bibinfo{author}{{Cortesi}, A.},
  \bibinfo{author}{{Delgado}, R.G.}, \bibinfo{author}{{Diaz Garcia}, L.},
  \bibinfo{author}{{Espinosa}, J.M.R.}, \bibinfo{author}{{Galliano}, E.},
  \bibinfo{author}{{Gonzalez-Serrano}, J.I.},
  \bibinfo{author}{{Falcon-Barroso}, J.}, \bibinfo{author}{{Fritz}, J.},
  \bibinfo{author}{{Fernandes}, C.}, \bibinfo{author}{{Gorgas}, J.},
  \bibinfo{author}{{Hoyos}, C.}, \bibinfo{author}{{Jimenez-Teja}, Y.},
  \bibinfo{author}{{Lopez-Aguerri}, J.A.}, \bibinfo{author}{{Lopez-San Juan},
  C.}, \bibinfo{author}{{Mateus}, A.}, \bibinfo{author}{{Molino}, A.},
  \bibinfo{author}{{Novais}, P.}, \bibinfo{author}{{OMill}, A.},
  \bibinfo{author}{{Oteo}, I.}, \bibinfo{author}{{Perez-Gonzalez}, P.G.},
  \bibinfo{author}{{Poggianti}, B.}, \bibinfo{author}{{Proctor}, R.},
  \bibinfo{author}{{Ricciardelli}, E.}, \bibinfo{author}{{Sanchez-Blazquez},
  P.}, \bibinfo{author}{{Storchi-Bergmann}, T.}, \bibinfo{author}{{Telles},
  E.}, \bibinfo{author}{{Schoennell}, W.}, \bibinfo{author}{{Trujillo}, N.},
  \bibinfo{author}{{Vazdekis}, A.}, \bibinfo{author}{{Viironen}, K.},
  \bibinfo{author}{{Daflon}, S.}, \bibinfo{author}{{Aparicio-Villegas}, T.},
  \bibinfo{author}{{Rocha}, D.}, \bibinfo{author}{{Ribeiro}, T.},
  \bibinfo{author}{{Borges}, M.}, \bibinfo{author}{{Martins}, S.L.},
  \bibinfo{author}{{Marcolino}, W.}, \bibinfo{author}{{Martinez-Delgado}, D.},
  \bibinfo{author}{{Perez-Torres}, M.A.}, \bibinfo{author}{{Siffert}, B.B.},
  \bibinfo{author}{{Calvao}, M.O.}, \bibinfo{author}{{Sako}, M.},
  \bibinfo{author}{{Kessler}, R.}, \bibinfo{author}{{Alvarez-Candal}, A.},
  \bibinfo{author}{{De Pra}, M.}, \bibinfo{author}{{Roig}, F.},
  \bibinfo{author}{{Lazzaro}, D.}, \bibinfo{author}{{Gorosabel}, J.},
  \bibinfo{author}{{Lopes de Oliveira}, R.}, \bibinfo{author}{{Lima-Neto},
  G.B.}, \bibinfo{author}{{Irwin}, J.}, \bibinfo{author}{{Liu}, J.F.},
  \bibinfo{author}{{Alvarez}, E.}, \bibinfo{author}{{Balmes}, I.},
  \bibinfo{author}{{Chueca}, S.}, \bibinfo{author}{{Costa-Duarte}, M.V.},
  \bibinfo{author}{{da Costa}, A.A.}, \bibinfo{author}{{Dantas}, M.L.L.},
  \bibinfo{author}{{Diaz}, A.Y.}, \bibinfo{author}{{Fabregat}, J.},
  \bibinfo{author}{{Ferrari}, F.}, \bibinfo{author}{{Gavela}, B.},
  \bibinfo{author}{{Gracia}, S.G.}, \bibinfo{author}{{Gruel}, N.},
  \bibinfo{author}{{Gutierrez}, J.L.L.}, \bibinfo{author}{{Guzman}, R.},
  \bibinfo{author}{{Hernandez-Fernandez}, J.D.}, \bibinfo{author}{{Herranz},
  D.}, \bibinfo{author}{{Hurtado-Gil}, L.}, \bibinfo{author}{{Jablonsky}, F.},
  \bibinfo{author}{{Laporte}, R.}, \bibinfo{author}{{Le Tiran}, L.L.},
  \bibinfo{author}{{Licandro}, J.}, \bibinfo{author}{{Lima}, M.},
  \bibinfo{author}{{Martin}, E.}, \bibinfo{author}{{Martinez}, V.},
  \bibinfo{author}{{Montero}, J.J.C.}, \bibinfo{author}{{Penteado}, P.},
  \bibinfo{author}{{Pereira}, C.B.}, \bibinfo{author}{{Peris}, V.},
  \bibinfo{author}{{Quilis}, V.}, \bibinfo{author}{{Sanchez-Portal}, M.},
  \bibinfo{author}{{Soja}, A.C.}, \bibinfo{author}{{Solano}, E.},
  \bibinfo{author}{{Torra}, J.}, \bibinfo{author}{{Valdivielso}, L.},
  \bibinfo{year}{2014}.
\newblock \bibinfo{title}{{J-PAS: The Javalambre-Physics of the Accelerated
  Universe Astrophysical Survey}}.
\newblock \bibinfo{journal}{arXiv e-prints}
  \href{http://arxiv.org/abs/1403.5237}{\tt arXiv:1403.5237}.
\bibitem[{{Ben{\'\i}tez} et~al.(2009){Ben{\'\i}tez}, {Gazta{\~n}aga}, {Miquel},
  {Castander}, {Moles}, {Crocce}, {Fern{\'a}ndez-Soto}, {Fosalba},
  {Ballesteros}, {Campa}, {Cardiel-Sas}, {Castilla}, {Crist{\'o}bal-Hornillos},
  {Delfino}, {Fern{\'a}ndez}, {Fern{\'a}ndez-Sopuerta}, {Garc{\'\i}a-Bellido},
  {Lobo}, {Mart{\'\i}nez}, {Ortiz}, {Pacheco}, {Paredes}, {Pons-Border{\'\i}a},
  {S{\'a}nchez}, {S{\'a}nchez}, {Varela} and {de Vicente}}]{BenitezBAO}
\bibinfo{author}{{Ben{\'\i}tez}, N.}, \bibinfo{author}{{Gazta{\~n}aga}, E.},
  \bibinfo{author}{{Miquel}, R.}, \bibinfo{author}{{Castander}, F.},
  \bibinfo{author}{{Moles}, M.}, \bibinfo{author}{{Crocce}, M.},
  \bibinfo{author}{{Fern{\'a}ndez-Soto}, A.}, \bibinfo{author}{{Fosalba}, P.},
  \bibinfo{author}{{Ballesteros}, F.}, \bibinfo{author}{{Campa}, J.},
  \bibinfo{author}{{Cardiel-Sas}, L.}, \bibinfo{author}{{Castilla}, J.},
  \bibinfo{author}{{Crist{\'o}bal-Hornillos}, D.}, \bibinfo{author}{{Delfino},
  M.}, \bibinfo{author}{{Fern{\'a}ndez}, E.},
  \bibinfo{author}{{Fern{\'a}ndez-Sopuerta}, C.},
  \bibinfo{author}{{Garc{\'\i}a-Bellido}, J.}, \bibinfo{author}{{Lobo}, J.A.},
  \bibinfo{author}{{Mart{\'\i}nez}, V.J.}, \bibinfo{author}{{Ortiz}, A.},
  \bibinfo{author}{{Pacheco}, A.}, \bibinfo{author}{{Paredes}, S.},
  \bibinfo{author}{{Pons-Border{\'\i}a}, M.J.}, \bibinfo{author}{{S{\'a}nchez},
  E.}, \bibinfo{author}{{S{\'a}nchez}, S.F.}, \bibinfo{author}{{Varela}, J.},
  \bibinfo{author}{{de Vicente}, J.F.}, \bibinfo{year}{2009}.
\newblock \bibinfo{title}{{Measuring Baryon Acoustic Oscillations Along the
  Line of Sight with Photometric Redshifts: The PAU Survey}}.
\newblock \bibinfo{journal}{\apj} \bibinfo{volume}{691},
  \bibinfo{pages}{241--260}.
\newblock \DOIprefix\doi{10.1088/0004-637X/691/1/241},
  \href{http://arxiv.org/abs/0807.0535}{\tt arXiv:0807.0535}.
\bibitem[{{Bertin} and {Arnouts}(1996)}]{SExtractor}
\bibinfo{author}{{Bertin}, E.}, \bibinfo{author}{{Arnouts}, S.},
  \bibinfo{year}{1996}.
\newblock \bibinfo{title}{{SExtractor: Software for source extraction.}}
\newblock \bibinfo{journal}{\aaps} \bibinfo{volume}{117},
  \bibinfo{pages}{393--404}.
\newblock \DOIprefix\doi{10.1051/aas:1996164}.
\bibitem[{Bishop(1994)}]{MDN1994}
\bibinfo{author}{Bishop, C.}, \bibinfo{year}{1994}.
\newblock \bibinfo{title}{Mixture density networks}.
\newblock \bibinfo{type}{WorkingPaper}. Aston University.
\bibitem[{{Blanton} et~al.(2017){Blanton}, {Bershady}, {Abolfathi}, {Albareti},
  {Allende Prieto}, {Almeida}, {Alonso-Garc{\'{\i}}a}, {Anders}, {Anderson},
  {Andrews} and et~al.}]{SDSS15}
\bibinfo{author}{{Blanton}, M.R.}, \bibinfo{author}{{Bershady}, M.A.},
  \bibinfo{author}{{Abolfathi}, B.}, \bibinfo{author}{{Albareti}, F.D.},
  \bibinfo{author}{{Allende Prieto}, C.}, \bibinfo{author}{{Almeida}, A.},
  \bibinfo{author}{{Alonso-Garc{\'{\i}}a}, J.}, \bibinfo{author}{{Anders}, F.},
  \bibinfo{author}{{Anderson}, S.F.}, \bibinfo{author}{{Andrews}, B.},
  \bibinfo{author}{et~al.}, \bibinfo{year}{2017}.
\newblock \bibinfo{title}{{Sloan Digital Sky Survey IV: Mapping the Milky Way,
  Nearby Galaxies, and the Distant Universe}}.
\newblock \bibinfo{journal}{The Astronomical Journal} \bibinfo{volume}{154},
  \bibinfo{pages}{28}.
\newblock \DOIprefix\doi{10.3847/1538-3881/aa7567},
  \href{http://arxiv.org/abs/1703.00052}{\tt arXiv:1703.00052}.
\bibitem[{{Bonoli} et~al.(2020){Bonoli}, {Mar{\'\i}n-Franch}, {Varela},
  {V{\'a}zquez Rami{\'o}}, {Abramo}, {Cenarro}, {Dupke}, {V{\'\i}lchez},
  {Crist{\'o}bal-Hornillos}, {Gonz{\'a}lez Delgado},
  {Hern{\'a}ndez-Monteagudo}, {L{\'o}pez-Sanjuan}, {Muniesa}, {Civera},
  {Ederoclite}, {Hern{\'a}n-Caballero}, {Marra}, {Baqui}, {Cortesi},
  {Cypriano}, {Daflon}, {de Amorim}, {D{\'\i}az-Garc{\'\i}a}, {Diego},
  {Mart{\'\i}nez-Solaeche}, {P{\'e}rez}, {Placco}, {Prada}, {Queiroz},
  {Alcaniz}, {Alvarez-Candal}, {Cepa}, {Maroto}, {Roig}, {Siffert}, {Taylor},
  {Benitez}, {Moles}, {Sodr{\'e}}, {Carneiro}, {Mendes de Oliveira}, {Abdalla},
  {Angulo}, {Aparicio Resco}, {Balaguera-Antol{\'\i}nez}, {Ballesteros},
  {Brito-Silva}, {Broadhurst}, {Carrasco}, {Castro}, {Cid Fernandes}, {Coelho},
  {de Melo}, {Doubrawa}, {Fernandez-Soto}, {Ferrari}, {Finoguenov},
  {Garc{\'\i}a-Benito}, {Iglesias-P{\'a}ramo}, {Jim{\'e}nez-Teja}, {Kitaura},
  {Laur}, {Lopes}, {Lucatelli}, {Mart{\'\i}nez}, {Maturi}, {Quartin},
  {Pigozzo}, {Rodr{\`\i}guez-Mart{\`\i}n}, {Salzano}, {Tamm}, {Tempel},
  {Umetsu}, {Valdivielso}, {von Marttens}, {Zitrin}, {D{\'\i}az-Mart{\'\i}n},
  {L{\'o}pez-Alegre}, {L{\'o}pez-Sainz}, {Yanes-D{\'\i}az}, {Rueda-Teruel},
  {Rueda-Teruel}, {Abril Iba{\~n}ez}, {Ant{\'o}n Bravo}, {Bello Ferrer},
  {Bielsa}, {Casino}, {Castillo}, {Chueca}, {Cuesta}, {Garzar{\'a}n Calderaro},
  {Iglesias-Marzoa}, {{\'I}niguez}, {Lamadrid Gutierrez}, {Lopez-Martinez},
  {Lozano-P{\'e}rez}, {Ma{\'\i}cas Sacrist{\'a}n}, {Molina-Ib{\'a}{\~n}ez},
  {Moreno-Signes}, {Rodr{\'\i}guez Llano}, {Royo Navarro}, {Tilve Rua},
  {Andrade}, {Alfaro}, {Akras}, {Arnalte-Mur}, {Ascaso}, {Barbosa},
  {Beltr{\'a}n Jim{\'e}nez}, {Benetti}, {Bengaly}, {Bernui}, {Blanco-Pillado},
  {Borges Fernandes}, {Bregman}, {Bruzual}, {Calderone}, {Carvano}, {Casarini},
  {Chies-Santos}, {Coutinho de Carvalho}, {Dimauro}, {Duarte Puertas},
  {Figueruelo}, {Gonz{\'a}lez-Serrano}, {Guerrero}, {Gurung-L{\'o}pez},
  {Herranz}, {Huertas-Company}, {Irwin}, {Izquierdo-Villalba}, {Kanaan},
  {Kehrig}, {Kirkpatrick}, {Lim}, {Lopes}, {Lopes de Oliveira},
  {Marcos-Caballero}, {Mart{\'\i}nez-Delgado}, {Mart{\'\i}nez-Gonz{\'a}lez},
  {Mart{\'\i}nez-Somonte}, {Oliveira}, {Orsi}, {Overzier}, {Penna-Lima},
  {Reis}, {Spinoso}, {Tsujikawa}, {Vielva}, {Vitorelli}, {Xia}, {Yuan},
  {Arroyo-Polonio}, {Dantas}, {Galarza}, {Gon{\c{c}}alves}, {Gon{\c{c}}alves},
  {Gonzalez}, {Gonzalez}, {Greisel}, {Landim}, {Lazzaro}, {Magris},
  {Monteiro-Oliveira}, {Pereira}, {Rebou{\c{c}}as}, {Rodriguez-Espinosa},
  {Santos da Costa} and {Telles}}]{MiniJPAS}
\bibinfo{author}{{Bonoli}, S.}, \bibinfo{author}{{Mar{\'\i}n-Franch}, A.},
  \bibinfo{author}{{Varela}, J.}, \bibinfo{author}{{V{\'a}zquez Rami{\'o}},
  H.}, \bibinfo{author}{{Abramo}, L.R.}, \bibinfo{author}{{Cenarro}, A.J.},
  \bibinfo{author}{{Dupke}, R.A.}, \bibinfo{author}{{V{\'\i}lchez}, J.M.},
  \bibinfo{author}{{Crist{\'o}bal-Hornillos}, D.},
  \bibinfo{author}{{Gonz{\'a}lez Delgado}, R.M.},
  \bibinfo{author}{{Hern{\'a}ndez-Monteagudo}, C.},
  \bibinfo{author}{{L{\'o}pez-Sanjuan}, C.}, \bibinfo{author}{{Muniesa}, D.J.},
  \bibinfo{author}{{Civera}, T.}, \bibinfo{author}{{Ederoclite}, A.},
  \bibinfo{author}{{Hern{\'a}n-Caballero}, A.}, \bibinfo{author}{{Marra}, V.},
  \bibinfo{author}{{Baqui}, P.O.}, \bibinfo{author}{{Cortesi}, A.},
  \bibinfo{author}{{Cypriano}, E.S.}, \bibinfo{author}{{Daflon}, S.},
  \bibinfo{author}{{de Amorim}, A.L.},
  \bibinfo{author}{{D{\'\i}az-Garc{\'\i}a}, L.A.}, \bibinfo{author}{{Diego},
  J.M.}, \bibinfo{author}{{Mart{\'\i}nez-Solaeche}, G.},
  \bibinfo{author}{{P{\'e}rez}, E.}, \bibinfo{author}{{Placco}, V.M.},
  \bibinfo{author}{{Prada}, F.}, \bibinfo{author}{{Queiroz}, C.},
  \bibinfo{author}{{Alcaniz}, J.}, \bibinfo{author}{{Alvarez-Candal}, A.},
  \bibinfo{author}{{Cepa}, J.}, \bibinfo{author}{{Maroto}, A.L.},
  \bibinfo{author}{{Roig}, F.}, \bibinfo{author}{{Siffert}, B.B.},
  \bibinfo{author}{{Taylor}, K.}, \bibinfo{author}{{Benitez}, N.},
  \bibinfo{author}{{Moles}, M.}, \bibinfo{author}{{Sodr{\'e}}, L., J.},
  \bibinfo{author}{{Carneiro}, S.}, \bibinfo{author}{{Mendes de Oliveira}, C.},
  \bibinfo{author}{{Abdalla}, E.}, \bibinfo{author}{{Angulo}, R.E.},
  \bibinfo{author}{{Aparicio Resco}, M.},
  \bibinfo{author}{{Balaguera-Antol{\'\i}nez}, A.},
  \bibinfo{author}{{Ballesteros}, F.J.}, \bibinfo{author}{{Brito-Silva}, D.},
  \bibinfo{author}{{Broadhurst}, T.}, \bibinfo{author}{{Carrasco}, E.R.},
  \bibinfo{author}{{Castro}, T.}, \bibinfo{author}{{Cid Fernandes}, R.},
  \bibinfo{author}{{Coelho}, P.}, \bibinfo{author}{{de Melo}, R.B.},
  \bibinfo{author}{{Doubrawa}, L.}, \bibinfo{author}{{Fernandez-Soto}, A.},
  \bibinfo{author}{{Ferrari}, F.}, \bibinfo{author}{{Finoguenov}, A.},
  \bibinfo{author}{{Garc{\'\i}a-Benito}, R.},
  \bibinfo{author}{{Iglesias-P{\'a}ramo}, J.},
  \bibinfo{author}{{Jim{\'e}nez-Teja}, Y.}, \bibinfo{author}{{Kitaura}, F.S.},
  \bibinfo{author}{{Laur}, J.}, \bibinfo{author}{{Lopes}, P.A.A.},
  \bibinfo{author}{{Lucatelli}, G.}, \bibinfo{author}{{Mart{\'\i}nez}, V.J.},
  \bibinfo{author}{{Maturi}, M.}, \bibinfo{author}{{Quartin}, M.},
  \bibinfo{author}{{Pigozzo}, C.},
  \bibinfo{author}{{Rodr{\`\i}guez-Mart{\`\i}n}, J.E.},
  \bibinfo{author}{{Salzano}, V.}, \bibinfo{author}{{Tamm}, A.},
  \bibinfo{author}{{Tempel}, E.}, \bibinfo{author}{{Umetsu}, K.},
  \bibinfo{author}{{Valdivielso}, L.}, \bibinfo{author}{{von Marttens}, R.},
  \bibinfo{author}{{Zitrin}, A.}, \bibinfo{author}{{D{\'\i}az-Mart{\'\i}n},
  M.C.}, \bibinfo{author}{{L{\'o}pez-Alegre}, G.},
  \bibinfo{author}{{L{\'o}pez-Sainz}, A.}, \bibinfo{author}{{Yanes-D{\'\i}az},
  A.}, \bibinfo{author}{{Rueda-Teruel}, F.}, \bibinfo{author}{{Rueda-Teruel},
  S.}, \bibinfo{author}{{Abril Iba{\~n}ez}, J.}, \bibinfo{author}{{Ant{\'o}n
  Bravo}, J.L.}, \bibinfo{author}{{Bello Ferrer}, R.},
  \bibinfo{author}{{Bielsa}, S.}, \bibinfo{author}{{Casino}, J.M.},
  \bibinfo{author}{{Castillo}, J.}, \bibinfo{author}{{Chueca}, S.},
  \bibinfo{author}{{Cuesta}, L.}, \bibinfo{author}{{Garzar{\'a}n Calderaro},
  J.}, \bibinfo{author}{{Iglesias-Marzoa}, R.}, \bibinfo{author}{{{\'I}niguez},
  C.}, \bibinfo{author}{{Lamadrid Gutierrez}, J.L.},
  \bibinfo{author}{{Lopez-Martinez}, F.}, \bibinfo{author}{{Lozano-P{\'e}rez},
  D.}, \bibinfo{author}{{Ma{\'\i}cas Sacrist{\'a}n}, N.},
  \bibinfo{author}{{Molina-Ib{\'a}{\~n}ez}, E.L.},
  \bibinfo{author}{{Moreno-Signes}, A.}, \bibinfo{author}{{Rodr{\'\i}guez
  Llano}, S.}, \bibinfo{author}{{Royo Navarro}, M.}, \bibinfo{author}{{Tilve
  Rua}, V.}, \bibinfo{author}{{Andrade}, U.}, \bibinfo{author}{{Alfaro}, E.J.},
  \bibinfo{author}{{Akras}, S.}, \bibinfo{author}{{Arnalte-Mur}, P.},
  \bibinfo{author}{{Ascaso}, B.}, \bibinfo{author}{{Barbosa}, C.E.},
  \bibinfo{author}{{Beltr{\'a}n Jim{\'e}nez}, J.}, \bibinfo{author}{{Benetti},
  M.}, \bibinfo{author}{{Bengaly}, C.A.P.}, \bibinfo{author}{{Bernui}, A.},
  \bibinfo{author}{{Blanco-Pillado}, J.J.}, \bibinfo{author}{{Borges
  Fernandes}, M.}, \bibinfo{author}{{Bregman}, J.N.},
  \bibinfo{author}{{Bruzual}, G.}, \bibinfo{author}{{Calderone}, G.},
  \bibinfo{author}{{Carvano}, J.M.}, \bibinfo{author}{{Casarini}, L.},
  \bibinfo{author}{{Chies-Santos}, A.L.}, \bibinfo{author}{{Coutinho de
  Carvalho}, G.}, \bibinfo{author}{{Dimauro}, P.}, \bibinfo{author}{{Duarte
  Puertas}, S.}, \bibinfo{author}{{Figueruelo}, D.},
  \bibinfo{author}{{Gonz{\'a}lez-Serrano}, J.I.}, \bibinfo{author}{{Guerrero},
  M.A.}, \bibinfo{author}{{Gurung-L{\'o}pez}, S.}, \bibinfo{author}{{Herranz},
  D.}, \bibinfo{author}{{Huertas-Company}, M.}, \bibinfo{author}{{Irwin},
  J.A.}, \bibinfo{author}{{Izquierdo-Villalba}, D.}, \bibinfo{author}{{Kanaan},
  A.}, \bibinfo{author}{{Kehrig}, C.}, \bibinfo{author}{{Kirkpatrick}, C.C.},
  \bibinfo{author}{{Lim}, J.}, \bibinfo{author}{{Lopes}, A.R.},
  \bibinfo{author}{{Lopes de Oliveira}, R.},
  \bibinfo{author}{{Marcos-Caballero}, A.},
  \bibinfo{author}{{Mart{\'\i}nez-Delgado}, D.},
  \bibinfo{author}{{Mart{\'\i}nez-Gonz{\'a}lez}, E.},
  \bibinfo{author}{{Mart{\'\i}nez-Somonte}, G.}, \bibinfo{author}{{Oliveira},
  N.}, \bibinfo{author}{{Orsi}, A.A.}, \bibinfo{author}{{Overzier}, R.A.},
  \bibinfo{author}{{Penna-Lima}, M.}, \bibinfo{author}{{Reis}, R.R.R.},
  \bibinfo{author}{{Spinoso}, D.}, \bibinfo{author}{{Tsujikawa}, S.},
  \bibinfo{author}{{Vielva}, P.}, \bibinfo{author}{{Vitorelli}, A.Z.},
  \bibinfo{author}{{Xia}, J.Q.}, \bibinfo{author}{{Yuan}, H.B.},
  \bibinfo{author}{{Arroyo-Polonio}, A.}, \bibinfo{author}{{Dantas}, M.L.L.},
  \bibinfo{author}{{Galarza}, C.A.}, \bibinfo{author}{{Gon{\c{c}}alves}, D.R.},
  \bibinfo{author}{{Gon{\c{c}}alves}, R.S.}, \bibinfo{author}{{Gonzalez},
  J.E.}, \bibinfo{author}{{Gonzalez}, A.H.}, \bibinfo{author}{{Greisel}, N.},
  \bibinfo{author}{{Landim}, R.G.}, \bibinfo{author}{{Lazzaro}, D.},
  \bibinfo{author}{{Magris}, G.}, \bibinfo{author}{{Monteiro-Oliveira}, R.},
  \bibinfo{author}{{Pereira}, C.B.}, \bibinfo{author}{{Rebou{\c{c}}as}, M.J.},
  \bibinfo{author}{{Rodriguez-Espinosa}, J.M.}, \bibinfo{author}{{Santos da
  Costa}, S.}, \bibinfo{author}{{Telles}, E.}, \bibinfo{year}{2020}.
\newblock \bibinfo{title}{{The miniJPAS survey: a preview of the Universe in 56
  colours}}.
\newblock \bibinfo{journal}{arXiv e-prints} ,
  \bibinfo{pages}{arXiv:2007.01910}\href{http://arxiv.org/abs/2007.01910}{\tt
  arXiv:2007.01910}.
\bibitem[{Brammer et~al.(2008)Brammer, van Dokkum and Coppi}]{Brammer2008}
\bibinfo{author}{Brammer, G.B.}, \bibinfo{author}{van Dokkum, P.G.},
  \bibinfo{author}{Coppi, P.}, \bibinfo{year}{2008}.
\newblock \bibinfo{title}{{EAZY}: A fast, public photometric redshift code}.
\newblock \bibinfo{journal}{The Astrophysical Journal} \bibinfo{volume}{686},
  \bibinfo{pages}{1503--1513}.
\newblock \URLprefix \url{https://doi.org/10.1086/591786},
  \DOIprefix\doi{10.1086/591786}.
\bibitem[{Cenarro et~al.(2019)Cenarro, Moles, Crist{\'{o}}bal-Hornillos,
  Mar{\'{\i}}n-Franch, Ederoclite, Varela, L{\'{o}}pez-Sanjuan,
  Hern{\'{a}}ndez-Monteagudo, Angulo, Rami{\'{o}}, Viironen, Bonoli, Orsi,
  Hurier, Roman, Greisel, Vilella-Rojo, D{\'{\i}}az-Garc{\'{\i}}a,
  Logro{\~{n}}o-Garc{\'{\i}}a, Gurung-L{\'{o}}pez, Spinoso, Izquierdo-Villalba,
  Aguerri, Prieto, Bonatto, Carvano, Chies-Santos, Daflon, Dupke,
  Falc{\'{o}}n-Barroso, Gon{\c{c}}alves, Jim{\'{e}}nez-Teja, Molino, Placco,
  Solano, Whitten, Abril, Ant{\'{o}}n, Bello, de~Toledo,
  Castillo-Ram{\'{\i}}rez, Chueca, Civera, D{\'{\i}}az-Mart{\'{\i}}n,
  Dom{\'{\i}}nguez-Mart{\'{\i}}nez, Garzar{\'{a}}n-Calderaro,
  Hern{\'{a}}ndez-Fuertes, Iglesias-Marzoa, I{\~{n}}iguez, Ruiz, Kruuse,
  Lamadrid, Lasso-Cabrera, L{\'{o}}pez-Alegre, L{\'{o}}pez-Sainz,
  Ma{\'{\i}}cas, Moreno-Signes, Muniesa, Rodr{\'{\i}}guez-Llano, Rueda-Teruel,
  Rueda-Teruel, Soriano-Lagu{\'{\i}}a, Tilve, Valdivielso, Yanes-D{\'{\i}}az,
  Alcaniz, de~Oliveira, Sodr{\'{e}}, Coelho, de~Oliveira, Tamm, Xavier, Abramo,
  Akras, Alfaro, Alvarez-Candal, Ascaso, Beasley, Beers, Fernandes, Bruzual,
  Buzzo, Carrasco, Cepa, Cortesi, Costa-Duarte, Pr{\'{a}}, Favole, Galarza,
  Galbany, Garcia, Delgado, Gonz{\'{a}}lez-Serrano, Guti{\'{e}}rrez-Soto,
  Hernandez-Jimenez, Kanaan, Kuncarayakti, Landim, Laur, Licandro, Neto, Lyman,
  Apell{\'{a}}niz, Miralda-Escud{\'{e}}, Morate, Nogueira-Cavalcante, Novais,
  Oncins, Oteo, Overzier, Pereira, Rebassa-Mansergas, Reis, Roig, Sako,
  Salvador-Rusi{\~{n}}ol, Sampedro, S{\'{a}}nchez-Bl{\'{a}}zquez, Santos,
  Schmidtobreick, Siffert, Telles and Vilchez}]{JPLUS}
\bibinfo{author}{Cenarro, A.J.}, \bibinfo{author}{Moles, M.},
  \bibinfo{author}{Crist{\'{o}}bal-Hornillos, D.},
  \bibinfo{author}{Mar{\'{\i}}n-Franch, A.}, \bibinfo{author}{Ederoclite, A.},
  \bibinfo{author}{Varela, J.}, \bibinfo{author}{L{\'{o}}pez-Sanjuan, C.},
  \bibinfo{author}{Hern{\'{a}}ndez-Monteagudo, C.}, \bibinfo{author}{Angulo,
  R.E.}, \bibinfo{author}{Rami{\'{o}}, H.V.}, \bibinfo{author}{Viironen, K.},
  \bibinfo{author}{Bonoli, S.}, \bibinfo{author}{Orsi, A.A.},
  \bibinfo{author}{Hurier, G.}, \bibinfo{author}{Roman, I.S.},
  \bibinfo{author}{Greisel, N.}, \bibinfo{author}{Vilella-Rojo, G.},
  \bibinfo{author}{D{\'{\i}}az-Garc{\'{\i}}a, L.A.},
  \bibinfo{author}{Logro{\~{n}}o-Garc{\'{\i}}a, R.},
  \bibinfo{author}{Gurung-L{\'{o}}pez, S.}, \bibinfo{author}{Spinoso, D.},
  \bibinfo{author}{Izquierdo-Villalba, D.}, \bibinfo{author}{Aguerri, J.A.L.},
  \bibinfo{author}{Prieto, C.A.}, \bibinfo{author}{Bonatto, C.},
  \bibinfo{author}{Carvano, J.M.}, \bibinfo{author}{Chies-Santos, A.L.},
  \bibinfo{author}{Daflon, S.}, \bibinfo{author}{Dupke, R.A.},
  \bibinfo{author}{Falc{\'{o}}n-Barroso, J.}, \bibinfo{author}{Gon{\c{c}}alves,
  D.R.}, \bibinfo{author}{Jim{\'{e}}nez-Teja, Y.}, \bibinfo{author}{Molino,
  A.}, \bibinfo{author}{Placco, V.M.}, \bibinfo{author}{Solano, E.},
  \bibinfo{author}{Whitten, D.D.}, \bibinfo{author}{Abril, J.},
  \bibinfo{author}{Ant{\'{o}}n, J.L.}, \bibinfo{author}{Bello, R.},
  \bibinfo{author}{de~Toledo, S.B.}, \bibinfo{author}{Castillo-Ram{\'{\i}}rez,
  J.}, \bibinfo{author}{Chueca, S.}, \bibinfo{author}{Civera, T.},
  \bibinfo{author}{D{\'{\i}}az-Mart{\'{\i}}n, M.C.},
  \bibinfo{author}{Dom{\'{\i}}nguez-Mart{\'{\i}}nez, M.},
  \bibinfo{author}{Garzar{\'{a}}n-Calderaro, J.},
  \bibinfo{author}{Hern{\'{a}}ndez-Fuertes, J.},
  \bibinfo{author}{Iglesias-Marzoa, R.}, \bibinfo{author}{I{\~{n}}iguez, C.},
  \bibinfo{author}{Ruiz, J.M.J.}, \bibinfo{author}{Kruuse, K.},
  \bibinfo{author}{Lamadrid, J.L.}, \bibinfo{author}{Lasso-Cabrera, N.},
  \bibinfo{author}{L{\'{o}}pez-Alegre, G.}, \bibinfo{author}{L{\'{o}}pez-Sainz,
  A.}, \bibinfo{author}{Ma{\'{\i}}cas, N.}, \bibinfo{author}{Moreno-Signes,
  A.}, \bibinfo{author}{Muniesa, D.J.},
  \bibinfo{author}{Rodr{\'{\i}}guez-Llano, S.}, \bibinfo{author}{Rueda-Teruel,
  F.}, \bibinfo{author}{Rueda-Teruel, S.},
  \bibinfo{author}{Soriano-Lagu{\'{\i}}a, I.}, \bibinfo{author}{Tilve, V.},
  \bibinfo{author}{Valdivielso, L.}, \bibinfo{author}{Yanes-D{\'{\i}}az, A.},
  \bibinfo{author}{Alcaniz, J.S.}, \bibinfo{author}{de~Oliveira, C.M.},
  \bibinfo{author}{Sodr{\'{e}}, L.}, \bibinfo{author}{Coelho, P.},
  \bibinfo{author}{de~Oliveira, R.L.}, \bibinfo{author}{Tamm, A.},
  \bibinfo{author}{Xavier, H.S.}, \bibinfo{author}{Abramo, L.R.},
  \bibinfo{author}{Akras, S.}, \bibinfo{author}{Alfaro, E.J.},
  \bibinfo{author}{Alvarez-Candal, A.}, \bibinfo{author}{Ascaso, B.},
  \bibinfo{author}{Beasley, M.A.}, \bibinfo{author}{Beers, T.C.},
  \bibinfo{author}{Fernandes, M.B.}, \bibinfo{author}{Bruzual, G.R.},
  \bibinfo{author}{Buzzo, M.L.}, \bibinfo{author}{Carrasco, J.M.},
  \bibinfo{author}{Cepa, J.}, \bibinfo{author}{Cortesi, A.},
  \bibinfo{author}{Costa-Duarte, M.V.}, \bibinfo{author}{Pr{\'{a}}, M.D.},
  \bibinfo{author}{Favole, G.}, \bibinfo{author}{Galarza, A.},
  \bibinfo{author}{Galbany, L.}, \bibinfo{author}{Garcia, K.},
  \bibinfo{author}{Delgado, R.M.G.}, \bibinfo{author}{Gonz{\'{a}}lez-Serrano,
  J.I.}, \bibinfo{author}{Guti{\'{e}}rrez-Soto, L.A.},
  \bibinfo{author}{Hernandez-Jimenez, J.A.}, \bibinfo{author}{Kanaan, A.},
  \bibinfo{author}{Kuncarayakti, H.}, \bibinfo{author}{Landim, R.C.G.},
  \bibinfo{author}{Laur, J.}, \bibinfo{author}{Licandro, J.},
  \bibinfo{author}{Neto, G.B.L.}, \bibinfo{author}{Lyman, J.D.},
  \bibinfo{author}{Apell{\'{a}}niz, J.M.},
  \bibinfo{author}{Miralda-Escud{\'{e}}, J.}, \bibinfo{author}{Morate, D.},
  \bibinfo{author}{Nogueira-Cavalcante, J.P.}, \bibinfo{author}{Novais, P.M.},
  \bibinfo{author}{Oncins, M.}, \bibinfo{author}{Oteo, I.},
  \bibinfo{author}{Overzier, R.A.}, \bibinfo{author}{Pereira, C.B.},
  \bibinfo{author}{Rebassa-Mansergas, A.}, \bibinfo{author}{Reis, R.R.R.},
  \bibinfo{author}{Roig, F.}, \bibinfo{author}{Sako, M.},
  \bibinfo{author}{Salvador-Rusi{\~{n}}ol, N.}, \bibinfo{author}{Sampedro, L.},
  \bibinfo{author}{S{\'{a}}nchez-Bl{\'{a}}zquez, P.}, \bibinfo{author}{Santos,
  W.A.}, \bibinfo{author}{Schmidtobreick, L.}, \bibinfo{author}{Siffert, B.B.},
  \bibinfo{author}{Telles, E.}, \bibinfo{author}{Vilchez, J.M.},
  \bibinfo{year}{2019}.
\newblock \bibinfo{title}{J-{PLUS}: The javalambre photometric local universe
  survey}.
\newblock \bibinfo{journal}{Astronomy {\&} Astrophysics} \bibinfo{volume}{622},
  \bibinfo{pages}{A176}.
\newblock \URLprefix \url{https://doi.org/10.1051/0004-6361/201833036},
  \DOIprefix\doi{10.1051/0004-6361/201833036}.
\bibitem[{Chollet(2017)}]{CholletDL}
\bibinfo{author}{Chollet, F.}, \bibinfo{year}{2017}.
\newblock \bibinfo{title}{Deep Learning with Python}.
\newblock \bibinfo{publisher}{Manning}.
\bibitem[{Chollet et~al.(2015)}]{Keras}
\bibinfo{author}{Chollet, F.}, et~al., \bibinfo{year}{2015}.
\newblock \bibinfo{title}{Keras}.
\newblock \bibinfo{howpublished}{\url{https://keras.io}}.
\bibitem[{{Costa-Duarte} et~al.(2019){Costa-Duarte}, {Sampedro}, {Molino},
  {Xavier}, {Herpich}, {Chies-Santos}, {Barbosa}, {Cortesi}, {Schoenell},
  {Kanaan}, {Ribeiro}, {Mendes de Oliveira}, {Akras}, {Alvarez-Candal},
  {Barbosa}, {Castell{\'o}n}, {Coelho}, {Dantas}, {Dupke}, {Ederoclite},
  {Galarza}, {Gon{\c{c}}alves}, {Hernandez-Jimenez}, {Jim{\'e}nez-Teja},
  {Lopes}, {Lopes}, {Lopes de Oliveira}, {Melo de Azevedo}, {Nakazono},
  {Perottoni}, {Queiroz}, {Saha}, {Sodr{\'e}}, {Telles} and {Thom de
  Souza}}]{MarcusSep}
\bibinfo{author}{{Costa-Duarte}, M.V.}, \bibinfo{author}{{Sampedro}, L.},
  \bibinfo{author}{{Molino}, A.}, \bibinfo{author}{{Xavier}, H.S.},
  \bibinfo{author}{{Herpich}, F.R.}, \bibinfo{author}{{Chies-Santos}, A.L.},
  \bibinfo{author}{{Barbosa}, C.E.}, \bibinfo{author}{{Cortesi}, A.},
  \bibinfo{author}{{Schoenell}, W.}, \bibinfo{author}{{Kanaan}, A.},
  \bibinfo{author}{{Ribeiro}, T.}, \bibinfo{author}{{Mendes de Oliveira}, C.},
  \bibinfo{author}{{Akras}, S.}, \bibinfo{author}{{Alvarez-Candal}, A.},
  \bibinfo{author}{{Barbosa}, C.L.}, \bibinfo{author}{{Castell{\'o}n}, J.L.N.},
  \bibinfo{author}{{Coelho}, P.}, \bibinfo{author}{{Dantas}, M.L.L.},
  \bibinfo{author}{{Dupke}, R.}, \bibinfo{author}{{Ederoclite}, A.},
  \bibinfo{author}{{Galarza}, A.}, \bibinfo{author}{{Gon{\c{c}}alves}, T.S.},
  \bibinfo{author}{{Hernandez-Jimenez}, J.A.},
  \bibinfo{author}{{Jim{\'e}nez-Teja}, Y.}, \bibinfo{author}{{Lopes}, A.},
  \bibinfo{author}{{Lopes}, P.A.A.}, \bibinfo{author}{{Lopes de Oliveira}, R.},
  \bibinfo{author}{{Melo de Azevedo}, J.L.}, \bibinfo{author}{{Nakazono},
  L.M.}, \bibinfo{author}{{Perottoni}, H.D.}, \bibinfo{author}{{Queiroz}, C.},
  \bibinfo{author}{{Saha}, K.}, \bibinfo{author}{{Sodr{\'e}}, L., J.},
  \bibinfo{author}{{Telles}, E.}, \bibinfo{author}{{Thom de Souza}, R.C.},
  \bibinfo{year}{2019}.
\newblock \bibinfo{title}{{The S-PLUS: a star/galaxy classification based on a
  Machine Learning approach}}.
\newblock \bibinfo{journal}{arXiv e-prints} ,
  \bibinfo{pages}{arXiv:1909.08626}\href{http://arxiv.org/abs/1909.08626}{\tt
  arXiv:1909.08626}.
\bibitem[{{Dillon} et~al.(2017){Dillon}, {Langmore}, {Tran}, {Brevdo},
  {Vasudevan}, {Moore}, {Patton}, {Alemi}, {Hoffman} and {Saurous}}]{TFP}
\bibinfo{author}{{Dillon}, J.V.}, \bibinfo{author}{{Langmore}, I.},
  \bibinfo{author}{{Tran}, D.}, \bibinfo{author}{{Brevdo}, E.},
  \bibinfo{author}{{Vasudevan}, S.}, \bibinfo{author}{{Moore}, D.},
  \bibinfo{author}{{Patton}, B.}, \bibinfo{author}{{Alemi}, A.},
  \bibinfo{author}{{Hoffman}, M.}, \bibinfo{author}{{Saurous}, R.A.},
  \bibinfo{year}{2017}.
\newblock \bibinfo{title}{{TensorFlow Distributions}}.
\newblock \bibinfo{journal}{arXiv e-prints} ,
  \bibinfo{pages}{arXiv:1711.10604}\href{http://arxiv.org/abs/1711.10604}{\tt
  arXiv:1711.10604}.
\bibitem[{Dozat(2016)}]{Nadam}
\bibinfo{author}{Dozat, T.}, \bibinfo{year}{2016}.
\newblock \bibinfo{title}{Incorporating nesterov momentum into adam}.
\bibitem[{Feldmann et~al.(2006)Feldmann, Carollo, Porciani, Lilly, Capak,
  Taniguchi, F{\`{e}}vre, Renzini, Scoville, Ajiki, Aussel, Contini, McCracken,
  Mobasher, Murayama, Sanders, Sasaki, Scarlata, Scodeggio, Shioya, Silverman,
  Takahashi, Thompson and Zamorani}]{Feldmann2006}
\bibinfo{author}{Feldmann, R.}, \bibinfo{author}{Carollo, C.M.},
  \bibinfo{author}{Porciani, C.}, \bibinfo{author}{Lilly, S.J.},
  \bibinfo{author}{Capak, P.}, \bibinfo{author}{Taniguchi, Y.},
  \bibinfo{author}{F{\`{e}}vre, O.L.}, \bibinfo{author}{Renzini, A.},
  \bibinfo{author}{Scoville, N.}, \bibinfo{author}{Ajiki, M.},
  \bibinfo{author}{Aussel, H.}, \bibinfo{author}{Contini, T.},
  \bibinfo{author}{McCracken, H.}, \bibinfo{author}{Mobasher, B.},
  \bibinfo{author}{Murayama, T.}, \bibinfo{author}{Sanders, D.},
  \bibinfo{author}{Sasaki, S.}, \bibinfo{author}{Scarlata, C.},
  \bibinfo{author}{Scodeggio, M.}, \bibinfo{author}{Shioya, Y.},
  \bibinfo{author}{Silverman, J.}, \bibinfo{author}{Takahashi, M.},
  \bibinfo{author}{Thompson, D.}, \bibinfo{author}{Zamorani, G.},
  \bibinfo{year}{2006}.
\newblock \bibinfo{title}{The zurich extragalactic bayesian redshift analyzer
  and its first application: {COSMOS}}.
\newblock \bibinfo{journal}{Monthly Notices of the Royal Astronomical Society}
  \bibinfo{volume}{372}, \bibinfo{pages}{565--577}.
\newblock \URLprefix \url{https://doi.org/10.1111/j.1365-2966.2006.10930.x},
  \DOIprefix\doi{10.1111/j.1365-2966.2006.10930.x}.
\bibitem[{Firth et~al.(2003)Firth, Lahav and Somerville}]{Firth2003}
\bibinfo{author}{Firth, A.E.}, \bibinfo{author}{Lahav, O.},
  \bibinfo{author}{Somerville, R.S.}, \bibinfo{year}{2003}.
\newblock \bibinfo{title}{Estimating photometric redshifts with artificial
  neural networks}.
\newblock \bibinfo{journal}{Monthly Notices of the Royal Astronomical Society}
  \bibinfo{volume}{339}, \bibinfo{pages}{1195--1202}.
\newblock \URLprefix \url{https://doi.org/10.1046/j.1365-8711.2003.06271.x},
  \DOIprefix\doi{10.1046/j.1365-8711.2003.06271.x}.
\bibitem[{Gehrels and Spergel(2015)}]{WFIRST}
\bibinfo{author}{Gehrels, N.}, \bibinfo{author}{Spergel, D.},
  \bibinfo{year}{2015}.
\newblock \bibinfo{title}{Wide-field {InfraRed} survey telescope ({WFIRST})
  mission and synergies with {LISA} and {LIGO}-virgo}.
\newblock \bibinfo{journal}{Journal of Physics: Conference Series}
  \bibinfo{volume}{610}, \bibinfo{pages}{012007}.
\newblock \URLprefix \url{https://doi.org/10.1088/1742-6596/610/1/012007},
  \DOIprefix\doi{10.1088/1742-6596/610/1/012007}.
\bibitem[{Glorot et~al.(2011)Glorot, Bordes and Bengio}]{ReLU}
\bibinfo{author}{Glorot, X.}, \bibinfo{author}{Bordes, A.},
  \bibinfo{author}{Bengio, Y.}, \bibinfo{year}{2011}.
\newblock \bibinfo{title}{Deep sparse rectifier neural networks}, in:
  \bibinfo{editor}{Gordon, G.}, \bibinfo{editor}{Dunson, D.},
  \bibinfo{editor}{Dudík, M.} (Eds.), \bibinfo{booktitle}{Proceedings of the
  Fourteenth International Conference on Artificial Intelligence and
  Statistics}, \bibinfo{publisher}{PMLR}, \bibinfo{address}{Fort Lauderdale,
  FL, USA}. pp. \bibinfo{pages}{315--323}.
\newblock \URLprefix \url{http://proceedings.mlr.press/v15/glorot11a.html}.
\bibitem[{Gomes et~al.(2017)Gomes, Jarvis, Almosallam and Roberts}]{Gomes2017}
\bibinfo{author}{Gomes, Z.}, \bibinfo{author}{Jarvis, M.J.},
  \bibinfo{author}{Almosallam, I.A.}, \bibinfo{author}{Roberts, S.J.},
  \bibinfo{year}{2017}.
\newblock \bibinfo{title}{Improving photometric redshift estimation using
  {GPz}: size information, post processing, and improved photometry}.
\newblock \bibinfo{journal}{Monthly Notices of the Royal Astronomical Society}
  \bibinfo{volume}{475}, \bibinfo{pages}{331--342}.
\newblock \URLprefix \url{https://doi.org/10.1093/mnras/stx3187},
  \DOIprefix\doi{10.1093/mnras/stx3187}.
\bibitem[{Goodfellow et~al.(2016)Goodfellow, Bengio and
  Courville}]{Goodfellow-et-al-2016}
\bibinfo{author}{Goodfellow, I.}, \bibinfo{author}{Bengio, Y.},
  \bibinfo{author}{Courville, A.}, \bibinfo{year}{2016}.
\newblock \bibinfo{title}{Deep Learning}.
\newblock \bibinfo{publisher}{MIT Press}.
\newblock \bibinfo{note}{\url{http://www.deeplearningbook.org}}.
\bibitem[{{Gruel} et~al.(2012){Gruel}, {Moles}, {Varela},
  {Crist{\'o}bal-Hornillos}, {Ederoclite}, {Cenarro}, {Mar{\'\i}n-Franch},
  {Chueca}, {Yanes-D{\'\i}az}, {Rueda-Teruel}, {Rueda-Teruel}, {Luis-Simoes},
  {L{\'o}pez-Sainz} and {Hern{\'a}ndez Fuertes}}]{Gruel2012}
\bibinfo{author}{{Gruel}, N.}, \bibinfo{author}{{Moles}, M.},
  \bibinfo{author}{{Varela}, J.}, \bibinfo{author}{{Crist{\'o}bal-Hornillos},
  D.}, \bibinfo{author}{{Ederoclite}, A.}, \bibinfo{author}{{Cenarro}, A.J.},
  \bibinfo{author}{{Mar{\'\i}n-Franch}, A.}, \bibinfo{author}{{Chueca}, S.},
  \bibinfo{author}{{Yanes-D{\'\i}az}, A.}, \bibinfo{author}{{Rueda-Teruel},
  F.}, \bibinfo{author}{{Rueda-Teruel}, S.}, \bibinfo{author}{{Luis-Simoes},
  R.}, \bibinfo{author}{{L{\'o}pez-Sainz}, A.}, \bibinfo{author}{{Hern{\'a}ndez
  Fuertes}, J.}, \bibinfo{year}{2012}.
\newblock \bibinfo{title}{{Calibration plan for J-PAS and J-PLUS surveys}}, in:
  \bibinfo{editor}{{Peck}, A.B.}, \bibinfo{editor}{{Seaman}, R.L.},
  \bibinfo{editor}{{Comeron}, F.} (Eds.), \bibinfo{booktitle}{Observatory
  Operations: Strategies, Processes, and Systems IV}, p.
  \bibinfo{pages}{84481V}.
\newblock \DOIprefix\doi{10.1117/12.925581}.
\bibitem[{Hersbach(2000)}]{CRPS}
\bibinfo{author}{Hersbach, H.}, \bibinfo{year}{2000}.
\newblock \bibinfo{title}{Decomposition of the continuous ranked probability
  score for ensemble prediction systems}.
\newblock \bibinfo{journal}{Weather and Forecasting} \bibinfo{volume}{15},
  \bibinfo{pages}{559 -- 570}.
\newblock \URLprefix
  \url{https://journals.ametsoc.org/view/journals/wefo/15/5/1520-0434_2000_015_0559_dotcrp_2_0_co_2.xml},
  \DOIprefix\doi{10.1175/1520-0434(2000)015<0559:DOTCRP>2.0.CO;2}.
\bibitem[{Hildebrandt et~al.(2012)Hildebrandt, Erben, Kuijken, van Waerbeke,
  Heymans, Coupon, Benjamin, Bonnett, Fu, Hoekstra, Kitching, Mellier, Miller,
  Velander, Hudson, Rowe, Schrabback, Semboloni and
  Ben{\'{\i}}tez}]{Hildebrandt2012}
\bibinfo{author}{Hildebrandt, H.}, \bibinfo{author}{Erben, T.},
  \bibinfo{author}{Kuijken, K.}, \bibinfo{author}{van Waerbeke, L.},
  \bibinfo{author}{Heymans, C.}, \bibinfo{author}{Coupon, J.},
  \bibinfo{author}{Benjamin, J.}, \bibinfo{author}{Bonnett, C.},
  \bibinfo{author}{Fu, L.}, \bibinfo{author}{Hoekstra, H.},
  \bibinfo{author}{Kitching, T.D.}, \bibinfo{author}{Mellier, Y.},
  \bibinfo{author}{Miller, L.}, \bibinfo{author}{Velander, M.},
  \bibinfo{author}{Hudson, M.J.}, \bibinfo{author}{Rowe, B.T.P.},
  \bibinfo{author}{Schrabback, T.}, \bibinfo{author}{Semboloni, E.},
  \bibinfo{author}{Ben{\'{\i}}tez, N.}, \bibinfo{year}{2012}.
\newblock \bibinfo{title}{{CFHTLenS}: improving the quality of photometric
  redshifts with precision photometry}.
\newblock \bibinfo{journal}{Monthly Notices of the Royal Astronomical Society}
  \bibinfo{volume}{421}, \bibinfo{pages}{2355--2367}.
\newblock \URLprefix \url{https://doi.org/10.1111/j.1365-2966.2012.20468.x},
  \DOIprefix\doi{10.1111/j.1365-2966.2012.20468.x}.
\bibitem[{Hildebrandt et~al.(2016)Hildebrandt, Viola, Heymans, Joudaki,
  Kuijken, Blake, Erben, Joachimi, Klaes, Miller, Morrison, Nakajima, Kleijn,
  Amon, Choi, Covone, de~Jong, Dvornik, Conti, Grado, Harnois-D{\'{e}}raps,
  Herbonnet, Hoekstra, K\"{o}hlinger, McFarland, Mead, Merten, Napolitano,
  Peacock, Radovich, Schneider, Simon, Valentijn, van~den Busch, van Uitert and
  Waerbeke}]{KIDS}
\bibinfo{author}{Hildebrandt, H.}, \bibinfo{author}{Viola, M.},
  \bibinfo{author}{Heymans, C.}, \bibinfo{author}{Joudaki, S.},
  \bibinfo{author}{Kuijken, K.}, \bibinfo{author}{Blake, C.},
  \bibinfo{author}{Erben, T.}, \bibinfo{author}{Joachimi, B.},
  \bibinfo{author}{Klaes, D.}, \bibinfo{author}{Miller, L.},
  \bibinfo{author}{Morrison, C.B.}, \bibinfo{author}{Nakajima, R.},
  \bibinfo{author}{Kleijn, G.V.}, \bibinfo{author}{Amon, A.},
  \bibinfo{author}{Choi, A.}, \bibinfo{author}{Covone, G.},
  \bibinfo{author}{de~Jong, J.T.A.}, \bibinfo{author}{Dvornik, A.},
  \bibinfo{author}{Conti, I.F.}, \bibinfo{author}{Grado, A.},
  \bibinfo{author}{Harnois-D{\'{e}}raps, J.}, \bibinfo{author}{Herbonnet, R.},
  \bibinfo{author}{Hoekstra, H.}, \bibinfo{author}{K\"{o}hlinger, F.},
  \bibinfo{author}{McFarland, J.}, \bibinfo{author}{Mead, A.},
  \bibinfo{author}{Merten, J.}, \bibinfo{author}{Napolitano, N.},
  \bibinfo{author}{Peacock, J.A.}, \bibinfo{author}{Radovich, M.},
  \bibinfo{author}{Schneider, P.}, \bibinfo{author}{Simon, P.},
  \bibinfo{author}{Valentijn, E.A.}, \bibinfo{author}{van~den Busch, J.L.},
  \bibinfo{author}{van Uitert, E.}, \bibinfo{author}{Waerbeke, L.V.},
  \bibinfo{year}{2016}.
\newblock \bibinfo{title}{{KiDS}-450: cosmological parameter constraints from
  tomographic weak gravitational lensing}.
\newblock \bibinfo{journal}{Monthly Notices of the Royal Astronomical Society}
  \bibinfo{volume}{465}, \bibinfo{pages}{1454--1498}.
\newblock \URLprefix \url{https://doi.org/10.1093/mnras/stw2805},
  \DOIprefix\doi{10.1093/mnras/stw2805}.
\bibitem[{Hoyle et~al.(2015)Hoyle, Rau, Bonnett, Seitz and Weller}]{Hoyle2015}
\bibinfo{author}{Hoyle, B.}, \bibinfo{author}{Rau, M.M.},
  \bibinfo{author}{Bonnett, C.}, \bibinfo{author}{Seitz, S.},
  \bibinfo{author}{Weller, J.}, \bibinfo{year}{2015}.
\newblock \bibinfo{title}{Data augmentation for machine learning redshifts
  applied to sloan digital sky survey galaxies}.
\newblock \bibinfo{journal}{Monthly Notices of the Royal Astronomical Society}
  \bibinfo{volume}{450}, \bibinfo{pages}{305--316}.
\newblock \URLprefix \url{https://doi.org/10.1093/mnras/stv599},
  \DOIprefix\doi{10.1093/mnras/stv599}.
\bibitem[{{H{\"u}llermeier} and {Waegeman}(2019)}]{Hullermeier2019}
\bibinfo{author}{{H{\"u}llermeier}, E.}, \bibinfo{author}{{Waegeman}, W.},
  \bibinfo{year}{2019}.
\newblock \bibinfo{title}{{Aleatoric and Epistemic Uncertainty in Machine
  Learning: An Introduction to Concepts and Methods}}.
\newblock \bibinfo{journal}{arXiv e-prints} ,
  \bibinfo{pages}{arXiv:1910.09457}\href{http://arxiv.org/abs/1910.09457}{\tt
  arXiv:1910.09457}.
\bibitem[{Ilbert et~al.(2006)Ilbert, Arnouts, McCracken, Bolzonella, Bertin,
  F{\`{e}}vre, Mellier, Zamorani, Pell{\`{o}}, Iovino, Tresse, Brun, Bottini,
  Garilli, Maccagni, Picat, Scaramella, Scodeggio, Vettolani, Zanichelli,
  Adami, Bardelli, Cappi, Charlot, Ciliegi, Contini, Cucciati, Foucaud,
  Franzetti, Gavignaud, Guzzo, Marano, Marinoni, Mazure, Meneux, Merighi,
  Paltani, Pollo, Pozzetti, Radovich, Zucca, Bondi, Bongiorno, Busarello,
  Torre, Gregorini, Lamareille, Mathez, Merluzzi, Ripepi, Rizzo and
  Vergani}]{Ilbert2006}
\bibinfo{author}{Ilbert, O.}, \bibinfo{author}{Arnouts, S.},
  \bibinfo{author}{McCracken, H.J.}, \bibinfo{author}{Bolzonella, M.},
  \bibinfo{author}{Bertin, E.}, \bibinfo{author}{F{\`{e}}vre, O.L.},
  \bibinfo{author}{Mellier, Y.}, \bibinfo{author}{Zamorani, G.},
  \bibinfo{author}{Pell{\`{o}}, R.}, \bibinfo{author}{Iovino, A.},
  \bibinfo{author}{Tresse, L.}, \bibinfo{author}{Brun, V.L.},
  \bibinfo{author}{Bottini, D.}, \bibinfo{author}{Garilli, B.},
  \bibinfo{author}{Maccagni, D.}, \bibinfo{author}{Picat, J.P.},
  \bibinfo{author}{Scaramella, R.}, \bibinfo{author}{Scodeggio, M.},
  \bibinfo{author}{Vettolani, G.}, \bibinfo{author}{Zanichelli, A.},
  \bibinfo{author}{Adami, C.}, \bibinfo{author}{Bardelli, S.},
  \bibinfo{author}{Cappi, A.}, \bibinfo{author}{Charlot, S.},
  \bibinfo{author}{Ciliegi, P.}, \bibinfo{author}{Contini, T.},
  \bibinfo{author}{Cucciati, O.}, \bibinfo{author}{Foucaud, S.},
  \bibinfo{author}{Franzetti, P.}, \bibinfo{author}{Gavignaud, I.},
  \bibinfo{author}{Guzzo, L.}, \bibinfo{author}{Marano, B.},
  \bibinfo{author}{Marinoni, C.}, \bibinfo{author}{Mazure, A.},
  \bibinfo{author}{Meneux, B.}, \bibinfo{author}{Merighi, R.},
  \bibinfo{author}{Paltani, S.}, \bibinfo{author}{Pollo, A.},
  \bibinfo{author}{Pozzetti, L.}, \bibinfo{author}{Radovich, M.},
  \bibinfo{author}{Zucca, E.}, \bibinfo{author}{Bondi, M.},
  \bibinfo{author}{Bongiorno, A.}, \bibinfo{author}{Busarello, G.},
  \bibinfo{author}{Torre, S.D.L.}, \bibinfo{author}{Gregorini, L.},
  \bibinfo{author}{Lamareille, F.}, \bibinfo{author}{Mathez, G.},
  \bibinfo{author}{Merluzzi, P.}, \bibinfo{author}{Ripepi, V.},
  \bibinfo{author}{Rizzo, D.}, \bibinfo{author}{Vergani, D.},
  \bibinfo{year}{2006}.
\newblock \bibinfo{title}{Accurate photometric redshifts for the {CFHT} legacy
  survey calibrated using the {VIMOS} {VLT} deep survey}.
\newblock \bibinfo{journal}{Astronomy {\&} Astrophysics} \bibinfo{volume}{457},
  \bibinfo{pages}{841--856}.
\newblock \URLprefix \url{https://doi.org/10.1051/0004-6361:20065138},
  \DOIprefix\doi{10.1051/0004-6361:20065138}.
\bibitem[{Ioffe and Szegedy(2015)}]{BatchNorm}
\bibinfo{author}{Ioffe, S.}, \bibinfo{author}{Szegedy, C.},
  \bibinfo{year}{2015}.
\newblock \bibinfo{title}{Batch normalization: Accelerating deep network
  training by reducing internal covariate shift}.
\newblock \bibinfo{journal}{CoRR} \bibinfo{volume}{abs/1502.03167}.
\newblock \URLprefix \url{http://arxiv.org/abs/1502.03167},
  \href{http://arxiv.org/abs/1502.03167}{\tt arXiv:1502.03167}.
\bibitem[{Ivezic et~al.(2008)Ivezic, Axelrod, Brandt, Burke, Claver, Connolly,
  Cook, Gee, Gilmore, Jacoby, Jones, Kahn, Kantor, Krabbendam, Lupton, Monet,
  Pinto, Saha, Schalk and Schneider}]{LSST}
\bibinfo{author}{Ivezic, Z.}, \bibinfo{author}{Axelrod, T.},
  \bibinfo{author}{Brandt, W.}, \bibinfo{author}{Burke, D.},
  \bibinfo{author}{Claver, C.}, \bibinfo{author}{Connolly, A.},
  \bibinfo{author}{Cook, K.}, \bibinfo{author}{Gee, P.},
  \bibinfo{author}{Gilmore, D.}, \bibinfo{author}{Jacoby, S.},
  \bibinfo{author}{Jones, R.}, \bibinfo{author}{Kahn, S.},
  \bibinfo{author}{Kantor, J.}, \bibinfo{author}{Krabbendam, V.},
  \bibinfo{author}{Lupton, R.}, \bibinfo{author}{Monet, D.},
  \bibinfo{author}{Pinto, P.}, \bibinfo{author}{Saha, A.},
  \bibinfo{author}{Schalk, T.}, \bibinfo{author}{Schneider, D.},
  \bibinfo{year}{2008}.
\newblock \bibinfo{title}{Large synoptic survey telescope: From science drivers
  to reference design}.
\newblock \bibinfo{journal}{Serbian Astronomical Journal} ,
  \bibinfo{pages}{1--13}\URLprefix \url{https://doi.org/10.2298/saj0876001i},
  \DOIprefix\doi{10.2298/saj0876001i}.
\bibitem[{Kilbinger et~al.(2013)Kilbinger, Fu, Heymans, Simpson, Benjamin,
  Erben, Harnois-D{\'{e}}raps, Hoekstra, Hildebrandt, Kitching, Mellier,
  Miller, Waerbeke, Benabed, Bonnett, Coupon, Hudson, Kuijken, Rowe,
  Schrabback, Semboloni, Vafaei and Velander}]{Kilbinger2013}
\bibinfo{author}{Kilbinger, M.}, \bibinfo{author}{Fu, L.},
  \bibinfo{author}{Heymans, C.}, \bibinfo{author}{Simpson, F.},
  \bibinfo{author}{Benjamin, J.}, \bibinfo{author}{Erben, T.},
  \bibinfo{author}{Harnois-D{\'{e}}raps, J.}, \bibinfo{author}{Hoekstra, H.},
  \bibinfo{author}{Hildebrandt, H.}, \bibinfo{author}{Kitching, T.D.},
  \bibinfo{author}{Mellier, Y.}, \bibinfo{author}{Miller, L.},
  \bibinfo{author}{Waerbeke, L.V.}, \bibinfo{author}{Benabed, K.},
  \bibinfo{author}{Bonnett, C.}, \bibinfo{author}{Coupon, J.},
  \bibinfo{author}{Hudson, M.J.}, \bibinfo{author}{Kuijken, K.},
  \bibinfo{author}{Rowe, B.}, \bibinfo{author}{Schrabback, T.},
  \bibinfo{author}{Semboloni, E.}, \bibinfo{author}{Vafaei, S.},
  \bibinfo{author}{Velander, M.}, \bibinfo{year}{2013}.
\newblock \bibinfo{title}{{CFHTLenS}: combined probe cosmological model
  comparison using 2d weak gravitational lensing}.
\newblock \bibinfo{journal}{Monthly Notices of the Royal Astronomical Society}
  \bibinfo{volume}{430}, \bibinfo{pages}{2200--2220}.
\newblock \URLprefix \url{https://doi.org/10.1093/mnras/stt041},
  \DOIprefix\doi{10.1093/mnras/stt041}.
\bibitem[{Kind and Brunner(2013)}]{CarrascoKind2013}
\bibinfo{author}{Kind, M.C.}, \bibinfo{author}{Brunner, R.J.},
  \bibinfo{year}{2013}.
\newblock \bibinfo{title}{{TPZ}: photometric redshift {PDFs} and ancillary
  information by using prediction trees and random forests}.
\newblock \bibinfo{journal}{Monthly Notices of the Royal Astronomical Society}
  \bibinfo{volume}{432}, \bibinfo{pages}{1483--1501}.
\newblock \URLprefix \url{https://doi.org/10.1093/mnras/stt574},
  \DOIprefix\doi{10.1093/mnras/stt574}.
\bibitem[{{Kitching} et~al.(2016){Kitching}, {Verde}, {Heavens} and
  {Jimenez}}]{Kitching2016}
\bibinfo{author}{{Kitching}, T.D.}, \bibinfo{author}{{Verde}, L.},
  \bibinfo{author}{{Heavens}, A.F.}, \bibinfo{author}{{Jimenez}, R.},
  \bibinfo{year}{2016}.
\newblock \bibinfo{title}{{Discrepancies between CFHTLenS cosmic shear and
  Planck: new physics or systematic effects?}}
\newblock \bibinfo{journal}{\mnras} \bibinfo{volume}{459},
  \bibinfo{pages}{971--981}.
\newblock \DOIprefix\doi{10.1093/mnras/stw707},
  \href{http://arxiv.org/abs/1602.02960}{\tt arXiv:1602.02960}.
\bibitem[{{Laigle} et~al.(2016){Laigle}, {McCracken}, {Ilbert}, {Hsieh},
  {Davidzon}, {Capak}, {Hasinger}, {Silverman}, {Pichon}, {Coupon}, {Aussel},
  {Le Borgne}, {Caputi}, {Cassata}, {Chang}, {Civano}, {Dunlop}, {Fynbo},
  {Kartaltepe}, {Koekemoer}, {Le F{\`e}vre}, {Le Floc'h}, {Leauthaud}, {Lilly},
  {Lin}, {Marchesi}, {Milvang-Jensen}, {Salvato}, {Sanders}, {Scoville},
  {Smolcic}, {Stockmann}, {Taniguchi}, {Tasca}, {Toft}, {Vaccari} and
  {Zabl}}]{Laigle2016}
\bibinfo{author}{{Laigle}, C.}, \bibinfo{author}{{McCracken}, H.J.},
  \bibinfo{author}{{Ilbert}, O.}, \bibinfo{author}{{Hsieh}, B.C.},
  \bibinfo{author}{{Davidzon}, I.}, \bibinfo{author}{{Capak}, P.},
  \bibinfo{author}{{Hasinger}, G.}, \bibinfo{author}{{Silverman}, J.D.},
  \bibinfo{author}{{Pichon}, C.}, \bibinfo{author}{{Coupon}, J.},
  \bibinfo{author}{{Aussel}, H.}, \bibinfo{author}{{Le Borgne}, D.},
  \bibinfo{author}{{Caputi}, K.}, \bibinfo{author}{{Cassata}, P.},
  \bibinfo{author}{{Chang}, Y.Y.}, \bibinfo{author}{{Civano}, F.},
  \bibinfo{author}{{Dunlop}, J.}, \bibinfo{author}{{Fynbo}, J.},
  \bibinfo{author}{{Kartaltepe}, J.S.}, \bibinfo{author}{{Koekemoer}, A.},
  \bibinfo{author}{{Le F{\`e}vre}, O.}, \bibinfo{author}{{Le Floc'h}, E.},
  \bibinfo{author}{{Leauthaud}, A.}, \bibinfo{author}{{Lilly}, S.},
  \bibinfo{author}{{Lin}, L.}, \bibinfo{author}{{Marchesi}, S.},
  \bibinfo{author}{{Milvang-Jensen}, B.}, \bibinfo{author}{{Salvato}, M.},
  \bibinfo{author}{{Sanders}, D.B.}, \bibinfo{author}{{Scoville}, N.},
  \bibinfo{author}{{Smolcic}, V.}, \bibinfo{author}{{Stockmann}, M.},
  \bibinfo{author}{{Taniguchi}, Y.}, \bibinfo{author}{{Tasca}, L.},
  \bibinfo{author}{{Toft}, S.}, \bibinfo{author}{{Vaccari}, M.},
  \bibinfo{author}{{Zabl}, J.}, \bibinfo{year}{2016}.
\newblock \bibinfo{title}{{The COSMOS2015 Catalog: Exploring the 1 < z < 6
  Universe with Half a Million Galaxies}}.
\newblock \bibinfo{journal}{\apjs} \bibinfo{volume}{224}, \bibinfo{pages}{24}.
\newblock \DOIprefix\doi{10.3847/0067-0049/224/2/24},
  \href{http://arxiv.org/abs/1604.02350}{\tt arXiv:1604.02350}.
\bibitem[{{Lang}(2014)}]{unWISE}
\bibinfo{author}{{Lang}, D.}, \bibinfo{year}{2014}.
\newblock \bibinfo{title}{{unWISE: Unblurred Coadds of the WISE Imaging}}.
\newblock \bibinfo{journal}{\aj} \bibinfo{volume}{147}, \bibinfo{pages}{108}.
\newblock \DOIprefix\doi{10.1088/0004-6256/147/5/108},
  \href{http://arxiv.org/abs/1405.0308}{\tt arXiv:1405.0308}.
\bibitem[{{Laureijs} et~al.(2011){Laureijs}, {Amiaux}, {Arduini},
  {Augu{\`e}res}, {Brinchmann}, {Cole}, {Cropper}, {Dabin}, {Duvet}, {Ealet},
  {Garilli}, {Gondoin}, {Guzzo}, {Hoar}, {Hoekstra}, {Holmes}, {Kitching},
  {Maciaszek}, {Mellier}, {Pasian}, {Percival}, {Rhodes}, {Saavedra Criado},
  {Sauvage}, {Scaramella}, {Valenziano}, {Warren}, {Bender}, {Castander},
  {Cimatti}, {Le F{\`e}vre}, {Kurki-Suonio}, {Levi}, {Lilje}, {Meylan},
  {Nichol}, {Pedersen}, {Popa}, {Rebolo Lopez}, {Rix}, {Rottgering},
  {Zeilinger}, {Grupp}, {Hudelot}, {Massey}, {Meneghetti}, {Miller}, {Paltani},
  {Paulin-Henriksson}, {Pires}, {Saxton}, {Schrabback}, {Seidel}, {Walsh},
  {Aghanim}, {Amendola}, {Bartlett}, {Baccigalupi}, {Beaulieu}, {Benabed},
  {Cuby}, {Elbaz}, {Fosalba}, {Gavazzi}, {Helmi}, {Hook}, {Irwin}, {Kneib},
  {Kunz}, {Mannucci}, {Moscardini}, {Tao}, {Teyssier}, {Weller}, {Zamorani},
  {Zapatero Osorio}, {Boulade}, {Foumond}, {Di Giorgio}, {Guttridge}, {James},
  {Kemp}, {Martignac}, {Spencer}, {Walton}, {Bl{\"u}mchen}, {Bonoli},
  {Bortoletto}, {Cerna}, {Corcione}, {Fabron}, {Jahnke}, {Ligori}, {Madrid},
  {Martin}, {Morgante}, {Pamplona}, {Prieto}, {Riva}, {Toledo}, {Trifoglio},
  {Zerbi}, {Abdalla}, {Douspis}, {Grenet}, {Borgani}, {Bouwens}, {Courbin},
  {Delouis}, {Dubath}, {Fontana}, {Frailis}, {Grazian}, {Koppenh{\"o}fer},
  {Mansutti}, {Melchior}, {Mignoli}, {Mohr}, {Neissner}, {Noddle}, {Poncet},
  {Scodeggio}, {Serrano}, {Shane}, {Starck}, {Surace}, {Taylor},
  {Verdoes-Kleijn}, {Vuerli}, {Williams}, {Zacchei}, {Altieri}, {Escudero
  Sanz}, {Kohley}, {Oosterbroek}, {Astier}, {Bacon}, {Bardelli}, {Baugh},
  {Bellagamba}, {Benoist}, {Bianchi}, {Biviano}, {Branchini}, {Carbone},
  {Cardone}, {Clements}, {Colombi}, {Conselice}, {Cresci}, {Deacon}, {Dunlop},
  {Fedeli}, {Fontanot}, {Franzetti}, {Giocoli}, {Garcia-Bellido}, {Gow},
  {Heavens}, {Hewett}, {Heymans}, {Holland}, {Huang}, {Ilbert}, {Joachimi},
  {Jennins}, {Kerins}, {Kiessling}, {Kirk}, {Kotak}, {Krause}, {Lahav}, {van
  Leeuwen}, {Lesgourgues}, {Lombardi}, {Magliocchetti}, {Maguire}, {Majerotto},
  {Maoli}, {Marulli}, {Maurogordato}, {McCracken}, {McLure}, {Melchiorri},
  {Merson}, {Moresco}, {Nonino}, {Norberg}, {Peacock}, {Pello}, {Penny},
  {Pettorino}, {Di Porto}, {Pozzetti}, {Quercellini}, {Radovich}, {Rassat},
  {Roche}, {Ronayette}, {Rossetti}, {Sartoris}, {Schneider}, {Semboloni},
  {Serjeant}, {Simpson}, {Skordis}, {Smadja}, {Smartt}, {Spano}, {Spiro},
  {Sullivan}, {Tilquin}, {Trotta}, {Verde}, {Wang}, {Williger}, {Zhao},
  {Zoubian} and {Zucca}}]{EUCLID}
\bibinfo{author}{{Laureijs}, R.}, \bibinfo{author}{{Amiaux}, J.},
  \bibinfo{author}{{Arduini}, S.}, \bibinfo{author}{{Augu{\`e}res}, J.L.},
  \bibinfo{author}{{Brinchmann}, J.}, \bibinfo{author}{{Cole}, R.},
  \bibinfo{author}{{Cropper}, M.}, \bibinfo{author}{{Dabin}, C.},
  \bibinfo{author}{{Duvet}, L.}, \bibinfo{author}{{Ealet}, A.},
  \bibinfo{author}{{Garilli}, B.}, \bibinfo{author}{{Gondoin}, P.},
  \bibinfo{author}{{Guzzo}, L.}, \bibinfo{author}{{Hoar}, J.},
  \bibinfo{author}{{Hoekstra}, H.}, \bibinfo{author}{{Holmes}, R.},
  \bibinfo{author}{{Kitching}, T.}, \bibinfo{author}{{Maciaszek}, T.},
  \bibinfo{author}{{Mellier}, Y.}, \bibinfo{author}{{Pasian}, F.},
  \bibinfo{author}{{Percival}, W.}, \bibinfo{author}{{Rhodes}, J.},
  \bibinfo{author}{{Saavedra Criado}, G.}, \bibinfo{author}{{Sauvage}, M.},
  \bibinfo{author}{{Scaramella}, R.}, \bibinfo{author}{{Valenziano}, L.},
  \bibinfo{author}{{Warren}, S.}, \bibinfo{author}{{Bender}, R.},
  \bibinfo{author}{{Castander}, F.}, \bibinfo{author}{{Cimatti}, A.},
  \bibinfo{author}{{Le F{\`e}vre}, O.}, \bibinfo{author}{{Kurki-Suonio}, H.},
  \bibinfo{author}{{Levi}, M.}, \bibinfo{author}{{Lilje}, P.},
  \bibinfo{author}{{Meylan}, G.}, \bibinfo{author}{{Nichol}, R.},
  \bibinfo{author}{{Pedersen}, K.}, \bibinfo{author}{{Popa}, V.},
  \bibinfo{author}{{Rebolo Lopez}, R.}, \bibinfo{author}{{Rix}, H.W.},
  \bibinfo{author}{{Rottgering}, H.}, \bibinfo{author}{{Zeilinger}, W.},
  \bibinfo{author}{{Grupp}, F.}, \bibinfo{author}{{Hudelot}, P.},
  \bibinfo{author}{{Massey}, R.}, \bibinfo{author}{{Meneghetti}, M.},
  \bibinfo{author}{{Miller}, L.}, \bibinfo{author}{{Paltani}, S.},
  \bibinfo{author}{{Paulin-Henriksson}, S.}, \bibinfo{author}{{Pires}, S.},
  \bibinfo{author}{{Saxton}, C.}, \bibinfo{author}{{Schrabback}, T.},
  \bibinfo{author}{{Seidel}, G.}, \bibinfo{author}{{Walsh}, J.},
  \bibinfo{author}{{Aghanim}, N.}, \bibinfo{author}{{Amendola}, L.},
  \bibinfo{author}{{Bartlett}, J.}, \bibinfo{author}{{Baccigalupi}, C.},
  \bibinfo{author}{{Beaulieu}, J.P.}, \bibinfo{author}{{Benabed}, K.},
  \bibinfo{author}{{Cuby}, J.G.}, \bibinfo{author}{{Elbaz}, D.},
  \bibinfo{author}{{Fosalba}, P.}, \bibinfo{author}{{Gavazzi}, G.},
  \bibinfo{author}{{Helmi}, A.}, \bibinfo{author}{{Hook}, I.},
  \bibinfo{author}{{Irwin}, M.}, \bibinfo{author}{{Kneib}, J.P.},
  \bibinfo{author}{{Kunz}, M.}, \bibinfo{author}{{Mannucci}, F.},
  \bibinfo{author}{{Moscardini}, L.}, \bibinfo{author}{{Tao}, C.},
  \bibinfo{author}{{Teyssier}, R.}, \bibinfo{author}{{Weller}, J.},
  \bibinfo{author}{{Zamorani}, G.}, \bibinfo{author}{{Zapatero Osorio}, M.R.},
  \bibinfo{author}{{Boulade}, O.}, \bibinfo{author}{{Foumond}, J.J.},
  \bibinfo{author}{{Di Giorgio}, A.}, \bibinfo{author}{{Guttridge}, P.},
  \bibinfo{author}{{James}, A.}, \bibinfo{author}{{Kemp}, M.},
  \bibinfo{author}{{Martignac}, J.}, \bibinfo{author}{{Spencer}, A.},
  \bibinfo{author}{{Walton}, D.}, \bibinfo{author}{{Bl{\"u}mchen}, T.},
  \bibinfo{author}{{Bonoli}, C.}, \bibinfo{author}{{Bortoletto}, F.},
  \bibinfo{author}{{Cerna}, C.}, \bibinfo{author}{{Corcione}, L.},
  \bibinfo{author}{{Fabron}, C.}, \bibinfo{author}{{Jahnke}, K.},
  \bibinfo{author}{{Ligori}, S.}, \bibinfo{author}{{Madrid}, F.},
  \bibinfo{author}{{Martin}, L.}, \bibinfo{author}{{Morgante}, G.},
  \bibinfo{author}{{Pamplona}, T.}, \bibinfo{author}{{Prieto}, E.},
  \bibinfo{author}{{Riva}, M.}, \bibinfo{author}{{Toledo}, R.},
  \bibinfo{author}{{Trifoglio}, M.}, \bibinfo{author}{{Zerbi}, F.},
  \bibinfo{author}{{Abdalla}, F.}, \bibinfo{author}{{Douspis}, M.},
  \bibinfo{author}{{Grenet}, C.}, \bibinfo{author}{{Borgani}, S.},
  \bibinfo{author}{{Bouwens}, R.}, \bibinfo{author}{{Courbin}, F.},
  \bibinfo{author}{{Delouis}, J.M.}, \bibinfo{author}{{Dubath}, P.},
  \bibinfo{author}{{Fontana}, A.}, \bibinfo{author}{{Frailis}, M.},
  \bibinfo{author}{{Grazian}, A.}, \bibinfo{author}{{Koppenh{\"o}fer}, J.},
  \bibinfo{author}{{Mansutti}, O.}, \bibinfo{author}{{Melchior}, M.},
  \bibinfo{author}{{Mignoli}, M.}, \bibinfo{author}{{Mohr}, J.},
  \bibinfo{author}{{Neissner}, C.}, \bibinfo{author}{{Noddle}, K.},
  \bibinfo{author}{{Poncet}, M.}, \bibinfo{author}{{Scodeggio}, M.},
  \bibinfo{author}{{Serrano}, S.}, \bibinfo{author}{{Shane}, N.},
  \bibinfo{author}{{Starck}, J.L.}, \bibinfo{author}{{Surace}, C.},
  \bibinfo{author}{{Taylor}, A.}, \bibinfo{author}{{Verdoes-Kleijn}, G.},
  \bibinfo{author}{{Vuerli}, C.}, \bibinfo{author}{{Williams}, O.R.},
  \bibinfo{author}{{Zacchei}, A.}, \bibinfo{author}{{Altieri}, B.},
  \bibinfo{author}{{Escudero Sanz}, I.}, \bibinfo{author}{{Kohley}, R.},
  \bibinfo{author}{{Oosterbroek}, T.}, \bibinfo{author}{{Astier}, P.},
  \bibinfo{author}{{Bacon}, D.}, \bibinfo{author}{{Bardelli}, S.},
  \bibinfo{author}{{Baugh}, C.}, \bibinfo{author}{{Bellagamba}, F.},
  \bibinfo{author}{{Benoist}, C.}, \bibinfo{author}{{Bianchi}, D.},
  \bibinfo{author}{{Biviano}, A.}, \bibinfo{author}{{Branchini}, E.},
  \bibinfo{author}{{Carbone}, C.}, \bibinfo{author}{{Cardone}, V.},
  \bibinfo{author}{{Clements}, D.}, \bibinfo{author}{{Colombi}, S.},
  \bibinfo{author}{{Conselice}, C.}, \bibinfo{author}{{Cresci}, G.},
  \bibinfo{author}{{Deacon}, N.}, \bibinfo{author}{{Dunlop}, J.},
  \bibinfo{author}{{Fedeli}, C.}, \bibinfo{author}{{Fontanot}, F.},
  \bibinfo{author}{{Franzetti}, P.}, \bibinfo{author}{{Giocoli}, C.},
  \bibinfo{author}{{Garcia-Bellido}, J.}, \bibinfo{author}{{Gow}, J.},
  \bibinfo{author}{{Heavens}, A.}, \bibinfo{author}{{Hewett}, P.},
  \bibinfo{author}{{Heymans}, C.}, \bibinfo{author}{{Holland}, A.},
  \bibinfo{author}{{Huang}, Z.}, \bibinfo{author}{{Ilbert}, O.},
  \bibinfo{author}{{Joachimi}, B.}, \bibinfo{author}{{Jennins}, E.},
  \bibinfo{author}{{Kerins}, E.}, \bibinfo{author}{{Kiessling}, A.},
  \bibinfo{author}{{Kirk}, D.}, \bibinfo{author}{{Kotak}, R.},
  \bibinfo{author}{{Krause}, O.}, \bibinfo{author}{{Lahav}, O.},
  \bibinfo{author}{{van Leeuwen}, F.}, \bibinfo{author}{{Lesgourgues}, J.},
  \bibinfo{author}{{Lombardi}, M.}, \bibinfo{author}{{Magliocchetti}, M.},
  \bibinfo{author}{{Maguire}, K.}, \bibinfo{author}{{Majerotto}, E.},
  \bibinfo{author}{{Maoli}, R.}, \bibinfo{author}{{Marulli}, F.},
  \bibinfo{author}{{Maurogordato}, S.}, \bibinfo{author}{{McCracken}, H.},
  \bibinfo{author}{{McLure}, R.}, \bibinfo{author}{{Melchiorri}, A.},
  \bibinfo{author}{{Merson}, A.}, \bibinfo{author}{{Moresco}, M.},
  \bibinfo{author}{{Nonino}, M.}, \bibinfo{author}{{Norberg}, P.},
  \bibinfo{author}{{Peacock}, J.}, \bibinfo{author}{{Pello}, R.},
  \bibinfo{author}{{Penny}, M.}, \bibinfo{author}{{Pettorino}, V.},
  \bibinfo{author}{{Di Porto}, C.}, \bibinfo{author}{{Pozzetti}, L.},
  \bibinfo{author}{{Quercellini}, C.}, \bibinfo{author}{{Radovich}, M.},
  \bibinfo{author}{{Rassat}, A.}, \bibinfo{author}{{Roche}, N.},
  \bibinfo{author}{{Ronayette}, S.}, \bibinfo{author}{{Rossetti}, E.},
  \bibinfo{author}{{Sartoris}, B.}, \bibinfo{author}{{Schneider}, P.},
  \bibinfo{author}{{Semboloni}, E.}, \bibinfo{author}{{Serjeant}, S.},
  \bibinfo{author}{{Simpson}, F.}, \bibinfo{author}{{Skordis}, C.},
  \bibinfo{author}{{Smadja}, G.}, \bibinfo{author}{{Smartt}, S.},
  \bibinfo{author}{{Spano}, P.}, \bibinfo{author}{{Spiro}, S.},
  \bibinfo{author}{{Sullivan}, M.}, \bibinfo{author}{{Tilquin}, A.},
  \bibinfo{author}{{Trotta}, R.}, \bibinfo{author}{{Verde}, L.},
  \bibinfo{author}{{Wang}, Y.}, \bibinfo{author}{{Williger}, G.},
  \bibinfo{author}{{Zhao}, G.}, \bibinfo{author}{{Zoubian}, J.},
  \bibinfo{author}{{Zucca}, E.}, \bibinfo{year}{2011}.
\newblock \bibinfo{title}{{Euclid Definition Study Report}}.
\newblock \bibinfo{journal}{arXiv e-prints} ,
  \bibinfo{pages}{arXiv:1110.3193}\href{http://arxiv.org/abs/1110.3193}{\tt
  arXiv:1110.3193}.
\bibitem[{{Mar{\'\i}n-Franch} et~al.(2012){Mar{\'\i}n-Franch}, {Chueca},
  {Moles}, {Benitez}, {Taylor}, {Cepa}, {Cenarro}, {Cristobal-Hornillos},
  {Ederoclite}, {Gruel}, {Hern{\'a}ndez-Fuertes}, {L{\'o}pez-Sainz},
  {Luis-Simoes}, {Rueda-Teruel}, {Rueda-Teruel}, {Varela}, {Yanes-D{\'\i}az},
  {Brauneck}, {Danielou}, {Dupke}, {Fern{\'a}ndez-Soto}, {Mendes de Oliveira}
  and {Sodr{\'e}}}]{Marin-Franch2012}
\bibinfo{author}{{Mar{\'\i}n-Franch}, A.}, \bibinfo{author}{{Chueca}, S.},
  \bibinfo{author}{{Moles}, M.}, \bibinfo{author}{{Benitez}, N.},
  \bibinfo{author}{{Taylor}, K.}, \bibinfo{author}{{Cepa}, J.},
  \bibinfo{author}{{Cenarro}, A.J.}, \bibinfo{author}{{Cristobal-Hornillos},
  D.}, \bibinfo{author}{{Ederoclite}, A.}, \bibinfo{author}{{Gruel}, N.},
  \bibinfo{author}{{Hern{\'a}ndez-Fuertes}, J.},
  \bibinfo{author}{{L{\'o}pez-Sainz}, A.}, \bibinfo{author}{{Luis-Simoes}, R.},
  \bibinfo{author}{{Rueda-Teruel}, F.}, \bibinfo{author}{{Rueda-Teruel}, S.},
  \bibinfo{author}{{Varela}, J.}, \bibinfo{author}{{Yanes-D{\'\i}az}, A.},
  \bibinfo{author}{{Brauneck}, U.}, \bibinfo{author}{{Danielou}, A.},
  \bibinfo{author}{{Dupke}, R.}, \bibinfo{author}{{Fern{\'a}ndez-Soto}, A.},
  \bibinfo{author}{{Mendes de Oliveira}, C.}, \bibinfo{author}{{Sodr{\'e}},
  L.}, \bibinfo{year}{2012}.
\newblock \bibinfo{title}{{Design of the J-PAS and J-PLUS filter systems}}, in:
  \bibinfo{editor}{{Navarro}, R.}, \bibinfo{editor}{{Cunningham}, C.R.},
  \bibinfo{editor}{{Prieto}, E.} (Eds.), \bibinfo{booktitle}{Modern
  Technologies in Space- and Ground-based Telescopes and Instrumentation II},
  p. \bibinfo{pages}{84503S}.
\newblock \DOIprefix\doi{10.1117/12.925430}.
\bibitem[{{Mendes de Oliveira} et~al.(2019){Mendes de Oliveira}, {Ribeiro},
  {Schoenell}, {Kanaan}, {Overzier}, {Molino}, {Sampedro}, {Coelho}, {Barbosa},
  {Cortesi}, {Costa-Duarte}, {Herpich}, {Hernand ez-Jimenez}, {Placco},
  {Xavier}, {Abramo}, {Saito}, {Chies-Santos}, {Ederoclite}, {Lopes de
  Oliveira}, {Gon{\c{c}}alves}, {Akras}, {Almeida}, {Almeida-Fernandes},
  {Beers}, {Bonatto}, {Bonoli}, {Cypriano}, {Vinicius-Lima}, {de Souza},
  {Fabiano de Souza}, {Ferrari}, {Gon{\c{c}}alves}, {Gonzalez},
  {Guti{\'e}rrez-Soto}, {Hartmann}, {Jaffe}, {Kerber}, {Lima-Dias}, {Lopes},
  {Menendez-Delmestre}, {Nakazono}, {Novais}, {Ortega-Minakata}, {Pereira},
  {Perottoni}, {Queiroz}, {Reis}, {Santos}, {Santos-Silva}, {Santucci},
  {Barbosa}, {Siffert}, {Sodr{\'e}}, {Torres-Flores}, {Westera}, {Whitten},
  {Alcaniz}, {Alonso-Garc{\'\i}a}, {Alencar}, {Alvarez-Cand al}, {Amram},
  {Azanha}, {Barb{\'a}}, {Bernardinelli}, {Borges Fernandes}, {Branco},
  {Brito-Silva}, {Buzzo}, {Caffer}, {Campillay}, {Cano}, {Carvano}, {Castejon},
  {Cid Fernandes}, {Dantas}, {Daflon}, {Damke}, {de la Reza}, {de Melo de
  Azevedo}, {De Paula}, {Diem}, {Donnerstein}, {Dors}, {Dupke}, {Eikenberry},
  {Escudero}, {Faifer}, {Far{\'\i}as}, {Fernandes}, {Fernandes}, {Fontes},
  {Galarza}, {Hirata}, {Katena}, {Gregorio-Hetem},
  {Hern{\'a}ndez-Fern{\'a}ndez}, {Izzo}, {Jaque Arancibia}, {Jatenco-Pereira},
  {Jim{\'e}nez-Teja}, {Kann}, {Krabbe}, {Labayru}, {Lazzaro}, {Lima Neto},
  {Lopes}, {Magalh{\~a}es}, {Makler}, {de Menezes}, {Miralda-Escud{\'e}},
  {Monteiro-Oliveira}, {Montero-Dorta}, {Mu{\~n}oz-Elgueta}, {Nemmen}, {Nilo
  Castell{\'o}n}, {Oliveira}, {Ort{\'\i}z}, {Pattaro}, {Pereira}, {Quint},
  {Riguccini}, {Rocha Pinto}, {Rodrigues}, {Roig}, {Rossi}, {Saha}, {Santos},
  {Schnorr M{\"u}ller}, {Sesto}, {Silva}, {Smith Castelli}, {Teixeira},
  {Telles}, {Thom de Souza}, {Th{\"o}ne}, {Trevisan}, {de Ugarte Postigo},
  {Urrutia-Viscarra}, {Veiga}, {Vika}, {Vitorelli}, {Werle}, {Werner} and
  {Zaritsky}}]{SPLUS}
\bibinfo{author}{{Mendes de Oliveira}, C.}, \bibinfo{author}{{Ribeiro}, T.},
  \bibinfo{author}{{Schoenell}, W.}, \bibinfo{author}{{Kanaan}, A.},
  \bibinfo{author}{{Overzier}, R.A.}, \bibinfo{author}{{Molino}, A.},
  \bibinfo{author}{{Sampedro}, L.}, \bibinfo{author}{{Coelho}, P.},
  \bibinfo{author}{{Barbosa}, C.E.}, \bibinfo{author}{{Cortesi}, A.},
  \bibinfo{author}{{Costa-Duarte}, M.V.}, \bibinfo{author}{{Herpich}, F.R.},
  \bibinfo{author}{{Hernand ez-Jimenez}, J.A.}, \bibinfo{author}{{Placco},
  V.M.}, \bibinfo{author}{{Xavier}, H.S.}, \bibinfo{author}{{Abramo}, L.R.},
  \bibinfo{author}{{Saito}, R.K.}, \bibinfo{author}{{Chies-Santos}, A.L.},
  \bibinfo{author}{{Ederoclite}, A.}, \bibinfo{author}{{Lopes de Oliveira},
  R.}, \bibinfo{author}{{Gon{\c{c}}alves}, D.R.}, \bibinfo{author}{{Akras},
  S.}, \bibinfo{author}{{Almeida}, L.A.}, \bibinfo{author}{{Almeida-Fernandes},
  F.}, \bibinfo{author}{{Beers}, T.C.}, \bibinfo{author}{{Bonatto}, C.},
  \bibinfo{author}{{Bonoli}, S.}, \bibinfo{author}{{Cypriano}, E.S.},
  \bibinfo{author}{{Vinicius-Lima}, E.}, \bibinfo{author}{{de Souza}, R.S.},
  \bibinfo{author}{{Fabiano de Souza}, G.}, \bibinfo{author}{{Ferrari}, F.},
  \bibinfo{author}{{Gon{\c{c}}alves}, T.S.}, \bibinfo{author}{{Gonzalez},
  A.H.}, \bibinfo{author}{{Guti{\'e}rrez-Soto}, L.A.},
  \bibinfo{author}{{Hartmann}, E.A.}, \bibinfo{author}{{Jaffe}, Y.},
  \bibinfo{author}{{Kerber}, L.O.}, \bibinfo{author}{{Lima-Dias}, C.},
  \bibinfo{author}{{Lopes}, P.A.A.}, \bibinfo{author}{{Menendez-Delmestre},
  K.}, \bibinfo{author}{{Nakazono}, L.M.I.}, \bibinfo{author}{{Novais}, P.M.},
  \bibinfo{author}{{Ortega-Minakata}, R.A.}, \bibinfo{author}{{Pereira}, E.S.},
  \bibinfo{author}{{Perottoni}, H.D.}, \bibinfo{author}{{Queiroz}, C.},
  \bibinfo{author}{{Reis}, R.R.R.}, \bibinfo{author}{{Santos}, W.A.},
  \bibinfo{author}{{Santos-Silva}, T.}, \bibinfo{author}{{Santucci}, R.M.},
  \bibinfo{author}{{Barbosa}, C.L.}, \bibinfo{author}{{Siffert}, B.B.},
  \bibinfo{author}{{Sodr{\'e}}, L.}, \bibinfo{author}{{Torres-Flores}, S.},
  \bibinfo{author}{{Westera}, P.}, \bibinfo{author}{{Whitten}, D.D.},
  \bibinfo{author}{{Alcaniz}, J.S.}, \bibinfo{author}{{Alonso-Garc{\'\i}a},
  J.}, \bibinfo{author}{{Alencar}, S.}, \bibinfo{author}{{Alvarez-Cand al},
  A.}, \bibinfo{author}{{Amram}, P.}, \bibinfo{author}{{Azanha}, L.},
  \bibinfo{author}{{Barb{\'a}}, R.H.}, \bibinfo{author}{{Bernardinelli}, P.H.},
  \bibinfo{author}{{Borges Fernandes}, M.}, \bibinfo{author}{{Branco}, V.},
  \bibinfo{author}{{Brito-Silva}, D.}, \bibinfo{author}{{Buzzo}, M.L.},
  \bibinfo{author}{{Caffer}, J.}, \bibinfo{author}{{Campillay}, A.},
  \bibinfo{author}{{Cano}, Z.}, \bibinfo{author}{{Carvano}, J.M.},
  \bibinfo{author}{{Castejon}, M.}, \bibinfo{author}{{Cid Fernandes}, R.},
  \bibinfo{author}{{Dantas}, M.L.L.}, \bibinfo{author}{{Daflon}, S.},
  \bibinfo{author}{{Damke}, G.}, \bibinfo{author}{{de la Reza}, R.},
  \bibinfo{author}{{de Melo de Azevedo}, L.J.}, \bibinfo{author}{{De Paula},
  D.F.}, \bibinfo{author}{{Diem}, K.G.}, \bibinfo{author}{{Donnerstein}, R.},
  \bibinfo{author}{{Dors}, O.L.}, \bibinfo{author}{{Dupke}, R.},
  \bibinfo{author}{{Eikenberry}, S.}, \bibinfo{author}{{Escudero}, C.G.},
  \bibinfo{author}{{Faifer}, F.R.}, \bibinfo{author}{{Far{\'\i}as}, H.},
  \bibinfo{author}{{Fernandes}, B.}, \bibinfo{author}{{Fernandes}, C.},
  \bibinfo{author}{{Fontes}, S.}, \bibinfo{author}{{Galarza}, A.},
  \bibinfo{author}{{Hirata}, N.S.T.}, \bibinfo{author}{{Katena}, L.},
  \bibinfo{author}{{Gregorio-Hetem}, J.},
  \bibinfo{author}{{Hern{\'a}ndez-Fern{\'a}ndez}, J.D.},
  \bibinfo{author}{{Izzo}, L.}, \bibinfo{author}{{Jaque Arancibia}, M.},
  \bibinfo{author}{{Jatenco-Pereira}, V.}, \bibinfo{author}{{Jim{\'e}nez-Teja},
  Y.}, \bibinfo{author}{{Kann}, D.A.}, \bibinfo{author}{{Krabbe}, A.C.},
  \bibinfo{author}{{Labayru}, C.}, \bibinfo{author}{{Lazzaro}, D.},
  \bibinfo{author}{{Lima Neto}, G.B.}, \bibinfo{author}{{Lopes}, A.a.R.},
  \bibinfo{author}{{Magalh{\~a}es}, R.}, \bibinfo{author}{{Makler}, M.},
  \bibinfo{author}{{de Menezes}, R.}, \bibinfo{author}{{Miralda-Escud{\'e}},
  J.}, \bibinfo{author}{{Monteiro-Oliveira}, R.},
  \bibinfo{author}{{Montero-Dorta}, A.D.},
  \bibinfo{author}{{Mu{\~n}oz-Elgueta}, N.}, \bibinfo{author}{{Nemmen}, R.S.},
  \bibinfo{author}{{Nilo Castell{\'o}n}, J.L.}, \bibinfo{author}{{Oliveira},
  A.S.}, \bibinfo{author}{{Ort{\'\i}z}, D.}, \bibinfo{author}{{Pattaro}, E.},
  \bibinfo{author}{{Pereira}, C.B.}, \bibinfo{author}{{Quint}, B.},
  \bibinfo{author}{{Riguccini}, L.}, \bibinfo{author}{{Rocha Pinto}, H.J.},
  \bibinfo{author}{{Rodrigues}, I.}, \bibinfo{author}{{Roig}, F.},
  \bibinfo{author}{{Rossi}, S.}, \bibinfo{author}{{Saha}, K.},
  \bibinfo{author}{{Santos}, R.}, \bibinfo{author}{{Schnorr M{\"u}ller}, A.},
  \bibinfo{author}{{Sesto}, L.A.}, \bibinfo{author}{{Silva}, R.},
  \bibinfo{author}{{Smith Castelli}, A.V.}, \bibinfo{author}{{Teixeira}, R.},
  \bibinfo{author}{{Telles}, E.}, \bibinfo{author}{{Thom de Souza}, R.C.},
  \bibinfo{author}{{Th{\"o}ne}, C.}, \bibinfo{author}{{Trevisan}, M.},
  \bibinfo{author}{{de Ugarte Postigo}, A.},
  \bibinfo{author}{{Urrutia-Viscarra}, F.}, \bibinfo{author}{{Veiga}, C.H.},
  \bibinfo{author}{{Vika}, M.}, \bibinfo{author}{{Vitorelli}, A.Z.},
  \bibinfo{author}{{Werle}, A.}, \bibinfo{author}{{Werner}, S.V.},
  \bibinfo{author}{{Zaritsky}, D.}, \bibinfo{year}{2019}.
\newblock \bibinfo{title}{{The Southern Photometric Local Universe Survey
  (S-PLUS): improved SEDs, morphologies, and redshifts with 12 optical
  filters}}.
\newblock \bibinfo{journal}{\mnras} \bibinfo{volume}{489},
  \bibinfo{pages}{241--267}.
\newblock \DOIprefix\doi{10.1093/mnras/stz1985},
  \href{http://arxiv.org/abs/1907.01567}{\tt arXiv:1907.01567}.
\bibitem[{Molino et~al.(2017)Molino, Ben{\'{\i}}tez, Ascaso, Coe, Postman,
  Jouvel, Host, Lahav, Seitz, Medezinski, Rosati, Schoenell, Koekemoer,
  Jimenez-Teja, Broadhurst, Melchior, Balestra, Bartelmann, Bouwens, Bradley,
  Czakon, Donahue, Ford, Graur, Graves, Grillo, Infante, Jha, Kelson, Lazkoz,
  Lemze, Maoz, Mercurio, Meneghetti, Merten, Moustakas, Nonino, Orgaz, Riess,
  Rodney, Sayers, Umetsu, Zheng and Zitrin}]{Molino2017}
\bibinfo{author}{Molino, A.}, \bibinfo{author}{Ben{\'{\i}}tez, N.},
  \bibinfo{author}{Ascaso, B.}, \bibinfo{author}{Coe, D.},
  \bibinfo{author}{Postman, M.}, \bibinfo{author}{Jouvel, S.},
  \bibinfo{author}{Host, O.}, \bibinfo{author}{Lahav, O.},
  \bibinfo{author}{Seitz, S.}, \bibinfo{author}{Medezinski, E.},
  \bibinfo{author}{Rosati, P.}, \bibinfo{author}{Schoenell, W.},
  \bibinfo{author}{Koekemoer, A.}, \bibinfo{author}{Jimenez-Teja, Y.},
  \bibinfo{author}{Broadhurst, T.}, \bibinfo{author}{Melchior, P.},
  \bibinfo{author}{Balestra, I.}, \bibinfo{author}{Bartelmann, M.},
  \bibinfo{author}{Bouwens, R.}, \bibinfo{author}{Bradley, L.},
  \bibinfo{author}{Czakon, N.}, \bibinfo{author}{Donahue, M.},
  \bibinfo{author}{Ford, H.}, \bibinfo{author}{Graur, O.},
  \bibinfo{author}{Graves, G.}, \bibinfo{author}{Grillo, C.},
  \bibinfo{author}{Infante, L.}, \bibinfo{author}{Jha, S.W.},
  \bibinfo{author}{Kelson, D.}, \bibinfo{author}{Lazkoz, R.},
  \bibinfo{author}{Lemze, D.}, \bibinfo{author}{Maoz, D.},
  \bibinfo{author}{Mercurio, A.}, \bibinfo{author}{Meneghetti, M.},
  \bibinfo{author}{Merten, J.}, \bibinfo{author}{Moustakas, L.},
  \bibinfo{author}{Nonino, M.}, \bibinfo{author}{Orgaz, S.},
  \bibinfo{author}{Riess, A.}, \bibinfo{author}{Rodney, S.},
  \bibinfo{author}{Sayers, J.}, \bibinfo{author}{Umetsu, K.},
  \bibinfo{author}{Zheng, W.}, \bibinfo{author}{Zitrin, A.},
  \bibinfo{year}{2017}.
\newblock \bibinfo{title}{{CLASH}: accurate photometric redshifts with 14 {HST}
  bands in massive galaxy cluster cores}.
\newblock \bibinfo{journal}{Monthly Notices of the Royal Astronomical Society}
  \bibinfo{volume}{470}, \bibinfo{pages}{95--113}.
\newblock \URLprefix \url{https://doi.org/10.1093/mnras/stx1243},
  \DOIprefix\doi{10.1093/mnras/stx1243}.
\bibitem[{{Molino} et~al.(2014){Molino}, {Ben{\'\i}tez}, {Moles},
  {Fern{\'a}ndez-Soto}, {Crist{\'o}bal-Hornillos}, {Ascaso},
  {Jim{\'e}nez-Teja}, {Schoenell}, {Arnalte-Mur}, {Povi{\'c}}, {Coe},
  {L{\'o}pez-Sanjuan}, {D{\'\i}az-Garc{\'\i}a}, {Varela}, {Stefanon},
  {Cenarro}, {Matute}, {Masegosa}, {M{\'a}rquez}, {Perea}, {Del Olmo},
  {Husillos}, {Alfaro}, {Aparicio-Villegas}, {Cervi{\~n}o}, {Huertas-Company},
  {Aguerri}, {Broadhurst}, {Cabrera-Ca{\~n}o}, {Cepa}, {Gonz{\'a}lez},
  {Infante}, {Mart{\'\i}nez}, {Prada} and {Quintana}}]{Molino2014}
\bibinfo{author}{{Molino}, A.}, \bibinfo{author}{{Ben{\'\i}tez}, N.},
  \bibinfo{author}{{Moles}, M.}, \bibinfo{author}{{Fern{\'a}ndez-Soto}, A.},
  \bibinfo{author}{{Crist{\'o}bal-Hornillos}, D.}, \bibinfo{author}{{Ascaso},
  B.}, \bibinfo{author}{{Jim{\'e}nez-Teja}, Y.}, \bibinfo{author}{{Schoenell},
  W.}, \bibinfo{author}{{Arnalte-Mur}, P.}, \bibinfo{author}{{Povi{\'c}}, M.},
  \bibinfo{author}{{Coe}, D.}, \bibinfo{author}{{L{\'o}pez-Sanjuan}, C.},
  \bibinfo{author}{{D{\'\i}az-Garc{\'\i}a}, L.A.}, \bibinfo{author}{{Varela},
  J.}, \bibinfo{author}{{Stefanon}, M.}, \bibinfo{author}{{Cenarro}, J.},
  \bibinfo{author}{{Matute}, I.}, \bibinfo{author}{{Masegosa}, J.},
  \bibinfo{author}{{M{\'a}rquez}, I.}, \bibinfo{author}{{Perea}, J.},
  \bibinfo{author}{{Del Olmo}, A.}, \bibinfo{author}{{Husillos}, C.},
  \bibinfo{author}{{Alfaro}, E.}, \bibinfo{author}{{Aparicio-Villegas}, T.},
  \bibinfo{author}{{Cervi{\~n}o}, M.}, \bibinfo{author}{{Huertas-Company}, M.},
  \bibinfo{author}{{Aguerri}, J.A.L.}, \bibinfo{author}{{Broadhurst}, T.},
  \bibinfo{author}{{Cabrera-Ca{\~n}o}, J.}, \bibinfo{author}{{Cepa}, J.},
  \bibinfo{author}{{Gonz{\'a}lez}, R.M.}, \bibinfo{author}{{Infante}, L.},
  \bibinfo{author}{{Mart{\'\i}nez}, V.J.}, \bibinfo{author}{{Prada}, F.},
  \bibinfo{author}{{Quintana}, J.M.}, \bibinfo{year}{2014}.
\newblock \bibinfo{title}{{The ALHAMBRA Survey: Bayesian photometric redshifts
  with 23 bands for 3 deg}}.
\newblock \bibinfo{journal}{\mnras} \bibinfo{volume}{441},
  \bibinfo{pages}{2891--2922}.
\newblock \DOIprefix\doi{10.1093/mnras/stu387},
  \href{http://arxiv.org/abs/1306.4968}{\tt arXiv:1306.4968}.
\bibitem[{{Molino} et~al.(2020){Molino}, {Costa-Duarte}, {Sampedro}, {Herpich},
  {Sodr{\'e}}, {Mendes de Oliveira}, {Schoenell}, {Barbosa}, {Queiroz}, {Lima},
  {Azanha}, {Mu{\~n}oz-Elgueta}, {Ribeiro}, {Kanaan}, {Hernandez-Jimenez},
  {Cortesi}, {Akras}, {Lopes de Oliveira}, {Torres-Flores}, {Lima-Dias},
  {Castellon}, {Damke}, {Alvarez-Candal}, {Jim{\'e}nez-Teja}, {Coelho},
  {Pereira}, {Montero-Dorta}, {Ben{\'\i}tez}, {Gon{\c{c}}alves},
  {Santana-Silva}, {Werner}, {Almeida}, {Lopes}, {Chies-Santos}, {Telles}, {de
  Souza}, {C}, {Gon{\c{c}}alves}, {de Souza}, {Makler}, {Buzzo}, {Placco},
  {Nakazono}, {Saito}, {Overzier} and {Abramo}}]{AlbertoSPLUS}
\bibinfo{author}{{Molino}, A.}, \bibinfo{author}{{Costa-Duarte}, M.V.},
  \bibinfo{author}{{Sampedro}, L.}, \bibinfo{author}{{Herpich}, F.R.},
  \bibinfo{author}{{Sodr{\'e}}, L., J.}, \bibinfo{author}{{Mendes de Oliveira},
  C.}, \bibinfo{author}{{Schoenell}, W.}, \bibinfo{author}{{Barbosa}, C.E.},
  \bibinfo{author}{{Queiroz}, C.}, \bibinfo{author}{{Lima}, E.V.R.},
  \bibinfo{author}{{Azanha}, L.}, \bibinfo{author}{{Mu{\~n}oz-Elgueta}, N.},
  \bibinfo{author}{{Ribeiro}, T.}, \bibinfo{author}{{Kanaan}, A.},
  \bibinfo{author}{{Hernandez-Jimenez}, J.A.}, \bibinfo{author}{{Cortesi}, A.},
  \bibinfo{author}{{Akras}, S.}, \bibinfo{author}{{Lopes de Oliveira}, R.},
  \bibinfo{author}{{Torres-Flores}, S.}, \bibinfo{author}{{Lima-Dias}, C.},
  \bibinfo{author}{{Castellon}, J.L.N.}, \bibinfo{author}{{Damke}, G.},
  \bibinfo{author}{{Alvarez-Candal}, A.}, \bibinfo{author}{{Jim{\'e}nez-Teja},
  Y.}, \bibinfo{author}{{Coelho}, P.}, \bibinfo{author}{{Pereira}, E.},
  \bibinfo{author}{{Montero-Dorta}, A.D.}, \bibinfo{author}{{Ben{\'\i}tez},
  N.}, \bibinfo{author}{{Gon{\c{c}}alves}, T.S.},
  \bibinfo{author}{{Santana-Silva}, L.}, \bibinfo{author}{{Werner}, S.V.},
  \bibinfo{author}{{Almeida}, L.A.}, \bibinfo{author}{{Lopes}, P.A.A.},
  \bibinfo{author}{{Chies-Santos}, A.L.}, \bibinfo{author}{{Telles}, E.},
  \bibinfo{author}{{de Souza}, T.}, \bibinfo{author}{{C}, R.},
  \bibinfo{author}{{Gon{\c{c}}alves}, D.R.}, \bibinfo{author}{{de Souza},
  R.S.}, \bibinfo{author}{{Makler}, M.}, \bibinfo{author}{{Buzzo}, M.L.},
  \bibinfo{author}{{Placco}, V.M.}, \bibinfo{author}{{Nakazono}, L.M.I.},
  \bibinfo{author}{{Saito}, R.K.}, \bibinfo{author}{{Overzier}, R.A.},
  \bibinfo{author}{{Abramo}, L.R.}, \bibinfo{year}{2020}.
\newblock \bibinfo{title}{{Assessing the photometric redshift precision of the
  S-PLUS survey: the Stripe-82 as a test-case}}.
\newblock \bibinfo{journal}{\mnras} \bibinfo{volume}{499},
  \bibinfo{pages}{3884--3908}.
\newblock \DOIprefix\doi{10.1093/mnras/staa1586},
  \href{http://arxiv.org/abs/1907.06315}{\tt arXiv:1907.06315}.
\bibitem[{Paszke et~al.(2017)Paszke, Gross, Chintala, Chanan, Yang, DeVito,
  Lin, Desmaison, Antiga and Lerer}]{PyTorch}
\bibinfo{author}{Paszke, A.}, \bibinfo{author}{Gross, S.},
  \bibinfo{author}{Chintala, S.}, \bibinfo{author}{Chanan, G.},
  \bibinfo{author}{Yang, E.}, \bibinfo{author}{DeVito, Z.},
  \bibinfo{author}{Lin, Z.}, \bibinfo{author}{Desmaison, A.},
  \bibinfo{author}{Antiga, L.}, \bibinfo{author}{Lerer, A.},
  \bibinfo{year}{2017}.
\newblock \bibinfo{title}{Automatic differentiation in pytorch}, in:
  \bibinfo{booktitle}{NIPS-W}.
\bibitem[{Pharo et~al.(2018)Pharo, Malhotra, Rhoads, Ryan, Tilvi, Pirzkal,
  Finkelstein, Windhorst, Grogin, Koekemoer, Zheng, Hathi, Kim, Joshi, Yang,
  Christensen, Cimatti, Gardner, Zakamska, Ferreras, Hibon and
  Pasquali}]{Pharo2018}
\bibinfo{author}{Pharo, J.}, \bibinfo{author}{Malhotra, S.},
  \bibinfo{author}{Rhoads, J.}, \bibinfo{author}{Ryan, R.},
  \bibinfo{author}{Tilvi, V.}, \bibinfo{author}{Pirzkal, N.},
  \bibinfo{author}{Finkelstein, S.}, \bibinfo{author}{Windhorst, R.},
  \bibinfo{author}{Grogin, N.}, \bibinfo{author}{Koekemoer, A.},
  \bibinfo{author}{Zheng, Z.}, \bibinfo{author}{Hathi, N.},
  \bibinfo{author}{Kim, K.}, \bibinfo{author}{Joshi, B.},
  \bibinfo{author}{Yang, H.}, \bibinfo{author}{Christensen, L.},
  \bibinfo{author}{Cimatti, A.}, \bibinfo{author}{Gardner, J.P.},
  \bibinfo{author}{Zakamska, N.}, \bibinfo{author}{Ferreras, I.},
  \bibinfo{author}{Hibon, P.}, \bibinfo{author}{Pasquali, A.},
  \bibinfo{year}{2018}.
\newblock \bibinfo{title}{Spectrophotometric redshifts in the faint infrared
  grism survey: Finding overdensities of faint galaxies}.
\newblock \bibinfo{journal}{The Astrophysical Journal} \bibinfo{volume}{856},
  \bibinfo{pages}{116}.
\newblock \URLprefix \url{https://doi.org/10.3847/1538-4357/aaadfc},
  \DOIprefix\doi{10.3847/1538-4357/aaadfc}.
\bibitem[{{Polsterer} et~al.(2016){Polsterer}, {D'Isanto} and
  {Gieseke}}]{Polsterer2016}
\bibinfo{author}{{Polsterer}, K.L.}, \bibinfo{author}{{D'Isanto}, A.},
  \bibinfo{author}{{Gieseke}, F.}, \bibinfo{year}{2016}.
\newblock \bibinfo{title}{{Uncertain Photometric Redshifts}}.
\newblock \bibinfo{journal}{arXiv e-prints} ,
  \bibinfo{pages}{arXiv:1608.08016}\href{http://arxiv.org/abs/1608.08016}{\tt
  arXiv:1608.08016}.
\bibitem[{Rosenblatt(1958)}]{Rosenblatt}
\bibinfo{author}{Rosenblatt, F.}, \bibinfo{year}{1958}.
\newblock \bibinfo{title}{The perceptron: A probabilistic model for information
  storage and organization in the brain}.
\newblock \bibinfo{journal}{Psychological Review} , \bibinfo{pages}{65--386}.
\bibitem[{Rozo et~al.(2009)Rozo, Wechsler, Rykoff, Annis, Becker, Evrard,
  Frieman, Hansen, Hao, Johnston, Koester, McKay, Sheldon and
  Weinberg}]{Rozo2009}
\bibinfo{author}{Rozo, E.}, \bibinfo{author}{Wechsler, R.H.},
  \bibinfo{author}{Rykoff, E.S.}, \bibinfo{author}{Annis, J.T.},
  \bibinfo{author}{Becker, M.R.}, \bibinfo{author}{Evrard, A.E.},
  \bibinfo{author}{Frieman, J.A.}, \bibinfo{author}{Hansen, S.M.},
  \bibinfo{author}{Hao, J.}, \bibinfo{author}{Johnston, D.E.},
  \bibinfo{author}{Koester, B.P.}, \bibinfo{author}{McKay, T.A.},
  \bibinfo{author}{Sheldon, E.S.}, \bibinfo{author}{Weinberg, D.H.},
  \bibinfo{year}{2009}.
\newblock \bibinfo{title}{Cosmological constrains from the sdss {MaxBCG}
  cluster catalog}.
\newblock \bibinfo{journal}{The Astrophysical Journal} \bibinfo{volume}{708},
  \bibinfo{pages}{645--660}.
\newblock \URLprefix \url{https://doi.org/10.1088/0004-637x/708/1/645},
  \DOIprefix\doi{10.1088/0004-637x/708/1/645}.
\bibitem[{Sadeh et~al.(2016)Sadeh, Abdalla and Lahav}]{Sadeh2016}
\bibinfo{author}{Sadeh, I.}, \bibinfo{author}{Abdalla, F.B.},
  \bibinfo{author}{Lahav, O.}, \bibinfo{year}{2016}.
\newblock \bibinfo{title}{{ANNz}2: Photometric redshift and probability
  distribution function estimation using machine learning}.
\newblock \bibinfo{journal}{Publications of the Astronomical Society of the
  Pacific} \bibinfo{volume}{128}, \bibinfo{pages}{104502}.
\newblock \URLprefix \url{https://doi.org/10.1088/1538-3873/128/968/104502},
  \DOIprefix\doi{10.1088/1538-3873/128/968/104502}.
\bibitem[{S{\'{a}}nchez(2010)}]{DES}
\bibinfo{author}{S{\'{a}}nchez, E.}, \bibinfo{year}{2010}.
\newblock \bibinfo{title}{The dark energy survey}.
\newblock \bibinfo{journal}{Journal of Physics: Conference Series}
  \bibinfo{volume}{259}, \bibinfo{pages}{012080}.
\newblock \URLprefix \url{https://doi.org/10.1088/1742-6596/259/1/012080},
  \DOIprefix\doi{10.1088/1742-6596/259/1/012080}.
\bibitem[{Speagle and Eisenstein(2017)}]{Speagle2017}
\bibinfo{author}{Speagle, J.S.}, \bibinfo{author}{Eisenstein, D.J.},
  \bibinfo{year}{2017}.
\newblock \bibinfo{title}{Deriving photometric redshifts using fuzzy archetypes
  and self-organizing maps {\textendash} i. methodology}.
\newblock \bibinfo{journal}{Monthly Notices of the Royal Astronomical Society}
  \bibinfo{volume}{469}, \bibinfo{pages}{1186--1204}.
\newblock \URLprefix \url{https://doi.org/10.1093/mnras/stw1485},
  \DOIprefix\doi{10.1093/mnras/stw1485}.
\bibitem[{{Taylor}(2005)}]{TOPCAT}
\bibinfo{author}{{Taylor}, M.B.}, \bibinfo{year}{2005}.
\newblock \bibinfo{title}{{TOPCAT \& STIL: Starlink Table/VOTable Processing
  Software}}, in: \bibinfo{editor}{{Shopbell}, P.}, \bibinfo{editor}{{Britton},
  M.}, \bibinfo{editor}{{Ebert}, R.} (Eds.), \bibinfo{booktitle}{Astronomical
  Data Analysis Software and Systems XIV}, p.~\bibinfo{pages}{29}.
\bibitem[{{The Theano Development Team} et~al.(2016){The Theano Development
  Team}, {Al-Rfou}, {Alain}, {Almahairi}, {Angermueller}, {Bahdanau}, {Ballas},
  {Bastien}, {Bayer}, {Belikov}, {Belopolsky}, {Bengio}, {Bergeron},
  {Bergstra}, {Bisson}, {Bleecher Snyder}, {Bouchard}, {Boulanger-Lewandowski},
  {Bouthillier}, {de Br{\'e}bisson}, {Breuleux}, {Carrier}, {Cho}, {Chorowski},
  {Christiano}, {Cooijmans}, {C{\^o}t{\'e}}, {C{\^o}t{\'e}}, {Courville},
  {Dauphin}, {Delalleau}, {Demouth}, {Desjardins}, {Dieleman}, {Dinh},
  {Ducoffe}, {Dumoulin}, {Ebrahimi Kahou}, {Erhan}, {Fan}, {Firat}, {Germain},
  {Glorot}, {Goodfellow}, {Graham}, {Gulcehre}, {Hamel}, {Harlouchet}, {Heng},
  {Hidasi}, {Honari}, {Jain}, {Jean}, {Jia}, {Korobov}, {Kulkarni}, {Lamb},
  {Lamblin}, {Larsen}, {Laurent}, {Lee}, {Lefrancois}, {Lemieux},
  {L{\'e}onard}, {Lin}, {Livezey}, {Lorenz}, {Lowin}, {Ma}, {Manzagol},
  {Mastropietro}, {McGibbon}, {Memisevic}, {van Merri{\"e}nboer}, {Michalski},
  {Mirza}, {Orlandi}, {Pal}, {Pascanu}, {Pezeshki}, {Raffel}, {Renshaw},
  {Rocklin}, {Romero}, {Roth}, {Sadowski}, {Salvatier}, {Savard},
  {Schl{\"u}ter}, {Schulman}, {Schwartz}, {Vlad Serban}, {Serdyuk},
  {Shabanian}, {Simon}, {Spieckermann}, {Ramana Subramanyam}, {Sygnowski},
  {Tanguay}, {van Tulder}, {Turian}, {Urban}, {Vincent}, {Visin}, {de Vries},
  {Warde-Farley}, {Webb}, {Willson}, {Xu}, {Xue}, {Yao}, {Zhang} and
  {Zhang}}]{Theano}
\bibinfo{author}{{The Theano Development Team}}, \bibinfo{author}{{Al-Rfou},
  R.}, \bibinfo{author}{{Alain}, G.}, \bibinfo{author}{{Almahairi}, A.},
  \bibinfo{author}{{Angermueller}, C.}, \bibinfo{author}{{Bahdanau}, D.},
  \bibinfo{author}{{Ballas}, N.}, \bibinfo{author}{{Bastien}, F.},
  \bibinfo{author}{{Bayer}, J.}, \bibinfo{author}{{Belikov}, A.},
  \bibinfo{author}{{Belopolsky}, A.}, \bibinfo{author}{{Bengio}, Y.},
  \bibinfo{author}{{Bergeron}, A.}, \bibinfo{author}{{Bergstra}, J.},
  \bibinfo{author}{{Bisson}, V.}, \bibinfo{author}{{Bleecher Snyder}, J.},
  \bibinfo{author}{{Bouchard}, N.}, \bibinfo{author}{{Boulanger-Lewandowski},
  N.}, \bibinfo{author}{{Bouthillier}, X.}, \bibinfo{author}{{de
  Br{\'e}bisson}, A.}, \bibinfo{author}{{Breuleux}, O.},
  \bibinfo{author}{{Carrier}, P.L.}, \bibinfo{author}{{Cho}, K.},
  \bibinfo{author}{{Chorowski}, J.}, \bibinfo{author}{{Christiano}, P.},
  \bibinfo{author}{{Cooijmans}, T.}, \bibinfo{author}{{C{\^o}t{\'e}}, M.A.},
  \bibinfo{author}{{C{\^o}t{\'e}}, M.}, \bibinfo{author}{{Courville}, A.},
  \bibinfo{author}{{Dauphin}, Y.N.}, \bibinfo{author}{{Delalleau}, O.},
  \bibinfo{author}{{Demouth}, J.}, \bibinfo{author}{{Desjardins}, G.},
  \bibinfo{author}{{Dieleman}, S.}, \bibinfo{author}{{Dinh}, L.},
  \bibinfo{author}{{Ducoffe}, M.}, \bibinfo{author}{{Dumoulin}, V.},
  \bibinfo{author}{{Ebrahimi Kahou}, S.}, \bibinfo{author}{{Erhan}, D.},
  \bibinfo{author}{{Fan}, Z.}, \bibinfo{author}{{Firat}, O.},
  \bibinfo{author}{{Germain}, M.}, \bibinfo{author}{{Glorot}, X.},
  \bibinfo{author}{{Goodfellow}, I.}, \bibinfo{author}{{Graham}, M.},
  \bibinfo{author}{{Gulcehre}, C.}, \bibinfo{author}{{Hamel}, P.},
  \bibinfo{author}{{Harlouchet}, I.}, \bibinfo{author}{{Heng}, J.P.},
  \bibinfo{author}{{Hidasi}, B.}, \bibinfo{author}{{Honari}, S.},
  \bibinfo{author}{{Jain}, A.}, \bibinfo{author}{{Jean}, S.},
  \bibinfo{author}{{Jia}, K.}, \bibinfo{author}{{Korobov}, M.},
  \bibinfo{author}{{Kulkarni}, V.}, \bibinfo{author}{{Lamb}, A.},
  \bibinfo{author}{{Lamblin}, P.}, \bibinfo{author}{{Larsen}, E.},
  \bibinfo{author}{{Laurent}, C.}, \bibinfo{author}{{Lee}, S.},
  \bibinfo{author}{{Lefrancois}, S.}, \bibinfo{author}{{Lemieux}, S.},
  \bibinfo{author}{{L{\'e}onard}, N.}, \bibinfo{author}{{Lin}, Z.},
  \bibinfo{author}{{Livezey}, J.A.}, \bibinfo{author}{{Lorenz}, C.},
  \bibinfo{author}{{Lowin}, J.}, \bibinfo{author}{{Ma}, Q.},
  \bibinfo{author}{{Manzagol}, P.A.}, \bibinfo{author}{{Mastropietro}, O.},
  \bibinfo{author}{{McGibbon}, R.T.}, \bibinfo{author}{{Memisevic}, R.},
  \bibinfo{author}{{van Merri{\"e}nboer}, B.}, \bibinfo{author}{{Michalski},
  V.}, \bibinfo{author}{{Mirza}, M.}, \bibinfo{author}{{Orlandi}, A.},
  \bibinfo{author}{{Pal}, C.}, \bibinfo{author}{{Pascanu}, R.},
  \bibinfo{author}{{Pezeshki}, M.}, \bibinfo{author}{{Raffel}, C.},
  \bibinfo{author}{{Renshaw}, D.}, \bibinfo{author}{{Rocklin}, M.},
  \bibinfo{author}{{Romero}, A.}, \bibinfo{author}{{Roth}, M.},
  \bibinfo{author}{{Sadowski}, P.}, \bibinfo{author}{{Salvatier}, J.},
  \bibinfo{author}{{Savard}, F.}, \bibinfo{author}{{Schl{\"u}ter}, J.},
  \bibinfo{author}{{Schulman}, J.}, \bibinfo{author}{{Schwartz}, G.},
  \bibinfo{author}{{Vlad Serban}, I.}, \bibinfo{author}{{Serdyuk}, D.},
  \bibinfo{author}{{Shabanian}, S.}, \bibinfo{author}{{Simon}, {\'E}.},
  \bibinfo{author}{{Spieckermann}, S.}, \bibinfo{author}{{Ramana Subramanyam},
  S.}, \bibinfo{author}{{Sygnowski}, J.}, \bibinfo{author}{{Tanguay}, J.},
  \bibinfo{author}{{van Tulder}, G.}, \bibinfo{author}{{Turian}, J.},
  \bibinfo{author}{{Urban}, S.}, \bibinfo{author}{{Vincent}, P.},
  \bibinfo{author}{{Visin}, F.}, \bibinfo{author}{{de Vries}, H.},
  \bibinfo{author}{{Warde-Farley}, D.}, \bibinfo{author}{{Webb}, D.J.},
  \bibinfo{author}{{Willson}, M.}, \bibinfo{author}{{Xu}, K.},
  \bibinfo{author}{{Xue}, L.}, \bibinfo{author}{{Yao}, L.},
  \bibinfo{author}{{Zhang}, S.}, \bibinfo{author}{{Zhang}, Y.},
  \bibinfo{year}{2016}.
\newblock \bibinfo{title}{{Theano: A Python framework for fast computation of
  mathematical expressions}}.
\newblock \bibinfo{journal}{arXiv e-prints} ,
  \bibinfo{pages}{arXiv:1605.02688}\href{http://arxiv.org/abs/1605.02688}{\tt
  arXiv:1605.02688}.
\bibitem[{Weinberg et~al.(2013)Weinberg, Mortonson, Eisenstein, Hirata, Riess
  and Rozo}]{Weinberg2013}
\bibinfo{author}{Weinberg, D.H.}, \bibinfo{author}{Mortonson, M.J.},
  \bibinfo{author}{Eisenstein, D.J.}, \bibinfo{author}{Hirata, C.},
  \bibinfo{author}{Riess, A.G.}, \bibinfo{author}{Rozo, E.},
  \bibinfo{year}{2013}.
\newblock \bibinfo{title}{Observational probes of cosmic acceleration}.
\newblock \bibinfo{journal}{Physics Reports} \bibinfo{volume}{530},
  \bibinfo{pages}{87--255}.
\newblock \URLprefix \url{https://doi.org/10.1016/j.physrep.2013.05.001},
  \DOIprefix\doi{10.1016/j.physrep.2013.05.001}.
\bibitem[{Williams and Rasmussen(1995)}]{GP1995}
\bibinfo{author}{Williams, C.K.I.}, \bibinfo{author}{Rasmussen, C.E.},
  \bibinfo{year}{1995}.
\newblock \bibinfo{title}{Gaussian processes for regression}, in:
  \bibinfo{booktitle}{Proceedings of the 8th International Conference on Neural
  Information Processing Systems}, \bibinfo{publisher}{MIT Press},
  \bibinfo{address}{Cambridge, MA, USA}. p. \bibinfo{pages}{514–520}.

\end{thebibliography}

\end{document}